\input harvmac
\input epsf

\lref\sen{A. Sen, ``F-Theory and Orientifolds,'' {\tt hep-th/9605150},
Nucl. Phys. B475 (1996) 562-578}

\lref\pol{J. Polchinski, S. Chaudhuri and C. Johnson, `` Notes on D-Branes,''
{\tt hep-th/9602052}}

\lref\scs{B.R.Greene, A. Shapere, C. Vafa and S.-T. Yau, ``Stringy
Cosmic Strings and Non-compact Calabi-Yau Manifolds,'' Nucl. Phys.
B337 (1990) 1}

\lref\lust{G. Cardoso, G. Curio, D. L\"{u}st and T. Mohaupt, ``On The
Duality Between The heterotic String and F-Theory in 8 Dimensions,'' {\tt
hep-th/9609111}, Phys. Lett. B389 (1996) 479-484}

\lref\lerche{W. Lerche and S. Stieberger, ``Prepotential, Mirror Map and
F-Theory on $K3$,'' {\tt hep-th/9804176}, Adv. Theor. Math. Phys. 2 (1998)
1105-1140}

\lref\yamada{Y. Yamada and S.K. Yang, ``Affine 7-brane Backgrounds and
Five-Dimensional $E_N$ Theories on $S^1$,'' {\tt hep-th/9907134}}

\lref\wittdu{E. Witten, ``String Theory Dynamics In Various
Dimensions,'' {\tt hep-th/9503124} Nucl. Phys. B443 (1995) 85-126}

\lref\mov{D. Morrison and C. Vafa, ``Compactifications of F-Theory on
Calabi-Yau Threefolds -- I,II,'' {\tt hep-th/9603161, 9603161},
Nucl. Phys. B473 (1996) 74-92, Nucl. Phys. B476 (1996) 437-469}

\lref\kuv{A. Kumar and C. Vafa, ``U-Manifolds,'' {\tt hep-th/9611007},
Phys. Lett. B396 (1997) 85-90}

\lref\nsv{K.Narain, M. Sarmadi and C. Vafa, ``Asymmetric Orbifolds,''
Nucl. Phys. B 288 (1987) 551}

\lref\dh{A. Dabholkar and J. A. Harvey, ``String Islands,'' {\tt hep-th/9809122}}

\lref\wittenfl{E. Witten, ``Toroidal Compactification Without Vector
Structure,'' {\tt hep-th/9712028}, JHEP 9802 (1998) 006}

\lref\bereta{M. Bershadsky, T. Pantev and V. Sadov,
``F-theory with quantized fluxes,'' {\tt
hep-th/9805056}}

\lref\kle{S. Kachru, A. Klemm and Y. Oz,``Calabi-Yau Duals for CHL Strings,''
{\tt  hep-th/9712035}, Nucl.Phys. B521 (1998) 58-70}

\lref\banet{T. Banks, W. Fischler and L. Motl, ``Duality versus
Singularities,'' {\tt hep-th/9811194}, JHEP 9901 (1999) 019}

\lref\kacv{S. Kachru and C. Vafa, ``Exact Results for N=2 Compactifications of
Heterotic Strings,'' {\tt hep-th/9505105}, Nucl. Phys. B450 (1995) 69-89 }

\lref\fhsv{S. Ferrara, J. A. Harvey, A. Strominger and C. Vafa,
``Second-Quantized Mirror Symmetry,'' {\tt
hep-th/9505162}, Phys.Lett. B361 (1995) 59-65}

\lref\klm{A. Klemm, W. Lerche and P. Mayr, ``K3 -- Fibrations and
heterotic-Type II String Duality,'' {\tt hep-th/9506091}, Phys. Lett. B357
(1995) 313-322}

\lref\vwi{C. Vafa and E. Witten, ``Dual String Pairs With N=1 And N=2
Supersymmetry In Four Dimensions,'' {\tt hep-th/9507050},
Nucl. Phys. Proc. Suppl. 46 (1996) 225-247}

\lref\Yauet{S.Hosono, B.H.Lian and S.-T.Yau, `` Calabi-Yau Varieties and
Pencils of K3 Surfaces,''
{\tt alg-geom/9603020}}

\lref\aspinwall{P. Aspinwall, ``M-Theory Versus F-Theory Pictures of the
Heterotic String,'' {\tt hep-th/9707014}, Adv. Theor. Math. Phys. 1 (1998) 127-147}

\lref\dall{D. Allcock, private communication.}

\lref\powi{J. Polchinski and E. Witten, ``Evidence for heterotic-Type I
String Duality,'' {\tt hep-th/9510169}, Nucl. Phys. B460 (1996) 525-540}

\lref\howi{P. Horava and E. Witten, ``heterotic and Type I String Dynamics
from Eleven Dimensions,'' {\tt hep-th/9510209}, Nucl. Phys. B460 (1996) 506-524 and ``Eleven-Dimensional
Supergravity on a Manifold with Boundary'',{\tt hep-th/9603142},
Nucl. Phys. B475 (1996) 94-114}

\lref\mohaup{T. Mohaupt, ``Critical Wilson Lines in Toroidal
Compactifications of Heterotic Strings,'' Int. J. Mod. Phys. A8 (1993) 3529-3552}

\lref\nase{K.S.Narain, ``New heterotic Theories in Uncompactified
Dimensions $<$ 10,'' Phys. Lett. 169B,41 (1986). K.S.Narain, M.H.Sarmadi, and
E. Witten, ``A Note on Toroidal Compactification of heterotic String
Theory,'' Nucl. Phys.B 279.  J.P. Serre, A course in Arithmetic (Springer,
Berlin, 1973)}

\lref\zerobrane{O. Bergman, M. Gaberdiel and G. Lifschytz, ``String
Creation and heterotic-Type I' Duality,'' {\tt hep-th/9711098},
Nucl. Phys. B524 (1998) 524-544}

\lref\seib{N. Seiberg, ``Five Dimensional SUSY Field Theories, Non-trivial
Fixed Points and String Dynamics,'' {\tt hep-th/9608111}, Phys. Lett. B388
(1996) 753-760}

\lref\dkv{M. Douglas, S. Katz and C. Vafa, ``Small Instantons, del Pezzo
Surfaces and Type I' theory,'' {\tt 9609071}, Nucl. Phys. B497 (1997) 155-172}

\lref\ms{D. Morrison and N. Seiberg, ``Extremal Transitions and
Five-Dimensional Supersymmetric Field Theories,'' {\tt hep-th/9609070},
Nucl. Phys. B483 (1997) 229-247}

\lref\greenet{C. Bachas, M. Green and A. Schwimmer, ``(8,0) Quantum mechanics
and symmetry enhancements in type I' superstrings,'' {\tt hep-th/9712086},
JHEP 9801 (1998) 006}

\lref\aspin{P. Aspinwall, ``Enhanced Gauge Symmetries and $K3$ Surfaces,''
{\tt hep-th/9507012}}

\lref\nik{V. Nikulin, in {\it Proc. Int. Congress of Mathematicians},
University of California, Berkeley, 1986, p. 654.}

\lref\bartetsen{A. Sen and B. Zwiebach, ``Stable Non-BPS States in
F-theory,'' {\tt hep-th/9907164}}

\lref\wolfe{O. De Wolfe, T. Hauer, A. Iqbal and B. Zwiebach,
``Uncovering the Symmetries on $[p,q]$ 7-branes: Beyond the Kodaira
Classification'', {\tt hep-th/9812028}. ``Uncovering Infinite Symmetries on
$[p,q]$ 7-branes: Kac-Moody Algebras and Beyond'', {\tt
hep-th/9812209}}

\lref\morrkatz{S. Katz and D.R. Morrison, ``Gorenstein Threefold
Singularities with Small Resolutions Via Invariant Theory for Weyl
Groups,'' Jour. Alg. Geom. 1 (1992) 449}

\lref\fth{C. Vafa, ``Evidence for F-Theory,''{\tt hep-th/9602022},
Nucl. Phys. B469 (1996) 403-418}

\lref\ginsparg{P. Ginsparg, ``On Toroidal Compactification of Heterotic
Superstrings,'' Phys. Rev. D35 (1987) 648-654}


\skip0=\baselineskip
\divide\skip0 by 2

\def\tmpsp{\the\skip0}

\let\linesp=\mylinesp

\def\skipthis#1{{}}

\def\a{\alpha}
\def\b{\beta}
\def\t{\theta}
\def\IR{\relax{\rm I\kern-.18em R}}
\def\IZ{\relax\ifmmode\hbox{Z\kern-.4em Z}\else{Z\kern-.4em Z}\fi}
\def\IQ{\relax{\rm I\kern-.40em Q}}
\def\IS{\relax{\rm I\kern-.18em S}}

\def\E{E_8 \times E_8}
\def\S{SO(32)}
\def\Sp{Spin(32)/Z_2}
\def\ep{\varepsilon}
\def\tr{{\rm Im}(\tau )}

\Title{\vbox{\baselineskip12pt\hbox{hep-th/0001029}\hbox{}
\hbox{HUTP-99/A058}}}{Type I' and Real Algebraic Geometry}

\centerline{Freddy Alexander Cachazo and Cumrun Vafa}
\bigskip\centerline{Jefferson Physical Laboratory}
\centerline{Harvard University}
\centerline{Cambridge, MA 02138}

\vskip .3in \centerline{\bf Abstract}
We revisit the duality between type I' and heterotic
strings in 9 dimensions.  We resolve a puzzle about
the validity of type I' perturbation theory and show
that there are regions in moduli which are not
within the reach of type I' perturbation theory. 
 We find 
however, that all regions of moduli are described by
a special class of real elliptic $K3$'s in the limit
where the $K3$ shrinks to a one dimensional interval.  We
 find a precise  map
between the 
geometry of dilaton and branes of type I' 
on the one hand and the geometry of real elliptic
$K3$ on the other.  We
also argue more generally that strong coupling limits of string
compactifications generically do not have a weakly coupled dual in terms
of any known theory (as is exemplified by the strong coupling
limit of heterotic strings in 9 dimensions for certain range of parameters).
\smallskip
\Date{January 2000}
\newsec{Introduction}

Thanks to the discovery of duality symmetries in string theory
we now understand in many cases what the light degrees of freedom
in a string theory 
are in various regimes of coupling constant and compactification
geometry of string theory.  In this way one has been able
to connect various theories to each other in an unexpected way.
In many cases this leads to a unified picture of string theory
suggesting there is a unique underlying theory with different manifestations
in various regimes, unifying Type IIA, B, type I, heterotic and 11
dimensional M-theory, in a single framework.

Indeed, in the case of maximal number of supersymmetries ($N=32$)
in various dimensions, Witten raised the following question \wittdu\ :
 If we consider strong coupling regime of type II strings
compactified on tori, then in principle we can discover new consistent
theories.  It was very surprising that by considering various
limits leaving at least 4 non-compact spacetime dimensions \wittdu\
one ended up with a theory which had a simple description
in terms of compactifications of known string theories or 11 dimensional
supergravity, M-theory.  In other words a consequence of \wittdu\
was the discovery of a single new 11 dimensional M-theory, which
together with other string theories in the case of maximal number of
supersymmetries gives a complete description of all boundaries of moduli
space of theories with $N=32$ supercharges.  

However this leaves open the possibility that if we consider
other cases, for example compactifications with less supersymmetry,
we may discover new theories by considering their strong coupling limit.
  In fact an example of this situation was discovered
in \fth\ where by considering heterotic string
compactified on $T^2$ one ended up
in the strong coupling regime with a theory which did not
have a well defined description in terms of a single string theory.
The new theory, F-theory, put together various $(p,q)$ type IIB
strings which are non-perturbative relative to each other in a
single compactification.  It involved using $(p,q)$ 
7-branes of type IIB taking advantage of the non-perturbative
U-duality group of type IIB, namely the $SL(2,{\bf Z})$. The
question raised was to come up with a complete low energy description
of this new theory.

The geometry of branes was encoded in terms of a limit of
elliptic $K3$ manifold suggesting a 12 dimensional origin.
However it is clear that a formulation in 12 dimension must
involve some new constructions (including the lack
of 12 dimensional Poincare invariance)
which, despite some progress, is still an open question.
The problem becomes acute when one considers compactifications
of F-theory with less supersymmetry.  For example to determine
even the massless degrees of freedom of F-theory on elliptic
Calabi-Yau threefold, one has to appeal to various consistency
conditions (including anomaly cancellations in the chiral
6-dimensional theory) to predict the spectrum \mov .  This clearly
is unsatisfactory and one would like to have a more direct
approach in finding the light degrees of freedom.  Thus the
new theory discovered is more mysterious than superstring theories
or M-theory.

The fact that different limits of various other string compactification may
exist which have no interpretation in terms of M-theory,
String theories, or F-theory, was already pointed out in \kuv\
where it was suggested that the strong coupling limit of asymmetric
orbifold compactifications of string theory \nsv\
 would provide such examples.
In fact many interesting such examples have been constructed
in \dh\ which in strong coupling
regimes may define new theories.  One can also use U-dualities
in the form of an orbifold to construct F-theory like
theories which in many cases correspond to new
theories
\kuv\  which do not have any conventional known dual
analog.  Various other F-theory like theories
which correspond to various specific $K3$  geometric duals has
also been considered which correspond to new theories
\wittenfl \bereta \kle .

To obtain limits which have no interpretation
in terms of M-theory or string theories, one can also
use conventional compactification geometries.  For example
even with $N=32$ supercharges, it was shown in \banet\ that if we
 consider toroidal compactification
leaving 2 or less non-compactified dimensions
similar thing happens.

In fact there are more such examples involving compactification
to 4 dimensions.  Consider Type IIA string theory compactified on
Calabi-Yau threefolds. Let us assume that the threefold is neither
elliptically fibered, nor K3 fibered.  Let us consider the
limit of strong coupling of this theory fixing the volume
of Calabi-Yau in string frame\foot{In terms
of M-theory this corresponds to compactification
on a large circle times a
Calabi-Yau threefold with infinitesimal volume.}.  Then we do
not have a candidate for a dual theory.  In fact
if there is a dual theory description
involving known theories, this will be a new duality which
cannot be related to other known dualities using adiabatic
principle.  This is because all the known dualities will
involve $K3$ or elliptic compactifications.  In particular the
known duality of type IIA on Calabi-Yau threefolds and
heterotic on $K3\times T^2$
\kacv\ (see also \fhsv )
goes through $K3$ or elliptic fibered Calabi-Yau
manifolds \klm\
and this can be related to string dualities in 6 dimensions
using the adiabatic principle \vwi .
Thus it is likely that for each Calabi-Yau threefold
which is not elliptic or $K3$ fibered, we end up defining
a new theory by considering a strong coupling limit
of type IIA compactification.  Given that in some sense
the generic Calabi-Yau threefold is not elliptic or
$K3$ fibered \Yauet\ we would conclude
that ``most'' type II compactifications with $N=2 $ supersymmetries
in $4d$ have strong coupling limits involving presumably
unknown theories.

It is thus apparent that the unifying framework
to consider all string theories will have various
unrecognizable corners in the moduli space of
various compactifications in  addition to the ones
already known.  The main aim of the present paper is to
consider one such corner.  This is the compactification
of heterotic strings on a circle.  The parameters
characterizing this compactification, in addition
to heterotic string coupling constant, involve the radius
of the circle and the sixteen parameters specifying the
expectation value of Wilson
loops around the circle.  For certain regions
of moduli at strong coupling limit of heterotic string
there is a dual description in terms of type I'
theory \powi\ on an interval
related to the Type I--heterotic ($SO(32)$) duality 
in 10 dimensions \wittdu .
We will show that for specific
choices of Wilson loop variables and radii, in the strong coupling
limit there is no perturbative Type I' description.  Moreover we show
that, just as in the case of F-theory, all regimes of parameters
can be usefully characterized by the geometry of a particular
class of {\it real} elliptic rational $K3$ surface, as was anticipated
in \mov .
We will also see that this real $K3$, has a natural and precise relation
with the geometry of dilaton and branes of type I'. 

The organization of this paper is as follows:
In section 2 we review the theories dual to heterotic strings
in 10,9 and 8 dimensions.  In section 3 we discuss a puzzle
in the case of type I' dual of heterotic string in 9 dimensions. 
In section 4 we
discuss global aspects of moduli of type I' (or heterotic) theory
in 9 dimensions.  In section 5 we resolve the puzzles raised
in section 3, and indicate why the regime
of validity of type I' perturbation theory misses some regions
of moduli space.  In section 6 we consider the limit
of F-theory corresponding to decompactifying one circle
and show how the relevant limit is captured by a particular
type of real elliptic $K3$'s.  In section 7 we discuss how
real elliptic $K3$'s fills the gap in moduli space where
the type I' perturbation breaks down.  In particular
we recover extra branes postulated
by Morrison and Seiberg \ms\ predicted from duality with heterotic
strings.  In section 8 we give various explicit examples.  Finally
some details of the computations are presented
in appendices A and B.

\newsec{Heterotic dual theories in 10, 9 and 8 dimensions Reviewed}  

In 10 dimensions there are two inequivalent heterotic theories, one with
$\S$ gauge group and the other with $E_8\times E_8$. The strong 
coupling limit of the
former has a complete description in terms of the weak coupling limit of
Type I theory \wittdu . However, the latter does not have a conventional 
string
theory as its strong coupling limit but is instead dual to 
a compactification
of M-theory on $S^1/Z_2$ \howi .

When we go down in dimensions compactifying on $T^k$ we find that both
heterotic theories can be connected continuously, i.e., they belong to the
same moduli space. This is due to the uniqueness of Lorentzian self-dual
lattices \nase .

Strong coupling limit of heterotic
strings in the 9 dimensional case follows from
compactifying the 10 dimensional duality
between heterotic $\Sp$ and Type I on an
$S^1$. In this case, if we study the heterotic theory at strong coupling
and radius close to the critical radius we end up with Type I' (which is
T-dual to Type I).  

In 8 dimensions, the heterotic dual description is given in terms of
F-theory compactified on an elliptic $K3$ surface. This captures the Type
IIB compactification on $P^1$ with 24 (p,q) 7-branes.  It is
natural to ask about the connection between the 8 dimensional description
and the 9 dimensional one.  In particular one would like
to take the large radius limit of the 8 dimensional dual theories
and see what one ends up in 9 dimensions.

It is the aim of this section to review the known descriptions
of these theories in 8,9 and 10 dimensions and develop
the necessary relations that will be useful in the rest of this work.

\subsec{10 and 9 Dimensions}

Let us start by considering the 10 dimensional low energy effective actions in
the string frame for the heterotic $Spin(32)/Z_2$ and Type I theories. 
Since these two theories have $N=1$
supersymmetry and the same gauge group, the two actions should just be related
by a field redefinition.

For the heterotic string we have,
\eqn\het{S_{het}= \int d^{10}x \sqrt{-g_h} e^{-2 \phi_h}\left[ R_h +
\partial_\mu\phi\partial^\mu\phi -|H_3|^2 - Tr_v(|F^2_2|) \right]}
and using the following field redefinition,
\eqn\rede{g_{I\mu\nu}=e^{-\phi_h}g_{h\mu\nu} \;\;\;\;\;\;\; \phi_I=-\phi_h \;\;\;\;\;\;\; F_3 = H_3
\;\;\;\;\;\;\; A_{I1} = A_{h1}}
we get the Type I effective action,
\eqn\tyI{S_{I}= \int d^{10}x \sqrt{-g_I} e^{-2 \phi_I}\left[ R_I +
\partial_\mu\phi\partial^\mu\phi - |F_3|^2 \right] -\int e^{-\phi_I} Tr_v(|F_2^2|)}

Compactifying on a circle the 10 dimensional duality should give us
information about the heterotic strong coupling limit in 9 dimensions, and in fact, for big
enough heterotic radius this is the case. Nevertheless one of the most interesting
features of the heterotic string is the enhancement of the gauge symmetries
at some points in the moduli space where the radius is not much larger than
the string length. Using the above field redefinitions we can see that 
\eqn\rely{\lambda_h^{10}={1 \over 
\lambda_I^{10}}\;\;\;\;\;\;\;\;\;\; R_I={R_h \over (\lambda_h^{10})^{1/2}}}
Therefore for strongly coupled heterotic string $\lambda_h^{10}\gg 1$ 
and $R_h^2\simeq
R^2_{hc}=2(1-A^2/2)$, where $R_{hc}$ is 
the critical radius\foot{The value of the Regge slope for the
heterotic $SO(32)$ is taken to be $\alpha_h'=2$}, at which point new
massless gauge bosons appear, we get $R_I \ll 1$. This implies that we need
to perform a T-duality in order to understand the physics clearly. The
theory thus obtained is called the Type I'.

In general, we can think about Type I as a theory in ten dimensions
containing one orientifold 9-plane and 32 D9-branes. When we compactify on
a circle and perform a T-duality we get a type IIA
theory on $S^1/Z_2$ with two
orientifold 8-planes located at the fixed points of the $Z_2$ action and 16
D-8 branes at generic positions on the interval. At a generic point in the
moduli space we have an $U(1)^{18}$ gauge group. Where $U(1)^{16}$
corresponds to the positions of the branes, one $U(1)$ from the graviphoton
and the last $U(1)$ is related to the R-R one form.  Using the fact that when $(n)$
D-branes are on the top of each other we get an $U(n)$ enhancement of
the gauge group
and if in addition they are located at one of the orientifolds we get
$SO(2n)$,  it is easy to see that $SO(32)$ and all its regular subalgebras
can be obtained in this fashion \foot{
The information about all possible
gauge symmetries allowed in heterotic strings is nicely encoded in the extended Dynkin diagram of
$SO(32)$ as we will
discuss it in detail in section 4. }.

The map between the moduli spaces of heterotic strings
and type I' was worked out in \powi\ for certain
regions of parameter space which we will now review.
In the heterotic theory we have the 16 Wilson lines $\theta^I , I=1,
\ldots 16$, the radius $R_h$
and the coupling constant $\Lambda_h=e^{\phi_h}$. On the Type I' side we
have the 16 positions of the branes $x^I, I=1\ldots 16$, 
where $x$ stands for the
coordinate along the 
interval and runs from $0$ to $2\pi$, $B$ and $C$ that control the
behavior of the type I' dilaton at the orientifolds and the physical length
of the interval respectively. 

It turns out to be convenient to define the following function,
\eqn\zeto{ z(x) = {3 \over \sqrt{2}} (B+8 x_{cm} - {1 \over
2}\sum_{I=1}^{16}|x-x_I|) }     
where $x_{cm}={1\over 16}\sum x_I$ is the position of the center of mass of the 16 
D-8 branes.
The metric in string frame is given by, 
$g_{MN}=\Omega^2(x)\eta_{MN}$, and the dilaton of Type I' are given by,
\eqn\abc{e^{\phi_{I'}}=(C z(x))^{-5/6}, \;\;\;\;\;\; \Omega (x)=C^{5/6} z(x)^{-1/6}}

Before writing down the explicit map between the heterotic and Type I'
moduli, let us express the Type I' dilaton as a function not of the
coordinate distance $x$ but of the proper distance measured from the
orientifold at $x=0$.

Let us call $\phi (\bar{x})$ the proper distance from $x=0$ to
$x=\bar{x}$. This is given by,
\eqn\proper{\phi (\bar{x})= {5\over 2^{3/2}} \int^{\bar{x}}_0 \Omega (y)dy}
where the numerical factor was introduced for later convenience. 

Let us define ${1 \over g(\phi)} = e^{-\phi_{I'}}$ to be
the coupling, $\phi_I$ to be the position of the $I^{th}$ brane in the
interval and ${1\over g_0} = (CB)^{5/6}$ to be the coupling at the
orientifold at $x=0$. 
The final answer is given by
\eqn\usual{{1 \over g(\phi )} = {1 \over g_0} + 8\phi_{cm}  - {1 \over
2}\sum_{I=1}^{16}|\phi -\phi_I| .}     

Let us now go back to the map of the moduli spaces between $SO(32)$
heterotic string and type I'.
The map was obtained in \powi\  by comparing the gravitational
 and gauge actions, the
mass of a K-K heterotic state and its corresponding dual type I' winding
state.  The heterotic radius is
given by, 
\eqn\eva{R_h = 2^{-3/4}\left( \int^{2\pi}_{0}dx z(x)^{1/3} \right)^{1/2} 
\left( \int^{2\pi}_{0}dx z(x)^{-1/3} \right)^{-1}} 
and the heterotic dilaton up to a numerical multiplicative 
constant\foot{The
constant contains some factors of $\alpha_{I'}'$.} is
\eqn\peet{e^{2 \phi_h}=C^{10/3}\left( \int^{2\pi}_{0}dx z(x)^{1/3}
 \right)^{3} \left( \int^{2\pi}_{0}dx z(x)^{-1/3} \right)^{-1}}

Finally, the Wilson lines and the positions of the branes can be related by
computing the mass of off-diagonal vector boson. Let $A=(\theta_1, \ldots
,\theta_{16})$ be the Wilson lines, then,
\eqn\Wilson{\theta_I = {1 \over 2}\left( \int^{x^I}_{0}dx z(x)^{-1/3} \right)
 \left( \int^{2\pi}_{0}dx z(x)^{-1/3} \right)^{-1}}

It is easy to see that for generic $x^I$ and $B$, the strong coupling limit
of the heterotic strings, i.e. $\lambda_h \gg 1$, can be obtained
 by taking $C \gg
1$. Moreover, the map allows us to compute $R_h$ and $\theta_I$
 only from $B$ and
$x^I$.  

Let us consider two examples that will be useful to illustrate how the map
works and how Type I' avoids possible contradictions at the points where
the heterotic is getting enhanced gauge symmetries.

Consider first the following set of Wilson lines $A =(0^n,
({1\over 2})^{16 - n})$ that was studied in \powi . This corresponds 
to having $n$ D8-branes at $x=0$
and $16-n$ D8-branes at $x=2\pi$. Using \eva\ we get,
\eqn\gracy{ R_h ={1\over 2}(8-n)^{1/2}{\;\;\;\left[ (B+2\pi
(8-n))^{4/3}-B^{4/3}\right]^{1/2}\over 
\left[ (B+2\pi (8-n))^{2/3}-B^{2/3}\right]}}
and from \abc\ the type I' dilaton is,
\eqn\dil{e^{\phi_{I'}} = \left[ B+(8-n)x_9 \right]^{-5/6}} 

As mentioned before, the behavior of the dilaton at $x=0$ is controlled by
$B$ and in particular it blows up for $B=0$. This is usually a sign that
something interesting should be happening on the dual heterotic theory. 
For $B=0$ we have, 
\eqn\Bo{e^{\phi_{I'}} \sim x_9^{-5/6} ~~~~~~~~~~~~~ R_h = 
{1\over 2}|n-8|^{1/2}}
But $R_h ={1\over 2}|n-8|^{1/2}$ is precisely the critical radius of the
heterotic string for the given Wilson line, i.e., $R_c^2 =2(1- A^2/2)$.  The gauge
group enhancements in each case are listed in the Table 1.

\bigskip
{\centerline{
\vbox{\offinterlineskip
\hrule
\halign{&\vrule#&
\strut\quad\hfil#\quad\cr
height2pt&\omit&&\omit&&\omit&&\omit&\cr
& \hfill   n   \hfill &&  \hfill  $G_o$  \hfill  && \hfill $R_c^2$  
\hfill && \hfill $G_{enhanced}$ \hfill & \cr
\noalign{\hrule}
height2pt&\omit&&\omit&&\omit&&\omit&\cr
& \hfill   7   \hfill &&  \hfill  $ SO(14)\times U(1)$  \hfill  && \hfill  1/4  \hfill && \hfill  $E_8$ \hfill & \cr
\noalign{\hrule}
height2pt&\omit&&\omit&&\omit&&\omit&\cr
& \hfill   6   \hfill &&  \hfill  $ SO(12)\times U(1)$  \hfill  && \hfill  1/2  \hfill && \hfill  $E_7$ \hfill & \cr
\noalign{\hrule}
height2pt&\omit&&\omit&&\omit&&\omit&\cr
&\hfill    5   \hfill &&  \hfill  $ SO(10)\times U(1)$  \hfill  && \hfill  3/4  \hfill && \hfill  $E_6$ \hfill & \cr
\noalign{\hrule}
height2pt&\omit&&\omit&&\omit&&\omit&\cr
&\hfill   4   \hfill &&  \hfill  $ SO(8)\times U(1)$  \hfill  && 
\hfill  1  \hfill && \hfill  $E_5 \simeq SO(10) $ \hfill & \cr
\noalign{\hrule}
height2pt&\omit&&\omit&&\omit&&\omit&\cr
& \hfill   3   \hfill &&  \hfill  $ SO(6)\times U(1)$  \hfill  && 
\hfill  5/4  \hfill && \hfill  $E_4 \simeq SU(5) $ \hfill & \cr
\noalign{\hrule}
height2pt&\omit&&\omit&&\omit&&\omit&\cr
& \hfill   2   \hfill &&  \hfill  $ SO(4)\times U(1)$  \hfill  && 
\hfill  3/2  \hfill && \hfill  $E_3 \simeq SU(3)\times SU(2) $ \hfill & \cr
\noalign{\hrule}
height2pt&\omit&&\omit&&\omit&&\omit&\cr
& \hfill   1   \hfill &&  \hfill  $ SO(2)\times U(1)$  \hfill  && \hfill 
 7/4  \hfill && \hfill  $E_2 \simeq SU(2)\times U(1) $ \hfill & \cr
\noalign{\hrule}
height2pt&\omit&&\omit&&\omit&&\omit&\cr
&\hfill   0   \hfill &&  \hfill  $U(1)$  \hfill  && \hfill  2  \hfill && \hfill  $E_1\simeq SU(2)$ \hfill & \cr
height2pt&\omit&&\omit&&\omit&&\omit&\cr}
\hrule}
}}
\noindent{\ninepoint\sl \baselineskip=8pt {\bf Table 1}: {\sl Gauge groups
$G_o$ at generic radius corresponding to Wilson lines of the form $A = (0^n,({1\over
2})^{16-n})$. Enhanced gauge groups $G_{enhanced}$ at the critical radius
$R_c^2 = {1\over 4} |n-8|$}.}
\bigskip

Therefore, we see that perturbation theory breaks down avoiding the
contradiction of having new massless states on the heterotic side that are
not in the perturbative spectrum of Type I'. It has been shown that the new
massless vector bosons of the heterotic string can be identified with
non-perturbative states of Type I'. In particular, we have D0-branes that
become massless at the orientifold with infinite
coupling \zerobrane \greenet .

The second example is given by the following Wilson line $A =
(0^{15},\lambda )$. This corresponds to 15 D8-branes at $x=0$ and one brane
whose position we denote by $x_1$. This is a particular case of the examples
studied in \zerobrane . The map is given by,
\eqn\oren{R_h = {1\over \sqrt{2}}{\left( {b^2-a^2 \over 7}+{a^2 -c^2 \over 8}\right) \over
\left( {b-a \over 7}+{a-c \over 8} \right)}^{1/2} ~~~~~ \lambda = \half
{ {(b-a)\over 7} \over {b-a \over 7}+{a-c \over 8} }}
where $a = (B-7x_1)^{2/3}$, $b = B^{2/3}$ and $c = (B+ x_1 -16\pi
)^{2/3}$. 

This configuration for generic $x_1$ and $B$ has $SO(30)$ as gauge
group. However, for special values of $x_1$ and $B$ enhancements of $SO(30)$ can be
obtained. This will be studied in detail in section 5. In particular,
for $x_1 = 0$ this is equivalent to the $n=0$ case of the first example.  

{\bf $E_8 \times E_8$ from Type I':}

Later in the paper we will need a more detailed description
for the map between heterotic string at the $E_8\times
E_8$ gauge symmetry enhancement point with the type I'
parameters.  In the above we discussed how one obtains
one extra $E_8$ symmetry by considering 7 branes on one orientifold
with infinite coupling.  If we did this on each orientifold
we would get $E_8\times E_8$.  In other words 
consider the following family of Wilson lines
$A =(0^7,{1\over 2}-\lambda , \lambda , ({1 \over 2})^7 )$ studied in
\pol . This corresponds
to having 7 D8-branes at $x=0$, 7 D8-branes at $x=2\pi$ and two more
D8-branes symmetrically located in the interval at positions $x_1$ and $x_2
= 2\pi - x_1$. This configuration generically corresponds to an unbroken
$SO(14)\times SO(14) \times U(1)^4$ gauge group.

The map in this case involves $R_h$, and $\lambda$ as 
functions of $B$ and $x^1$
and it is given by,
\eqn\eee{ R_h =2^{-3/2}3^{1/2}{\left[ 3(a^4-b^4)+4(\pi - x_1)a
                            \right]^{1/2}\over \left[ 3(a^2-b^2)+2(\pi
                            -x_1)a^{-1}\right] }}
where $a=(B+x_1)^{1/3}$ and $b=B^{1/3}$.

And,
\eqn\lamb{\lambda = {3\over 4}{a^2-b^2\over\left[ 3(a^2-b^2)+
2(\pi -x_1)a^{-1}\right] }}

The dilaton behaves as follows,
\eqn\fred{ e^{\phi_{I'}} = \left\{ \matrix{\hfill (B+x)^{-5/6} \hfill  &&& \hfill  0<x<x_1 \hfill \linesp 
                                          \hfill (B+x_1)^{-5/6} \hfill &&& \hfill x_1<x<x_2 \hfill \linesp                
                                          \hfill (B+2\pi -x)^{-5/6} \hfill
                                          &&& \hfill 
                                           x_2<x<2\pi \hfill } \right. }

It is clear that the $B \rightarrow 0$ limit is also very interesting in
this case. Indeed, for $B=0$ the dilaton blows up at both orientifold
points. This is a generic feature whenever the position of the center of
mass of the branes is in the middle of the interval, i.e., $x_{cm} = \pi$.

Let us see what the corresponding heterotic behavior is for $B=0$. From \eee\
and \lamb\ we get that,
\eqn\eeebo{ R^2_h={3\over 8}x_1{(4\pi -x_1)\over (2\pi + x_1)^2}
\;\;\;\;\;\;  \lambda = {3\over 4}\left[ 3+ 2{\pi -x_1\over x_1}\right]^{-1}} 

It is easy to invert the second equation and plug $x=x(\lambda )$ in the
first to get $R^2_h = 2 \lambda ({1 \over 2}-\lambda )$ that is precisely
the critical radius at which the heterotic string will have an $E_8 \times E_8$
gauge enhancement.  Also note that for $x_1=\pi$ we get $\lambda = {1 \over
4}$, and two branes in the middle are on top
of each other, and that corresponds to an extra $SU(2)$.

For unbroken $E_8\times E_8$ it is also natural to work with heterotic
$E_8\times E_8$ variables ($R_{E8}$, $\lambda_{E8} = e^{\phi_{E8}}$)
instead of the $SO(32)$ heterotic variables ($R_{SO}$ or $\lambda$,
$\lambda_{SO}=e^{\phi_{SO}}$) that we have been using, since the 
Wilson lines in the former are all
zero while in the latter they are functions of $R_{SO}$.
The map is worked out in Appendix A with the following
results,
\eqn\eso{R_{S0} = {R_{E8} \over (R^2_{E8}+2)} ~~~~~~~~~~  \lambda_{E8} =
(R^2_{E8}+2)^{1/2}\lambda_{SO}}

Now let us use the map from Type I' to the heterotic $SO(32)$ and \eso\
to find the map between the Type I' variables and the $E_8\times
E_8$ heterotic string variable.
{}From \eeebo\ and \eso\ we get,
\eqn\radii{R^2_{E8} = {2\over 3}\left( {4\pi - x_1 \over x_1}\right) ~~~
 {\rm or} ~~~  x_1
= 2\pi \left( {4 \over 3 R^2_{E8} +2}\right) }

Using \peet\ we can compute $C$ in terms of $R_{E8}$ and $\lambda_{E8}$
(remember that in \peet\ $e^{\phi_h}=\lambda_{SO}$ ) with the following
result,
\eqn\ceo{C^{5/3} = \lambda_{E8} {(3R^2_{E8}+2)^{5/3} \over R^3_{E8}}} 
Having done this we are ready to compute all the quantities that will be
relevant in section 8.2.
The Type I' dilaton is given by,
\eqn\edil{e^{\phi_{I'}} = C^{-5/6}z(x)^{-5/6} = \lambda_{E8}^{-1/2}
{R^{3/2}_{E8}\over (3R^2_{E8}+2)^{5/6}}z(x)^{-5/6}}
where $z(x) = {3 \over \sqrt{2}}\left[ \pi -\half |x-x_1|-\half |x-(2\pi
-x_1)|\right]$. This comes from \zeto\ by setting $B=0$ and $x_1$ is given
in \radii .
 
The metric is given by,
\eqn\metricee{ds^2 = \Omega^2(x)(\eta_{MN}dx^Mdx^N) =  \lambda_{E8}
 {(3R^2_{E8}+2)^{5/3} \over R^3_{E8}}z(x)^{-1/3}(\eta_{MN}dx^Mdx^N) }
Finally we need to compute the proper distances from $x=0$ to $x=x_1$ and
from $x=x_1$ to $x = 2\pi - x_1$. 

Let us start with $x=0$ to $x=x_1$, 
\eqn\properone{\Phi_1 = \int^{x_1}_0 \Omega (x)dx = {\lambda_{E8}^{1/2} \over
R_{E8}^{3/2} }}
and from $x=x_1$ to $x=2\pi - x_1$,
\eqn\propertwo{\Phi_2 = \int^{2\pi -x_1}_{x_1} \Omega (x)dx =
\lambda_{E8}^{1/2}{(R_{E8}^2 -2) \over R_{E8}^{3/2} }}

Notice that on the heterotic $E_8\times E_8$ we are {\bf not} at the critical
radius since $E_8\times E_8$ is not reached by an enhancement of the gauge
group. However, the extra $SU(2)$ we mentioned before that is perturbative
from Type I' since it corresponds to the two branes in the middle coinciding
at $x=\pi$ corresponds according to \propertwo\ to $R^2_{E8} =2$ that is
nothing but the critical radius for zero Wilson line.

This concludes our review of the 9 dimensional description using type I'.

\subsec{8 Dimensions}

If we try to extend the analysis of the previous section by further
compactifying on another $S^1$ in order to get a description of the
strongly coupled heterotic theory in 8 dimensions, it is easy to see that in general we will fail since
the two radii of the Type I theory will be small and we will be forced to
perform T-duality on both circles. This implies, for instance, in the case of unbroken
$SO(8)^4$, that the Type I' coupling behaves as follows,
\eqn\surprise{\lambda_{I'}={1 \over R_{h,1}R_{h,2}}}  
therefore if the two heterotic radii are of the order of critical radius
then we are out of the perturbative regime of type I'.

The full description of the heterotic moduli space is achieved by
considering F-theory compactified on an elliptic K3 \fth . 
The elliptic fibration
over $P^1$ is given by,
\eqn\fiber{y^2 = x^3 + f(z)x + g(z)}
where $z$ is the coordinate over the sphere, $f(z)$ and $g(z)$ are
polynomials of degree 8 and 12 respectively. 

The discriminant of this equation gives the location of the 24 singular fibers
over $P^1$ and is given by,  
\eqn\sing{\Delta = 4 f^3(z) + 27 g^2(z) }

The complex structure of the fiber located at a point $z$ is given by 
\eqn\tauf{j(\tau )= 1728 {4 f^3(z) \over \Delta } }
where $j(\tau )$ is the invariant modular function. This function can be
written as a Laurent series in $q=e^{2\pi i\tau}$ given by, 
\eqn\jtau{j(\tau) = q^{-1} + 744 + 196884 q + 21493760 q^2 + \ldots}

The F-theory geometry captures the Type IIB compactified on
$P^1$ with 24 (p,q)-7 branes transverse to the $P^1$ and located at the
positions of the singular fibers. The complexified IIB coupling constant
$\tau = \chi + i e^{-\phi}$ is identified with the complex structure of the
fibers and undergoes $SL(2,{\bf Z})$ monodromy.
The metric in the Einstein frame for this compactification is given by
\scs\

\eqn\metric{ ds^2 = k Im(\tau )\left|
{\eta^2(\tau)\over \Delta^{1/12}}dz\right|^2+\eta_{\mu\nu}dx^{\mu}dx^{\nu}}
where $\eta (\tau)= q^{1/24}\prod_{n=1}^{\infty}(1-q^n)$, and $k$ is an
overall constant controlling the volume of the sphere.

The last ingredient is the volume of the $P^1$ that is a positive real
number and is identified with the heterotic coupling constant in 8
dimensions. Therefore the strong coupling limit corresponds to a large
$P^1$ and the geometrical picture is a good description.

The possible gauge group on the heterotic side are reproduced on the
F-theory by developing ADE singularities on the $K3$. The possible fibers
that one can get when two or more singular fibers come to the same point
were classified by Kodaira and are given in  table 2 together with the
order of the zero that $f(z)$, $g(z)$ and $\Delta$ should have at those
points.

\bigskip
{\centerline{
\vbox{\offinterlineskip
\hrule
\halign{&\vrule#&
\strut\quad\hfil#\quad\cr
height2pt&\omit&&\omit&&\omit&&\omit&&\omit&\cr
& \hfill   orf($f(z)$)   \hfill &&  \hfill  ord($g(z)$)  \hfill  && \hfill
 ord($\Delta$)  \hfill && \hfill Fiber Type \hfill && \hfill Singularity
Type
 \hfill & \cr \noalign{\hrule}
height2pt&\omit&&\omit&&\omit&&\omit&&\omit&\cr
& \hfill   $\ge 0$       \hfill &&  \hfill   $\ge 0$     \hfill  && \hfill
$0$   \hfill && \hfill smooth \hfill && \hfill none  \hfill & \cr
\noalign{\hrule}
height2pt&\omit&&\omit&&\omit&&\omit&&\omit&\cr
& \hfill   $0$       \hfill &&  \hfill   $0$     \hfill  && \hfill   $n$
\hfill && \hfill $I_n$ \hfill  && \hfill $A_{n-1}$  \hfill & \cr
\noalign{\hrule}
height2pt&\omit&&\omit&&\omit&&\omit&&\omit&\cr
& \hfill   $\ge 1$       \hfill &&  \hfill   $1$     \hfill  && \hfill
$2$   \hfill && \hfill $II$ \hfill  && \hfill none  \hfill & \cr
\noalign{\hrule}
height2pt&\omit&&\omit&&\omit&&\omit&&\omit&\cr
& \hfill   $1$       \hfill &&  \hfill   $\ge 2$     \hfill  && \hfill
$3$   \hfill && \hfill $III$ \hfill && \hfill $A_1$  \hfill & \cr
\noalign{\hrule}
height2pt&\omit&&\omit&&\omit&&\omit&&\omit&\cr
& \hfill   $\ge 2$       \hfill &&  \hfill   $2$     \hfill  && \hfill
$4$   \hfill && \hfill $IV$ \hfill && \hfill $A_2$  \hfill & \cr
\noalign{\hrule}
height2pt&\omit&&\omit&&\omit&&\omit&&\omit&\cr
& \hfill   $2$       \hfill &&  \hfill   $\ge 3$     \hfill  && \hfill
$n+6$   \hfill && \hfill $I^*_n$ \hfill  && \hfill $D_{n+4}$  \hfill & \cr
\noalign{\hrule}
height2pt&\omit&&\omit&&\omit&&\omit&&\omit&\cr
& \hfill   $\ge 2$       \hfill &&  \hfill   $3$     \hfill  && \hfill    $n+6$
\hfill && \hfill $I^*_n$ \hfill  && \hfill $D_{n+4}$  \hfill & \cr
\noalign{\hrule}
height2pt&\omit&&\omit&&\omit&&\omit&&\omit&\cr
& \hfill   $\ge 3$       \hfill &&  \hfill   $4$     \hfill  && \hfill
$8$   \hfill && \hfill $IV^*$ \hfill && \hfill $E_6$  \hfill & \cr
\noalign{\hrule}
height2pt&\omit&&\omit&&\omit&&\omit&&\omit&\cr
& \hfill   $3$       \hfill &&  \hfill   $\ge 5$     \hfill  && \hfill
$9$   \hfill && \hfill $III^*$ \hfill && \hfill $E_7$  \hfill & \cr
\noalign{\hrule}
height2pt&\omit&&\omit&&\omit&&\omit&&\omit&\cr
& \hfill   $\ge 4$       \hfill &&  \hfill   $5$     \hfill  && \hfill
$10$   \hfill && \hfill $II^*$ \hfill && \hfill $E_8$  \hfill & \cr
height2pt&\omit&&\omit&&\omit&&\omit&&\omit&\cr}
\hrule}
}}
\noindent{\ninepoint\sl \baselineskip=8pt {\bf Table 2}: {\sl Kodaira
classification of singularities of an elliptic $K3$ according to the order
of vanishing of $f(z)$ , $g(z)$ and $\Delta (z)$.}}
\bigskip

The precise map between both moduli spaces is in general very complicated, but
it is known for several cases in which the IIB coupling $\tau$ is constant over
the sphere, for example $SO(8)^4$ \sen\ , and for the $E_8\times E_8$
unbroken point where $\tau$ is not constant \mov \lust . The map in the case of
$E_8\times E_8$ will be used in section 8 as an example of the limit 
to 9 dimensions.

\newsec{Puzzles in 9 Dimensions}

In the context of Type I', Seiberg
studied the theory seen by a D4 brane probe \seib\ and found evidence
for the existence of conformal quantum field theories when the
D4 brane probe was placed at the orientifold with infinite coupling.
The conformal theory flows to an $SU(2)$ supersymmetric gauge theory
by a deformation, where the $SU(2)$ is the gauge symmetry seen on the
probe.
That the string coupling be infinite at the orientifold was related to
the fact that the conformal theory with $SU(2)$ gauge symmetry on the probe
would need to come from a theory with inifnite coupling if it has a chance
of flowing from a conformal theory, because of simple dimensional
analysis
of Yang-Mills coupling constant in 5 dimensions.
Moreover the quantum field theory one obtains depends on how many D8 branes
are placed at the orientifold point.  If there are $n$ of them, one obtains
an $SU(2)$ gauge theory with $n$ massless hypermultiplets. Furthermore
 it was suggested
that these theories have a global $E_{n+1}$ symmetry. This follows
from the fact that the target space has the corresponding
gauge symmetry, as reviewed in the previous section, and the gauge
symmetry corresponds to global symmetries in the probe theory.

The same critical
theories were also obtained in a geometrical context by considering
M-theory compactification on Calabi-Yau threefolds, where the threefold
has a shrinking 4 dimensional submanifold corresponding to a  Del Pezzo
surface
\dkv \ms .  Del Pezzo surfaces
are 2 complex dimensional K\"{a}hler manifolds with positive $c_1$, and are obtained
by considering the blow up of $P^2$ at up to $m\leq 8$ points, and in
addition
$P^1\times P^1$.  The isomorphism with the probe picture required
identifying the number of blowup points of $P^2$, $m$  with $m=n+1$
where $n$ is the number of $D8$ branes at the orientifold.
However there was a discrepancy between the geometry and the probe
picture.  Namely for $P^2$ with no points blown up, there was
no brane probe description, as it would correspond to $n=-1$ D8 branes
at the orientifold!  Moreover for $n=0$, i.e. infinite coupling
at the orientifold plane without any D8 branes present,
 there were two possible choices for the geometry (rather
 than one anticipated from type I' probe picture), namely
$P^2$ blown up at one point or $P^1\times P^1$.
The probe in these two cases would have to give an $N=1$
supersymmetric $SU(2)$ gauge theory with no matter.  What distinguishes
the two choices is  a discrete $Z_2$ choice of $\theta$ angle
\dkv\ related to the non-triviality of $\pi_4 (SU(2))=Z_2$.
Moreover the two theories are distinguished by the condition
that for the case corresponding to $P^1\times P^1$ there is a
global $SU(2)$ symmetry for the conformal theory on the probe, whereas
for the case corresponding to $P^2$ blown up at one point (
corresponding to a non-trivial choice of the discrete theta angle), there is
no global symmetry on the probe conformal theory.

Type I' perturbation theory should break down as we approach
either of these two conformal theories, because they correspond
to $1/g=0$ at the orientifold.  But they could be viewed
as boundaries of regions where type I' perturbation theory is valid.
However, the same cannot be said for the conformal theory associated
to $P^2$.  Not only we do not have any type I' perturbative brane
picture in this regime, the probe gauge theory does not flow
to an $SU(2)$ but rather to a $U(1)$. This strongly
suggests that there are {\it regions} (not just boundaries)
in the moduli space where type I'
perturbation breaks down.

On the other hand aspects of BPS bound states and moduli space for type I'
 were studied in \greenet , with emphasis on
subloci in moduli space where heterotic
string predicts enhanced gauge symmetries.
 These
correspond to codimension one subspaces of moduli space.
 In other words
these loci correspond to ``walls'' in the moduli space.
If these walls decompose the moduli space into
disconnected components, then one would argue that Type I'
perturbation theory could potentially break down.  
In other words, the perturbative
type I' would describe the interior of only one region
in moduli space and the other regions cannot be reached
by changing moduli.
It was argued in \greenet\ that the domain walls do not
decompose the moduli space into disconnected components.
This was based on studying some examples, and the general
statement was suggested as
a conjecture.  As we will discuss in the next section, indeed
the conjecture is correct and {\it the moduli space is connected
even after removing the walls}. 

We thus seem to have two contradictory expectations:  Namely the
arguments 
in \greenet\ suggest that type I' pertrubation covers
the entire moduli space, whereas the
 probe picture suggests that type I' perturbation should
break down beyond some regime of parameters.  We will resolve this
puzzle in section 5 and show that the completion
of regions where type I' perturbation applies
does not cover the full moduli space.  However before
we do this, it is important to have a deeper understanding
of the global aspects of moduli space of type I' (or heterotic)
theory in 9 dimensions.  This is what we turn to in the
next section.

\newsec{Global Aspects of Moduli Space in 9 Dimensions}

Consider compactification of heterotic string or Type I theory
from 10 to 9, on a circle.  As discussed before, the
moduli space of this theory, in the heterotic language, corresponds
to varying the radius of the circle, the 16 Wilson lines
and the coupling constant.  The total space is
$${\cal M}={\bf R}^+\times \hat{\cal{M}}$$
$$\hat{\cal{M}}=SO(17,1;{\bf Z})\backslash SO(17,1;{\bf R})/SO(17,R)$$
where $R^+$ labels the coupling constant of heterotic string
and $\hat{\cal{M}}$ parameterizes the 17 dimensional space
of the radius of the circle and the 16 Wilson lines.
The T-duality group is given by $G=SO(17,1;{\bf Z})$.

Before quotienting by $G$, the 17 dimensional space
$SO(17,1;{\bf R})/SO(17,R)$
is simply the 17 dimensional Hyperbolic space, with constant
negative curvature.  Thus the global aspects of the moduli space
are completely encoded by the group $G$ and its action. 
We will  describe the known mathematical aspects of this moduli
space \dall\
as well as connect it to known facts about heterotic
string and its moduli.  This will in particular 
lead us to a concrete parametrization of the 
fundamental domain of the moduli space in terms of heterotic
string variables.

The group $G$ is intimately related to a generalized Dynkin diagram: 

\bigskip
\centerline{\epsfxsize=0.75\hsize\epsfbox{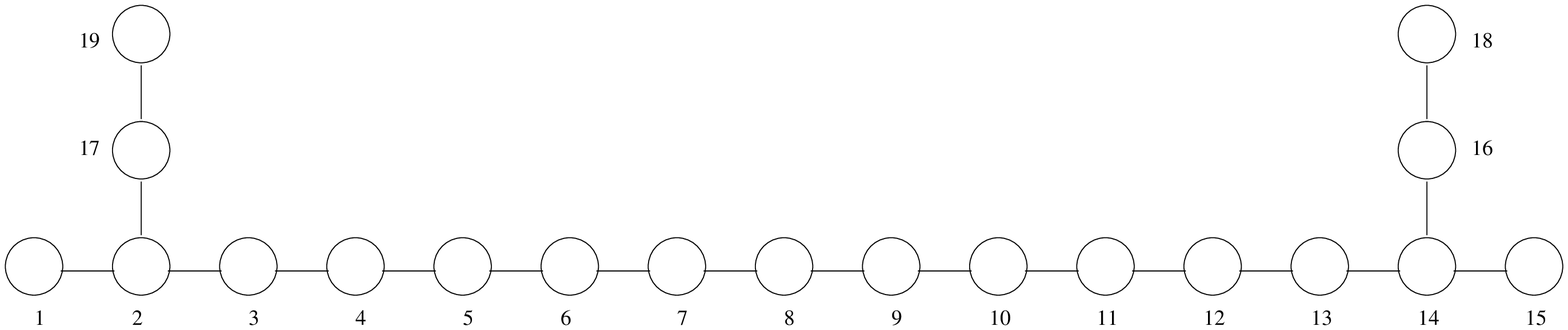}}
\noindent{\ninepoint\sl \baselineskip=8pt {\bf Figure 1}: {\sl Generalized
Dynkin diagram for $\Gamma^{17,1}$. The basis are chosen to show the
embedding of the Dynkin diagram of $SO(32)$ explicitly.}}
\bigskip

The meaning of this diagram is as follows:  $G$ is generated by elements
labeled by nodes of the diagram, $g_i$, satisfying
$$g_i^2=1$$
Moreover if the corresponding nodes are not connected by a line, then
the generators commute
\eqn\commu{g_i g_j=g_j g_i}
and if they are connected one gets the relation
\eqn\docom{(g_i g_j)^3=1}
In addition to get the full group $G$ we need a $Z_2$ involution
which conjugates the generators according to the outer atuomorphism
of the above Dynkin diagarm.  We will ignore this extra $Z_2$ in most
of this paper and
instead consider the double cover of the actual moduli space
(in the Type I' description this $Z_2$ corresponds to exchanging
the two ends of the interval, and in the heterotic string
description it is the outer automorphism exchanging the two $E_8$'s).

The elements $g_i$ can be also viewed as Weyl reflections in the
Narain lattice $\Gamma^{17,1}$.  In particular
for each node $g_i$ there is a vector $v_i\in \Gamma^{17,1}$
with the property that
$$v_i^2=2$$
(we are choosing the signature on $\Gamma^{17,1}$ corresponding
to $(+^{17},-^1)$)
and the Weyl reflection is given as
$$w\rightarrow w-(w\cdot v_i) v_i$$
This clearly is an automorphism of $\Gamma^{17,1}$ (as it preserves
the inner product) and so is an element of $G$.  The statement
is that $G$ is generated by 19 such Weyl reflections, given
by 19 vectors $v_i$. Moreover, the inner product of these
vectors is given by the above extended Dynkin diagram\foot{
For some aspects of the relation
between this extended Dynkin diagram and heterotic
strings in 9 dimensions see \ginsparg \mohaup .}.  In particular
$$v_i \cdot v_j=0 \qquad {\rm disconnected}\ {\rm nodes}$$
$$v_i \cdot v_j=-1 \qquad {\rm connected }\ {\rm nodes}$$
It is easy to check that the Weyl reflections generated
by such $v_i$'s satisfy the relations given in \commu\
and \docom .
In the Narain description of the vector, each $v_i$ corresponds
to 
$$v_i= (P_L^i,P_R^i)$$
where $P_L^i$ is a 17 dimensional vector and $P_R^i$ is a one dimensional
vector, and changing the moduli of the Narain lattice, corresponds
to a Lorentz $SO(17,1)$ rotation on the vector.  The inner product being given
by
$$v_i^2=(P_L^i)^2-(P_R^i)^2=2$$
If one chooses the Lorentz rotation so that $P_R^i=0$, this corresponds
to an enhanced gauge symmetry, where a $U(1)$ gets promoted to $SU(2)$.
Note that this involves one condition, and so it is a 16 dimensional
subspace of the 17 dimensional parameter space.
In this context the non-trivial Weyl reflection
symmetry of $SU(2)$ acts as a $Z_2$ on 
the parameters of the theory.  The fixed point of this
transformation on the Teichmuller space is exactly the
locus where we have (at least) an enhanced $SU(2)$ symmetry.
This is because at the $SU(2)$ point the Weyl symmetry
is a gauge symmetry of the theory and it maps the theory to
itself.  
Let us denote this 16 dimensional subspace by $D_i$.
The  $D_i$ divides the 17 dimensional space in two parts
mapped to each other by the $Z_2$ action, which is a symmetry
of the theory.  One can choose the moduli space to be
on one side of $D_i$. In particular the $D_i$ can
be viewed as boundaries of moduli space.
The statement that $G$ is generated
by Weyl reflection about $v_i$'s (modulo
the $Z_2$ outer automorphism
noted before) implies that all
the T-duality symmetries can be understood as Weyl symmetries
of some $SU(2)$ at some points on moduli space.

If we consider a collection of $N$ vectors $v_i$ and
consider the subspace of the moduli space given by the common
intersection locus of the corresponding $D_i$, this gives
a $17-N$ dimensional subspace.  Moreover on this subspace
the correponding $P_R^i=0$, and the heterotic string
will have an enhanced gauge symmetry of rank $N$ whose
Dynkin diagram (which may be disconnected) is given by the corresponding
nodes.  This is clear from the heterotic perspective
as the $P_L^i$'s will form the root lattice of the gauge
symmetry group. From this description it is also clear
that not all the $N$ loci $D_i$ interesect, otherwise
we would get Dynkin diagrams which do not correspond
to any group.  We thus conclude that the only $D_i$ that
have common intersection are the ones for which the corresponding
Dynkin nodes is that of an allowed group. This information
thus gives us the geometry of intersection of $D_i$'s.  It
also tells us all the allowed enhanced gauge symmetries
that we can obtain in this case.  In particular the maximal
gauge symmetry enhancements that we can have would correspond
to rank 17 groups whose Dynkin diagram is given by keeping
all the nodes of the extended Dynkin diagram of $G$ after deletion
of 2 of its nodes.

Now we are ready to describe the moduli space of
$\hat{\cal{M}}$.  The moduli space
can be chosen to be very similar to the fundamental domain
of $SL(2,{\bf Z})$ which is a subspace of the hyperbolic
2-space with three boundaries: two boundaries at $\tau_1={\pm 1/2}$
and the third corresponding to the sphere $\tau_1^2+\tau_2^2=1$.
The moduli space for $\hat{\cal{M}}$ can be chosen to be
given by a subspace of the 17 dimensional hyperbolic space
$B^{17}$ with $19$ boundaries correponding to $D_i$.  The
geometry resembles that of a higher dimensional chimney
(see Fig.2).

\bigskip 
\centerline{\epsfxsize=0.45\hsize\epsfbox{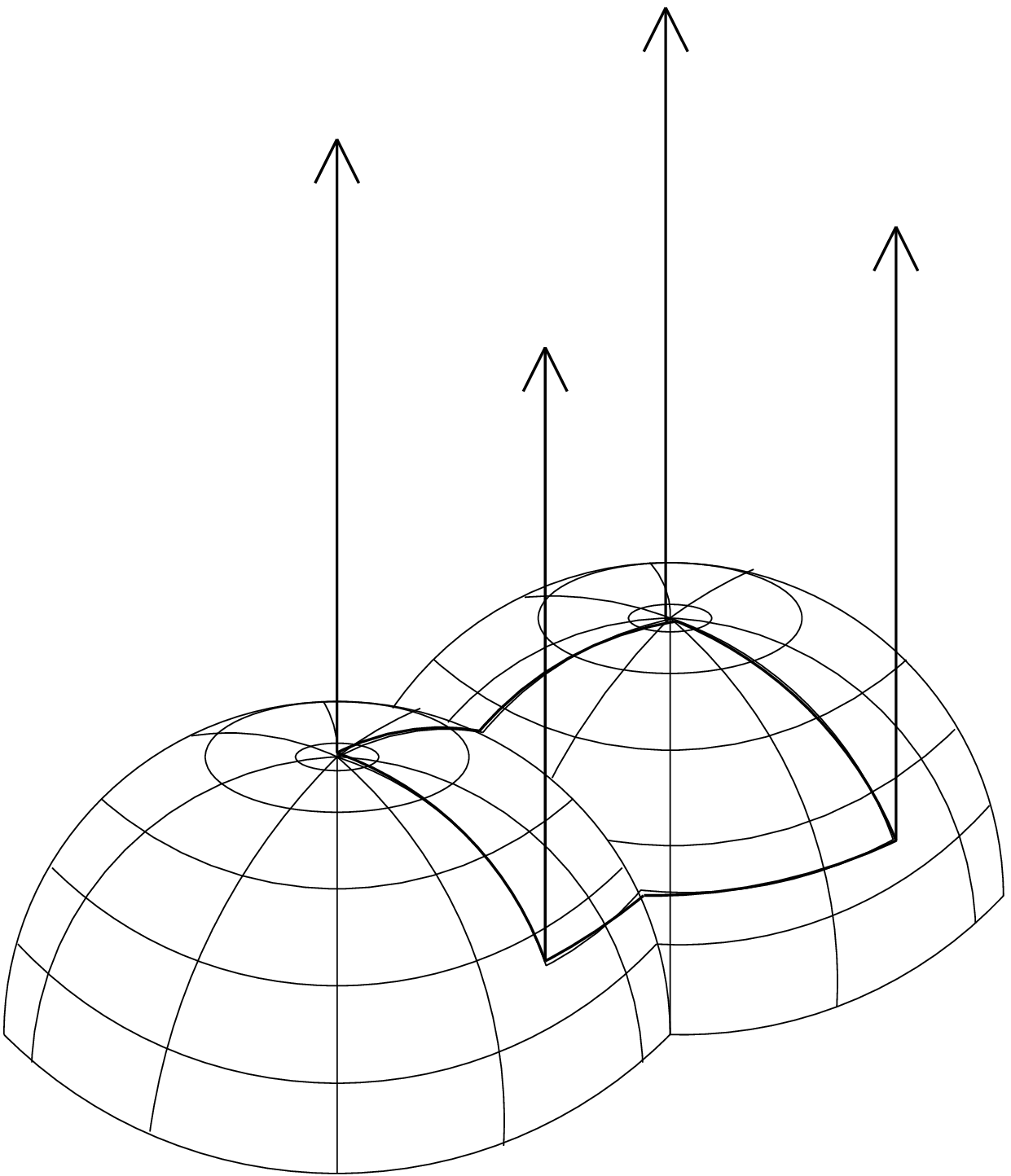}}
\noindent{\ninepoint\sl \baselineskip=8pt {\bf Figure 2}: {\sl
A chimney bounded by two
spherical walls on the bottom represents the Moduli Space $\hat{\cal{M}}$.}}
\bigskip

17 of the boundaries, correponding to the nodes of the
affine $SO(32)$ in the extended Dynkin diagram correspond
to 17 straight walls of the chimney and two of the boundaries,
corresponding to the two extra nodes of the Dynkin diagram
correspond to spherical ``bottom'' of the chimney.  The geometry
of their intersection is already discussed above and is in accordance
with allowed enhanced gauge symmetry points.  The direction
corresponding to increasing the radius of the 9-th direction
of the $SO(32)$ heterotic string is along
the linear direction of the chimney.
This geometry can be understood relatively simply:  Note that
the cross section of the chimney (the analog of $\tau_1$
for upper half-plane) is 16 dimensional. Moreover, for large $R$
it should be identified with the Wilson lines for the $SO(32)$ theory.
The choices of inequivalent Wilson lines for the $SO(32)$ theory are given by
choices of arbitrary 16 vevs $\theta_i$ in the Cartan of $SO(32)$ modulo
the action of the symmetries.  The symmetries in this case are
the shifts of $\theta_i$ and also the Weyl action.  The group
they form is the affine Weyl group, and that is why the extended
Dynkin diagram of $SO(32)$ enters the moduli space of flat
bundles on a circle.\foot{A similar statement is also
true for all other groups.}
 The fundamental domain for the $SO(32)$ Wilson lines
are given by the cross section of the chimney
enclosed by 17 walls which are in 1-1 correspondence
with the nodes of affine $SO(32)$.  This describes the cross section
of the chimney.  Let us be more explicit and give
a quantitative description of the cross section of this chimney,
parametrizing it in $R^{16}$ and identifying each of the boundaries with
the respective node in the Dynkin Diagram. 

Let $\theta_1 , \ldots \theta_{16}$ be coordinates of $R^{16}$, the
$Spin(32)$ wilson line can be chosen to be a diagonal matrix representing
the action on the fundamental representation, i.e. $W =
diag(e^{2\pi i\theta_1}, e^{-2\pi i \theta_1}, \ldots , e^{2\pi i\theta_{16}},
e^{-2\pi i\theta_{16}})$. Clearly, the Weyl group has as subgroup the
permutation group $S_{16}$, and therefore it is possible to introduce an
 ordering without loss of generality. Let $0 \le |\theta_1 |\le |\theta_2 |
 \le \ldots
\le  |\theta_{15} |\le  |\theta_{16} |\le 1$ be the region of $R^{16}$ that
would be expanded if no further elements of the Weyl group are considered. 

Moreover the Weyl group has more elements generated by
$\theta_i\rightarrow -\theta_i$ done for {\it pairs} of $\theta_i$'s
and similiarly for  $\theta_i \rightarrow 1 - \theta_i$ for a pair.
 Therefore we see
that the actual choice for the moduli can
be chosen to be $0 \le \theta_2 \le \ldots
\le  \theta_{15} \le {1 \over 2}$~, $~~|\theta_1 | \le \theta_2~$,$~~\theta_{15} \le \theta_{16} 
\le 1 - \theta_{15}$ (the
condition of having even pairs in the above Weyl action is 
what makes the first and last $\theta$'s have different
regimes). 
Thus, we can see the 17 boundaries defining the 17 codimension 1 walls in
$R^{16}$. There are 13 whenever any $\theta_i = \theta_{i+1}$ for
$i=2,\ldots , 14$, the last 4 are given when either $\theta_1$ or
$\theta_{16}$ meet any of their two boundaries.

Each of the first 13 boundaries given by $\theta_i =\theta_{i+1}$
corresponds to the node labelled by $i$ in the Dynkin diagram of figure 1.
The two boundaries given by $\theta_{1}=\theta_{2}$ and
$\theta_{1}=-\theta_2$ correspond to the nodes $1$ and $17$
respectively. Finally, the last two boundaries, $\theta_{16}=\theta_{15}$
and $\theta_{16}= 1-\theta_{15}$ correspond to the nodes $15$ 
and $16$ respectively.
On any of the 17 walls there is an $SU(2)$ symmetry
enhancement. Moreover, on the intersection of these hyperplanes we can
get the group given by taking the dots in the Dynkin diagram corresponding
to the intersecting hyperplanes, as discussed before.   

Having described explicitly the cross section of the chimney, we are only
left with the boundaries at the bottom when we introduce the radius
direction. These two spherical walls correspond to small radius enhancement
of gauge group by an extra $SU(2)$, and is already well
known in the context of heterotic strings.
Consider now, $R^{17}$, where the new coordinate is nothing but
$R_h$. The chimney is bounded from below by the following two $S^{16}$'s,
\eqn\esphere{R_h^2 + \sum^{16}_{i=1}\theta^2_i = 2
\;\;\;\;\;\;\;\;\;\;\;\;  R_h^2 + \sum^{16}_{i=1}\left( {1\over 2} -
 \theta_i\right)^2 = 2}
In terms of the Dynkin diagram of Figure 1, each of these boundaries
correspond to the nodes $18$ and $19$ respectively.

Now we are ready to give the complete parametrization of the fundamental
domain of the full moduli
space. In the coordinates of $R^{17}$ defined by $(\theta_1
,\ldots ,\theta_{16}, R_h )$, we have the following region,
\eqn\modulif{ \hat{\cal{M}} =
 \left\{ \matrix{\hfill -\theta_2\le\theta_1\le\theta_2 \hfill
   &&& \hfill  \hfill \linesp \hfill  0\le\theta_i\le {1\over
                                          2},
\qquad \theta_i\le \theta_{i+1}                                  
 \hfill &&& \hfill i=2\ldots 15
                                          \hfill \linesp \hfill \theta_{15} 
                                          \le \theta_{16} \le 1 - \theta_{15}
                                           \hfill &&&
                                          \hfill \hfill \linesp                                                          
                                          \hfill R_h^2 + 
                                          \sum^{16}_{i=1}\theta^2_i 
                                          \ge 2 \hfill
                                          &&& \hfill  \hfill 
                                          \linesp \hfill 
                                          R_h^2 + \sum^{16}_{i=1}({1\over 2} -
                                           \theta_i)^2 \ge 2
                                           \hfill &&& \hfill \hfill} \right. }
\noindent (the $Z_2$ outer automorphism noted
before acts on moduli space by taking all $\theta_i\rightarrow ({1\over 2}-
\theta_i)$)

{}From this explicit description and regions of enhanced gauge symmetry
we can now see exactly at which points we get which
enhanced gauge symmetries.

Incidentally, in terms of the coordinates we have
introduced for the hyperbolic
moduli space, its constant negative
curvature metric is given by
$$(ds)^2={1\over R^2_h}\left( dR_h^2+\sum_i d\theta_i^2\right)$$

There is another choice of the moduli space one can make
(by an $SO(17,1;{\bf Z})$ transformation) which is more
adaptable to the compactification of the $E_8\times E_8$ heterotic
string. In this case we again have 19 boundaries, but the
straight walls of the chimney correpond to the 18 nodes
of the extended Dynkin diagram of the two $E_8$'s.  The last
node corresponds to a sphere corresponding to the bottom
of the chimney.  The direction of increasing the ninth
radius for the $E_8\times E_8$ theory corresponds
to going along the linear direction of the chimney.  Again
the cross section of the chimney for large $R$ corresponds
to moduli of flat $E_8\times E_8$ connection on the
circle of fixed radius.

In Figure 3 we see how the Dynkin diagram of $\Gamma^{17,1}$ can be given
basis encoding the $E_8\times E_8$ structure. The nodes $1\ldots 8$ form
a Dynkin diagram of $E_8$ and so do the nodes $1' \ldots 8'$. Adding the
nodes $A$ and $A'$ to each of the $E_8$'s makes them affine
$\hat{E}_8$. These two affine versions of $E_8$ give the structure of the
section of the chimney. Finally, the node labelled by $B$ represents the
sphere bounding the bottom of the chimney in the 17-th direction
parametrized by $R_h$.  

\bigskip
\centerline{\epsfxsize=0.75\hsize\epsfbox{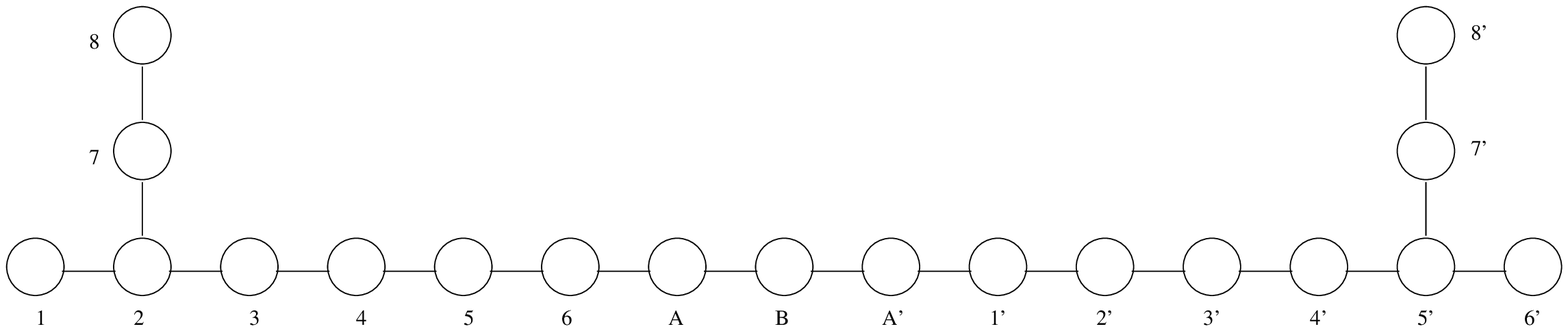}}
\noindent{\ninepoint\sl \baselineskip=8pt {\bf Figure 3}: {\sl Generalized
Dynkin diagram for $\Gamma^{17,1}$. The basis are chosen to show the
embedding of the Dynkin diagram of $E_8 \times E_8$ explicitly.}}
\bigskip

\newsec{Resolution of the Puzzles and Incompleteness of Type I'}

Let us start the analysis by considering a simple example that contains all
the important features of how the perturbative type I' description is
incomplete and resolve the apparent contradiction of section 3.

Consider the heterotic $SO(32)$ string compactified on $S^1$.
If we do not turn on any Wilson lines we obtain an $SO(32)$ gauge
symmetry in 9 dimensions (for sufficiently large radius).  There
is however another inequivalent choice of Wilson line
which also yields an $SO(32)$ gauge symmetry in 9 dimensions.
Consider acting by a $Z_2$ symmetry as we go around the circle,
where the $Z_2$ acts as $-1$ on the states which are
weights in the spinor of $ SO(32)$ and $+1$ on the other weights.  This
also preserves an $SO(32)$ gauge symmetry because the root lattice
is invariant under the $Z_2$.  From the viewpoint of type I (or type I')
theory, the two choices are the same at the perturbative level,
because there are no states in the perturbative
type I theory transforming according to the spinor of $SO(32)$. 
Let us connect these two classes of theories with a continuous
choice of Wilson line.  In particular consider the
Wilson line given by $\theta=(0^{15},\lambda)$. 
For generic $\lambda$ and generic
 radius the
unbroken gauge group is $SO(30)\times U(1)^3$. The
 $\lambda =0$ corresponds to turning on no Wilson line,
 leaving an $SO(32)$ gauge symmetry.  The choice $\lambda =1$
 correspond to the $Z_2$ Wilson line, which acts only on 
 the spinor degrees of freedom, again leaving an $SO(32)$
 gauge symmetry. The heterotic moduli space
for fixed coupling is a strip in the ($R_h$-$\lambda$) plane that 
is unbounded on
one side since $R_h$ can be arbitrarily large but bounded on the other 
by the condition that $R^2_h \ge R^2_{hc} = 2(1-\lambda^2/2)$. The width of
the strip is given by the condition that $0\le \lambda \le 1$ as discussed
before. This
moduli space is shown in Figure 4.  This is simply a 2-dimensional
slice of the chimney moduli space we discussed in the previous
section. 

\bigskip
\centerline{\epsfxsize=0.75\hsize\epsfbox{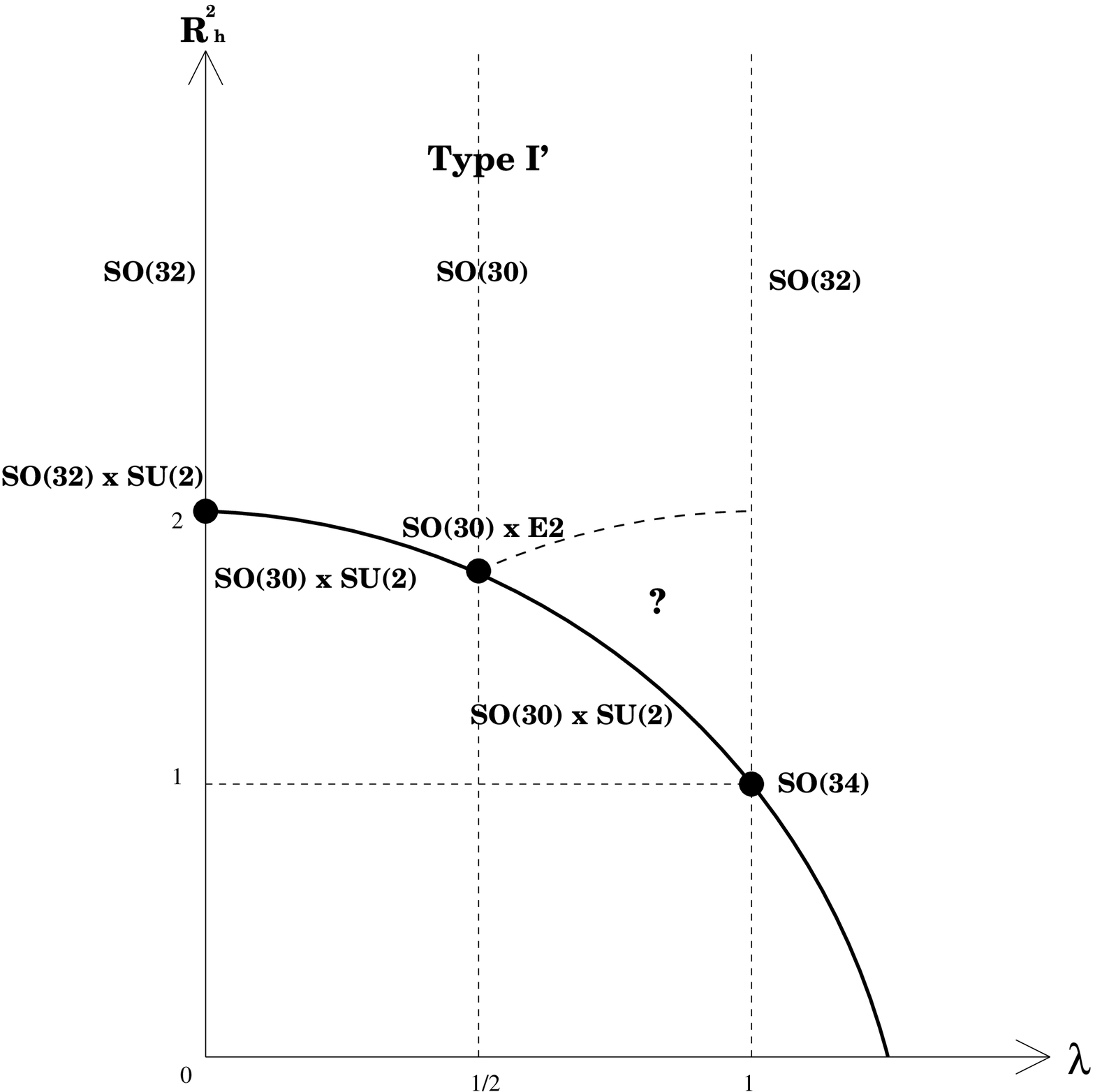}}
\noindent{\ninepoint\sl \baselineskip=8pt {\bf Figure 4}: {\sl
2-dimensional slice of the chimney moduli space parametrized by
$R_h$ and $\theta_I = (0^{15},\lambda )$. The solid curved line is the
critical radius for a given $\lambda$. The dashed curved line is the Type
I' boundary that has no extra massless
particles.}}
\bigskip

Starting at a generic point -- i.e., not in the boundary -- we can make
the radius smaller until we hit the $R^2_h = 2(1-\lambda^2/2)$ boundary at
which an $SU(2)$ gauge symmetry will appear in addition to the $SO(30)$ we
had. 
Starting again from the same generic point but moving in $\lambda$ we can hit
either the $\lambda =0$ or $\lambda =1$ boundaries, at any of them we get
an $SO(32)$. Now if we 
go down in the radius direction for $\lambda =0$ we will hit the $R^2_h=2$ boundary getting and
$SO(32)\times SU(2)$ gauge group while in the case 
$\lambda =1$ the critical
radius is at $R^2_h=1$ with an $SO(34)$ enhancement.  This follows from
our discussion of the global aspect of the moduli space and points
of gauge symmetry enhancement (the circle
corresponds to the 18-th node on the extended Dynkin
diagram, and the 16-th node in this case maps to line $\lambda =1$
which together with other nodes coming from $SO(30)$
will form $SO(34)$.

Let us try to follow the previous paths but from the Type I' view point.
Turning on $\lambda$ corresponds to moving one of the 16 D8 branes away from
one orientifold.  With the conventions we have used
this corresponds to the position $x$ of the D8 brane changing from
$0$ to $4\pi $ as we vary $\lambda$ from $0$ to $1$.
At $\lambda ={1\over 2}$ the D8 brane is
on the opposite orientifold plane.  Continuing to increase
$\lambda$ beyond this value from the viewpoint of the Type I' theory does not
change the perturbative theory at all, since it is equivalent
to taking the image D8 brane back to the orientifold we started with.

However now let us repeat the same process but tune the Type I' coupling
so that $1/g=0$ at the other orientifold. 
If $\lambda =0$ this corresponds to the gauge
symmetry enhancement $SO(32)\times SU(2)$.  However now consider
$1/g =0$ at the other orientifold but at $\lambda =1$.  
What gauge symmetry do we expect in this case?  To answer this
we have to know what is the map of the type I' parameters and heterotic
parameters in this regime. This is actually easy:  Turning on
or not turning on the $Z_2$ Wilson line acting only on the
spinors does not affect the map between the radius of the circle
viewed in the heterotic string $R_h$ and the coupling parameters
of Type I or its perturbative dual Type I'.  Thus again at $R_h^2=2$
we find that $1/g =0$ at the other orientifold. But for $\lambda =1$
and $R_h^2=2$ there is no gauge symmetry enhancement
expected on the heterotic side!  In fact in the whole
$(\lambda, R_h)$ plane the map between
heterotic and type I' variables, can be obtained by
restricting attention to the $\lambda <1/2$ because
of the perturbative $\lambda \rightarrow (1-\lambda)$ symmetry
of type I' and since the value of the coupling constant
in the type I' theory are fixed by supergravity solutions
and that also reflects only perturbative aspects of type I'.
Thus the region of validity of type I' perturbation is
in the interior of 
$$R_h^2\geq 2\left( 1-{\lambda ^2\over 2}\right) ~~~~ {\rm and} ~~~~ R_h^2
\geq 2\left( 1-{(1-\lambda) ^2\over 2}\right)$$
Thus in particular the region in the vicinity of where $SO(34)$ gauge symmetry
enhancement is to take place is not reachable by
type I' perturbation theory!

Now we come to the puzzle raised in \greenet :  The puzzle
they raised was that since all regions of moduli
space are reachable without passing through points
where extra massless particles appear, then perturbation
has no reason to break down.  However,
we are proposing here
that the Type I' perturbation is breaking down
without the apperance of extra massless particles,
namely all the points where $\lambda > \half$ and
$R_h^2=2(1-{(1-\lambda) ^2\over 2})$.
We now argue this is not very surprising and there
are already well known examples of this in quantum field theories.
Consider 2d supersymmetric sigma model on the blow up of an
$A_1$ singularity of $K3$.  This is parametrized in the
sigma model by a Kahler class, the size $r$ of ${\bf P}^1$ and a
B-field on ${\bf P}^1$, which is a $\theta$ angle.   The perturbative
description of sigma model corresponds to defining $g^2=1/r$.  In particular
if $r$ is large there is a well defined perturbative description
of the theory. Now go to the limit where $r\rightarrow 0$. In this limit
the perturbation breaks down, and one expects to end up with 
a singular theory with arbitrary light mass states.  This expectation
is borne out as long as $\theta =0$.  However if for example
$\theta =\pi$ this turns out not to be true.  In fact as was shown
by Aspinwall \aspin\
in this case one obtains the orbifold conformal theory on
$R^4/Z_2$, which is perfectly well behaved.  Mathematically
what is going on is roughly that in the correlation function we have
objects which behave as
$$1/(1-x)$$
where $x=exp(-r+i\theta)$.  The perturbative
regime corresponds to $x\sim 0$.  The radius
of convergence of pertubation expansion is $|x|<1$.
However if we put $x=-1$, which corresponds to
$\theta =\pi$ there is no singularity in the correlation,
but nevertheless the perturbative description breaks down.  
This is parallel to what we believe happens to perturbative
description of type I'.

Let us now give further evidence for this, related to D4
brane probe in the context of the example we just discussed.
Consider the point where $\lambda ={1\over 2}$ and take
$1/g=0$ on the other orientifold.  This corresponds
to having an $SU(2)$ symmetry at the other
orientifold point.  Now put the D4 brane probe also at
the orientifold.  Then on the D4 brane probe lives a superconformal
theory, which is equivalent to M-theory in a local
CY 3-fold geometry where we have a $P^2$ blown up
at 2 points shrunk to zero size.  This is called
the $E_2$ conformal theory, and flow upon deformation to
an $SU(2)$ with one massless fundamental flavor.  The mass of the 
fundamental field corresponds roughly to $m=\half-\lambda $.
It was shown that for $m>0$ and $m<0$ give rise
to two inequivalent conformal theories, corresponding
to an $SU(2)$ theory on the probe, with or without a discrete
$Z_2$ valued $\theta $ angle in the gauge theory. The theory
with the $Z_2$ valued $\theta$ angle turned on is expected
to have {\it no global symmetry} even though the D4 probe is
placed at the orientifold with the value of the coupling
$1/g=0$.  The absence of extra global symmetries in this
case, means in particular that the target space has no
extra gauge symmetries, {\it even with vanishing} $ 1/g=0$ 
at the orientifold.  This we identify with the
line emanating from $\lambda >{1\over 2}$ and
with $ R_h^2
= 2(1-{(1-\lambda) ^2\over 2})$, which has no extra gauge
symmetry.  
We can in fact do better.  Namely we can map the moduli
space expected from the transitions of $P^2 $ blown up
at 2 points, which is discussed in detail
in \ms\ (shown in Fig. 5)
with that given by the parameters
of the heterotic string near $\lambda ={\half}$ and
$R_h^2 ={7\over 4}$.

\bigskip
\centerline{\epsfxsize=0.70\hsize\epsfbox{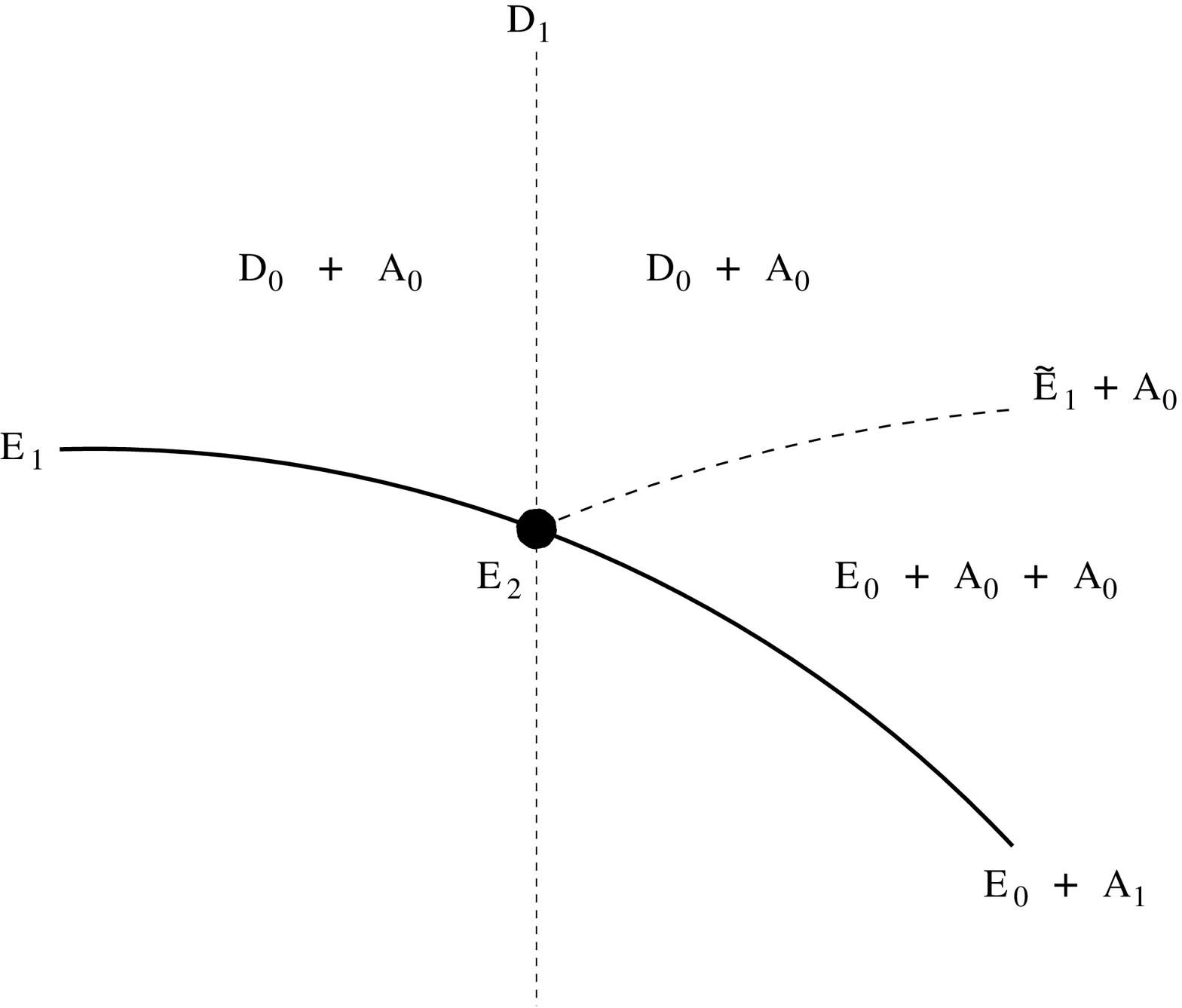}}
\noindent{\ninepoint\sl \baselineskip=8pt {\bf Figure 5}: {\sl Moduli space
around the $E_2$ point for vanishing Del Pezzo surfaces or 5-dimensional
field theories at non-trivial superconformal fixed points.}}
\bigskip

Here the $A,D,E$'s in the above figure represent global
symmetries expected from the del Pezzo description where $A_0, D_1, {\tilde
E}_1$
correspond to a $U(1)$ symmetry, $E_1, A_1$ corresponds to $SU(2)$ and
$E_0$ corresponds to no global symmetry.

The above figure should also represent (part of) the moduli
of Type I' theory if the conformal theory description
found from del Pezzos match parameters seen by the D4 brane probe.
  Moreover the global
symmetries of the conformal theory predicted
from the del Pezzos should correspond to gauge symmetries
of the bulk type I' theory, inherited by the D4 brane probe. 
Indeed, we see an isomorphism between the above figure
and the moduli of type I' given in figure 4 near $R_h^2={7\over 4}$
and $\lambda ={\half }$, suggesting a type I'-like
 description may be valid.
In fact the extra gauge symmetries anticiapated from the heterotic
string moduli exactly matches the global symmetries
anticipated from the del Pezzo description of the conformal
theory.  In particular the dashed line should correspond
to {\it no enhanced gauge symmetry} as ${\tilde E}_1$ has
only a $U(1)$ global symmetry.
The solid line represents on the heterotic side
a region with an extra $SU(2)$ symmetry and this also
matches the del Pezzo prediction, as on either side of the
$E_2$ point we have an $SU(2)$ global symmetry.
However, to the left of $E_2$ the global
symmetry is part of the symmetry seen by the
conformal theory, whereas on the right side the global
symmetry is a symmetry of the massive particles.
Moreover in the region below the dashed line
and above the solid line there is no enhanced
gauge symmetry in agreement with the fact that the
global symmetry there is expected to be $U(1)\times U(1)$.

If one tries to force a complete type I'-like description
in all regions of the above moduli space, one sees that
in bulk language, the $A_0$'s should correspond to D8-branes, but there is
a region between the dashed curve
and the solid line 
with $E_0+A_0+A_0$ where it seems that there are two branes instead
of one as we started with! This is precisely in the region
where we have argued Type I' perturbation does not apply.
There was a picture suggested by Morrison and Seiberg \ms , which
tries to extend the type I' description, beyond the regime of its
validity by forcing a type I'-like description.  This
involved the assumption that we can extract one extra
brane out of the orientifold at infinite coupling.  This
should be only possible when the orientifold with the
infinite coupling is correlated with the other choice
of the discrete Wilson line, so that it does
not give rise to an enhanced gauge symmetry.  For instance
going back to the region which was missing in type I' theory
they would assign a 17th D8-brane, whose position
is related to the $R_h$.  Let us recall that the horizontal direction is
controlled by the value of the Wilson line and the vertical direction by the
heterotic radius. Below the line with $\tilde{E}_1+A_0$ the value of the
Type I' coupling at the orientifold is frozen to be $1/g=0$, therefore
the radius should be controlling the position of the ``new brane'' that is
pulled out from the orientifold. More precisely, the radius is controlling the
relative position between the ``new brane'' and the old brane. For any  Wilson
line $\half < \lambda < 1$ we get that at the critical radius (solid line
in Figure 4) the relative
position is zero and the two $A_0$'s form an $A_1$ in the $E_0$ theory on
the D4-brane probe. Something especial happens at $\lambda=1$ since the
two branes reach the other orientifold with the 15 branes giving altogether 17
branes at the orientifold. This is the $SO(34)$ point.     

Having described the suggested picture for explaining the $SO(34)$ point
one could also ask about other possible enhanced gauge symmetry
points, for example $SU(18)$ which according to our discussion
is allowed. In this case if we pull one extra brane from
each orientifold, this can be achieved by 18 coincident
branes in the middle of the interval.  This was in fact
suggested to be possible in \ms .  

The fact that on the dashed line in figure 5 the type I'
breaks down without the appearance of massless
particles is somewhat novel.  It is natural to ask
if anything special happens there as viewed from the heterotic side.
As we have argued no extra massless states appear. However, the conformal
theory seen by the D4 brane is interpreted on the heterotic side
as the theory seen by a single 5 brane wrapped around $S^1$.  Thus
the heterotic theory also ``knows'' something special is happening
there:  There appears a non-trivial conformal theory on a single
wrapped 5 brane.  This is quite a novel effect.

Viewed from
the heterotic moduli we can describe exactly which piece of the
moduli space the Type I' perturbation misses.  In our global
description of moduli space the bottom of the chimney was described
by two spheres. All we have to do is to reflect the two
sphere by replacing $\theta_1 \rightarrow -\theta_1$ in the second sphere
of \esphere\ and  $\theta_{16} \rightarrow (1-\theta_{16})$ in the first sphere.  
This gives us altogether 4 spheres.  The Type I' perturbation
is exactly the top part of the chimney bounded by the first
sphere it encounters as one decreases $R_h$.  

Explicily, the Type I' moduli space will be the chimney bounded from below
by the following spheres,
\eqn\twosphere{{\rm Physical~Boundaries}:\qquad
R_h^2 + \sum^{16}_{i=1}\theta^2_i = 2
\;\;\;\;\; R_h^2 + \sum^{16}_{i=1}({1\over 2} - \theta_i)^2 = 2}
\eqn\foursphere{{\rm Pert.~Breakdown}:\qquad R_h^2 + 
(1- \theta_{16})^2 + \sum^{16}_{i=2}\theta^2_i = 2
\;\;\;\;\; R_h^2 + \sum^{15}_{i=1}({1\over 2} - \theta_i)^2 +
({1\over 2}+\theta_{1})^2 = 2}

It is interesting to see how the possibility
of having one or two extra branes fits with filling
the 3 missing regions of the moduli space.  
At a qualitative level, we have already explained how
this would arise.  In particular let us consider the region
where we have 2 extra branes. 
 Let us denote by $x_0, x_1,\ldots x_{16},x_{17}$ the positions
of the 18 branes, and let us say that the first 17 are independent. 
The two boundaries on the bottom of the ``chimney'' describing
the moduli space are now represented by $x_0=x_1$ and $x_{16}=x_{17}$, where
the mechanism for the generation of $SU(2)$ enhanced gauge symmetry
is the same as in the usual pertubative type I'.  This should
fill the region in moduli space given by
$$-\theta_2<\theta_1 <0,~~ 1/2<\theta_{16}<1-\theta_{15}$$
$$0<\theta_2<\theta_3<...<\theta_{15}<1/2$$
$$R_h^2 + \sum^{16}_{i=1}\theta^2_i > 2
\;\;\;\;\; R_h^2 + \sum^{16}_{i=1}({1\over 2} - \theta_i)^2 >2$$
$${\rm Pert.~Breakdown}:\qquad R_h^2 + (1- \theta_{16})^2 + 
\sum^{16}_{i=2}\theta^2_i <2
\;\;\;\;\; R_h^2 + \sum^{15}_{i=1}({1\over 2} - \theta_i)^2 +
({1\over 2}+\theta_{1})^2 < 2$$
Note that there are 19 boundaries in this region which match
with the nineteen boundaries for of $0<x_0<x_1...<x_{17}<2\pi $.
Similar statement can be made about the other regions
where only one extra brane appears.

Now we can try to generalize \zeto\ from the Type I' 
analysis of section 2 in order to
describe this situation in a more quantitative form. For instance, in the
case where we get two extra branes and the inverse coupling is frozen to
zero at both orientifolds, the most natural thing to write is,
\eqn\zeta{z(x)= {3 \over \sqrt{2}}({1 \over 2}\sum^{17}_{I=0}x_I -{1\over
2}\sum^{17}_{I=0}|x-x_I| )}
This reproduces the D4-brane expectations for the coupling constant when it
is expressed in terms of the proper distances as in \usual \foot{
We can try to go even further and try to match the parameter
of the moduli of type I' with those of heterotic strings. It is
natural in the region we just discussed, to introduce two new
$\theta$'s, namely $\theta_0$ and $\theta_{17}$. We define them
by
$$\theta_0=-{1\over 2}\left[ R_h^2+({1\over 2} +\theta_1)^2 +
\sum_{i=2}^{16}({1\over 2} -\theta_i)^2 -2\right]$$
$$\theta_{17}={1\over 2}\left[ R_h^2+(1-\theta_{16})^2+\sum_{i=1}^{15}\theta_i^2-
1\right]$$
The choice of these values are motivated by the
condition that when $\theta_0=0$  and $\theta_{17}={1\over 2}$ 
should correspond to the boundaries \foursphere\ and $\theta_0 =-\theta_1$
and $\theta_{17}=(1-\theta_{16})$ should correspond to \twosphere .
Then using the map between brane positions and the Wilson loop
expectation values given by \Wilson\ we can relate the 18 $x_i$'s
brane positions with the 18 $\theta$'s, which are in turn
captured by the 17 moduli parameters of heterotic strings.}

Even though this completion of type I' is compelling
and matches various aspects of heterotic gauge theory
enhancement one expects, it is clearly beyond the regime
of the type I' perturbation theory.  For example on the
D4 brane probe after we pull the extra brane off the
orientifold we do not expect to have an $SU(2)$ gauge symmetry.
Also we have no good explanation of these extra branes, and apart
from matching with expected behavior from the heterotic theory,
it seems ad hoc.
We will try to give a unified
description of all regimes of parameters of type
I' based on geometry of real elliptic $K3$ in the remaining sections.

\newsec{Real Elliptic $K3$ as a limit of F-Theory}

Type IIA on $K3$ is dual to heterotic string on $T^4$.  This
duality can be pushed up one dimension by considering
the strong coupling limit of type IIA, where we obtain
M-theory on $K3$ being dual to heterotic on $T^3$.
If we view $K3$ as an elliptic manifold over ${\bf P}^1$
and consider the limit where the elliptic fiber goes
to zero size, we should obtain a description involving
type IIB on ${\bf P}^1$ with 24 various $(p,q)$ type 7-branes.
This is the F-theory description.  This can also be viewed as type 
IIB compactified on $T^2$ modded out by a $Z_2$ with an orientifold
action \sen\ .  If the branes are not equally
distributed among the orientifold planes the theory become
non-perturbative in the type IIB language and has a description
which is captured by the geometry of the elliptic $K3$. 

The question is whether this geometric description can
be continued one more step to provide a strong coupling
description of heterotic strings in 9 dimensions. 
As we have found in the previous sections, type I' description
is inadequate, and it would be interesting to see if
geometry sheds any light on some of the questions raised
in that context.

What we should do is to ask how
the radius of the 8-th direction is encoded in the geometry
of elliptic $K3$ and use this to obtain the limiting geometry
as the radius in the $8$-th direction goes to infinity.  Before
doing this, it turns out to be convenient first to review
the situation in going from 7 dimension to 8 dimension; i.e.
in going from M-theory on $K3$ to F-theory on elliptic $K3$.

Consider heterotic strings on $T^3$.  Its moduli space is
captured, in addition to the string coupling, by the moduli of
the $\Gamma^{19,3}$ lattice.  Note that this is directly
related to the geometry of $K3$, namely, the lattice is identified
with the $H_2$ lattice on $K3$ and the choice of the metric on $K3$
determines the splitting to left and right part of the lattice 
by the action of $*$-duality induced by the metric.  The overall
radius of $K3$ does not enter the duality operation and is
related to the inverse of heterotic string coupling constant.

Now if we are interested in going to 8 dimensions, we consider
the limit where $T^3$ is given by $T^2\times S^1$ with
a large radius for $S^1$. Let us consider the metric
(including the anti-symmetric B-field) on $T^3$
in a block diagonal form, respecting this decomposition.  
We can also turn on Wilson lines on $S^1$, but clearly in the
limit of going to 8 dimensions, they are irrelevant.  So let
us consider turning them off around the $S^1$.  The moduli space
of this subset of $7$ dimensional compactifications is given by the
moduli of polarizations on $\Gamma^{18,2}+\Gamma^{1,1}$ respecting
this decomposition.  The moduli of $\Gamma^{18,2}$ is parametrized
by 18 complex parameters, and that of $\Gamma^{1,1}$ by
one real parameter.  This real parameter is in fact identified
in the usual way with the radius of the seventh circle.  Thus
in going to 8 dimensions the geometry of $\Gamma^{18,2}$ remains
intact.  Now we ask how this is realized in the geometry of $K3$.
The most natural way to say this is as follows:  Consider
the inversion map on the 7-th circle in the heterotic side $x^7
\rightarrow -x^7$.  This is a symmetry
of heterotic strings if there is no Wilson line on the circle
as well as if the metric is block diagonal on $T^2\times S^1$. In particular
it acts on the $\Gamma^{19,3}$ lattice.  Let us combine this, with
an overall $Z_2$ inversion of the full $\Gamma^{19,3}$ lattice.
We thus see that we have a $Z_2$ symmetry which acts as
\eqn\actz{\Gamma^{19,3}\rightarrow \Gamma^{18,2}_{(-)} + \Gamma^{1,1}_{(+)}}
On the $K3$ side this symmetry should be realized as a $Z_2$ symmetry
on the geometry acting on the 2-cycles of $K3$ exactly according to this
decomposition.  Moreover one should choose a metric on $K3$ respecting this
$Z_2$ symmetry.  In the context of F-theory, this $Z_2$
is realized by
\eqn\zact{y\rightarrow -y, \quad x\rightarrow x, \quad z\rightarrow z}
which is a symmetry of elliptic $K3$ given by
$$y^2 =x^3+f(z) x+g(z)$$
with $f$ and $g$ being polynomials of degree $8$ and $12$ respectively.
The elements of $H_2$ invariant under the $Z_2$ correspond to the class
of the elliptic fiber $E$ and the base $B=P^1$.  They have an intersection
given by
$$E^2=0,\quad B^2=-2, \quad E\cdot B=1$$
which defines the $\Gamma ^{1,1}$ lattice where if we define
$$e_1=E+B,\quad e_2=E$$
we get the standard description of inner product for $\Gamma^{1,1}$.
Note that under the $Z_2$ defined in \zact\ the classes E and B are
invariant.  That the base class is invariant is obvious, as that
is parametrized by $z$.  That E is invariant follows from the
fact that the $(1,1)$ form corresponding to it, has even number
of $y$ and ${\overline y}$'s (recall the $(1,1)$ form is given by
$|dx/y|^2$).  It is also possible to show that the other 20 classes
in $H_2$ are mapped to minus themselves.  This follows
by viewing them roughly speaking as products of 1-cycle in the base
and 1-cycle in the fiber.
Note in this context that the radius of the
heterotic string is related to the sizes of $E$ and
$B$ through the map given above. Namely, if we identify
$e_1$ with the winding vector $(P_L,P_R)=(-R,R)$ and $e_2$ with
the momentum vector $(P_L,P_R)=(1/R,1/R)$, and use the BPS formula
for $P_R$ which is proportional to the integral
of the kahler form over the corresponding class we learn that
$${\int_{E+B} k\over \int_{E} k}=R^2$$
which leads to
\eqn\limfor{k(B)/k(E)=R^2-1}
In particular the limit $R\rightarrow \infty$ for a fixed
volume of $K3$, corresponds to taking $k(B)\sim R$ and $k(E)\sim
1/R$, which is the usual statement that in the limit
of zero size elliptic fiber, compared to the base, we obtain
a geometry which captures the moduli space of heterotic strings
on $T^2$.  

Now we repeat this same idea, but in taking the limit
from F-theory on elliptic K3 and follow the moduli
of K3 in the direction
of decompactifying a circle of heterotic string. Again
if we turn off the Wilson lines on the circle we are decompactifying,
this means we will have additional $Z_2$ symmetry in the theory,
which acts exactly as the one we discussed above in the context of going
from 7 to 8 dimensions, namely the action \actz\ on the $H_2$ lattice.
Taking into account both $Z_2$'s, we can thus decompose
the lattice as
$$\Gamma^{19,3}\rightarrow \Gamma^{17,1}_{--}+\Gamma^{1,1}_{+,-}+
\Gamma^{1,1}_{-,+}$$
Note in particular that the new $Z_2$ we want should act as inversion
of the $\Gamma^{1,1}$ lattice corresponding to the original
$E$ and $B$ of the F-theory.

The main question to ask now is which subclass of $K3$'s is relevant
for which there is such a symmetry?  The main hint comes as follows:
In going from 10 to 8 dimensions, the Wilson lines of
the heterotic string around the 2-circles correspond
to complex moduli:
$$u^i=A_{9}^i+iA_{8}^i$$
with real $A_8$ and $A_9$.  The complex parameters $u^i$
 should be identified with (some of) the complex coefficients defining
$f$ and $g$ in F-theory on $K3$. Turning off the Wilson lines
in the $8$-th direction, would make the parameters $u^i$ real.
We are thus led to look for elliptic $K3$'s with real
coefficient.  The $Z_2$ symmetry, which flip the 8-th
circle, will act on $A_8\rightarrow -A_8$, and so will
take $u^i\rightarrow u^{i*}$.  Thus the $Z_2$ symmetry
we would like to define on $K3$ should be a real involution symmetry,
which would require the coefficients $u^i$ to be real.  Thus we look 
for elliptic $K3$'s which are real, i.e. which are of the form
$$y^2=x^3+f_8(z)x +g_{12}(z)$$
where $f$ and $g$ are real polynomials respectively of degree 8 and 12 
in $z$, and the $Z_2$ symmetry we are after acts on $K3$ as
$$y\rightarrow y^* \quad x\rightarrow x^* \quad z\rightarrow z^*$$
Furthermore we would like to make sure that the $Z_2$ acts
on the $H_2$ homology according \actz .  That it acts on $E$ and
$B$ of F-theory in the right way, is easy to check. However, we
also need
to check that it acts correctly on the rest of the homology elements
of $K3$.
It turns out, as we will now discuss, this puts restrictions
on the coefficients of $f_8$ and $g_{12}$. 

All real involutions on $K3$ have been classified by Nikulin
\nik\ with the following conclusion:  Consider the
fixed locus of the $Z_2$ involution in $K3$.  Let us call
this the real $K3_R$. In other words
$$K3_R\subset K3 \quad (K3_R)^*=K3_R$$
It is clear by dimension count that $K3_R$ is a 2-dimensional
real subspace of $K3$.  It has been shown by Nikulin, that
$K3_R$ is in general not a connected surface.  In particular,
depending on the $Z_2$ real involution, it consists of
$k$ spheres, together with 1 genus $g$ surface.  Moreover not
all $k$ and $g$ can appear.  The allowed ones are shown in the figure
below:

\bigskip
\centerline{\epsfxsize=0.75\hsize\epsfbox{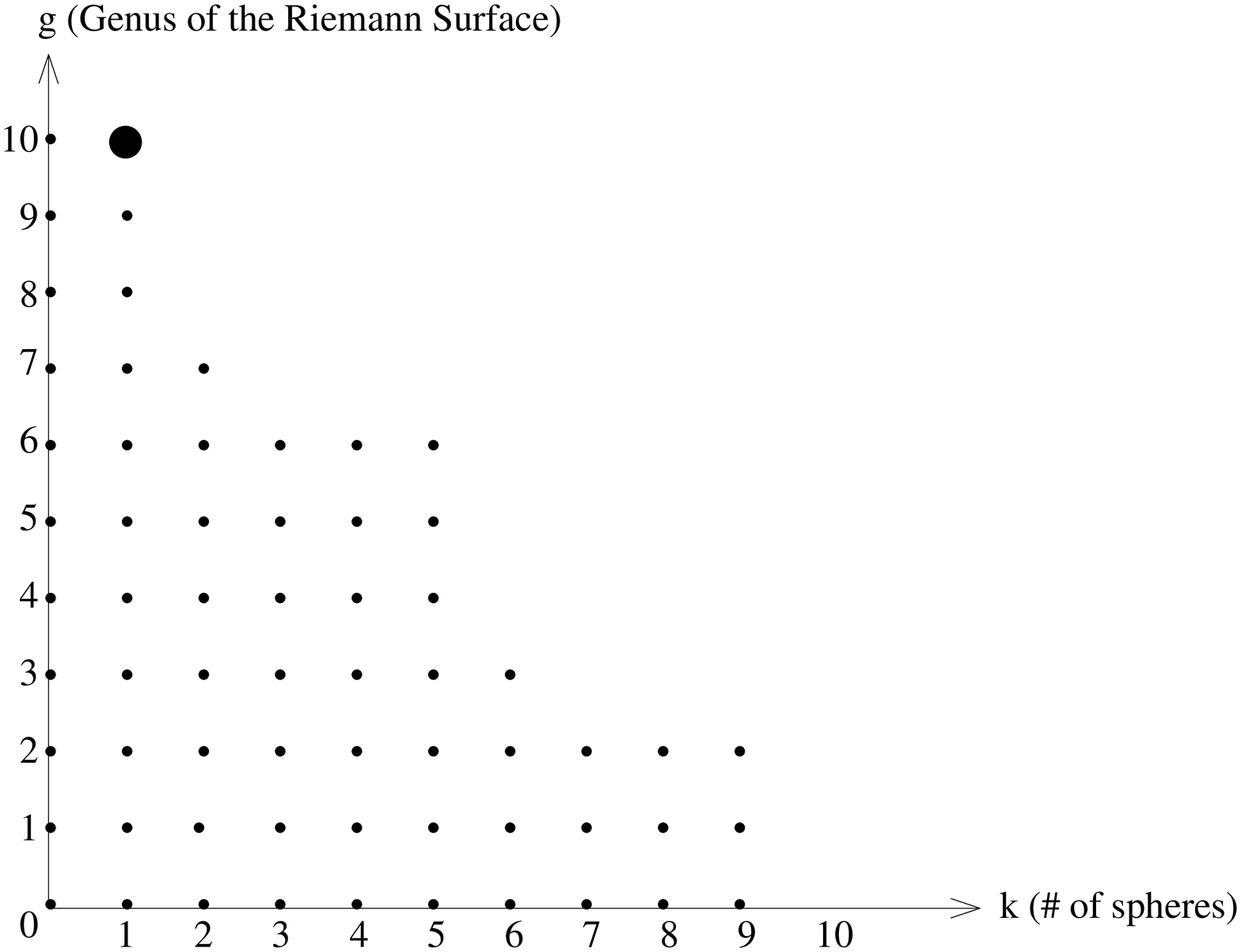}}
\noindent{\ninepoint\sl \baselineskip=8pt {\bf Figure 6}: {\sl
Classification of Real $K3$ surfaces according to the genus $(g)$ of the
Riemann Surface and the number $(k)$ of spheres. Small dots represent the
allowed values for $(k,g)$ and the big dot is the one corresponding to the
Real $K3$ related to the heterotic string in 9 dimensions.}}
\bigskip

The $k$ and the $g$ one gets is determined by
the $Z_2$ action on the $H_2$.  For the $Z_2$ action
we are interested, given by \actz\ we find
$k=1$ and $g=10$.  In other words we have
$$K3_R=S^2 + \Sigma_{10}\subset K3$$
where $\Sigma_{10} $ denotes a genus 10 Riemann surface.\foot{
There is a simple index theory argument which shows why, if we have
only one sphere, i.e. if $k=1$, the genus of the extra
component  $g=10$.  To see this 
let $h$ denote the $Z_2$ anti-holomorphic automorphism. 
Then $h$ acts on the cohomology
of $K3$ and by Lefshetz fixed point theorem we know
that its trace on the cohomology (weighted by $(-1)^{degree}$)
should be the Euler characteristic of the fixed point set.
This gives, since we know the action of $h$ on the mid-dimension
cohomology (as well as the fact that $h$ acts trivially on the
$H^0$ and $H^4$) that
$$2 +(2-2g)=4-20=-16$$
which implies that $g=10$.}
Nikulin's classification also applies to holomorphic
involution which takes the holomorphic 2-forms to minus
itself; in fact by an $SO(3)$ rotation in the choice of complex
structure, the two are equivalent.  In other words, even the involution
that we used in the context of F-theory, could be viewed
as a real involution for a particular choice of complex
structure on $K3$.  In that case, we can also check
that the structure of $K3_R$ i.e. the subspace
fixed by the $Z_2$ involution $y\rightarrow -y$ is as we have:
the fixed point of $y\rightarrow -y$ corresponds, for each
point on the z-plane to four points on the torus, the three
roots of the cubic, plus the point at infinity, corresponding
to $x=\infty$.  As we vary $z$ we get a surface.  The point
$x=\infty$ does not mix with the others, and gives a copy of the
base ${\bf P}^1$.  The other roots exchange and give a three fold
cover of the base, which turns out to have genus 10.  This is
easy to see.  The Riemann surface is given by the equation
$$x^3+f(z)x +g(z)=0$$
where $f$ and $g$ are polynomials of degree 8 and 12 respectively.
Its Euler characteristic $\chi$ can be computed by removing the 24
points
on $z$ sphere where roots of the cubic coincide, computing its Euler
characteristic (which is 3 times that of the sphere without 24 points)
and then adding two points for each removed point (because
two of the $x$'s meet at each of the 24 points).  We thus have
$$\chi =3(2-24)+2(24)=-18=2-2g\quad \rightarrow g=10$$

Here we want to give a concrete description of the real $K3$, which
also explains why we can have multiple components.
To describe $K3_R$ we consider real solutions 
(i.e. real $(x,y,z)$) of the real equation  $y^2 = x^3 + f(z) x + g(z)$. 
We can view the real $z$ (including infinity) as the equator
in the complex $z$-sphere base of F-theory, and real $x,y$ subject
to the above equation as real circle or circles in the elliptic
fiber of F-theory.  Note that there can be one or two real circles for
each real value of $z$ depending on the sign of
the discriminant of this equation 
 $\Delta =4 f^3 + 27 g^2 $.  If $\Delta$ is positive, the solution is
homeomorphic to a single circle (including the point $\infty$ to
the $(x,y)$ plane). However, if $\Delta$ is negative, the solution is
homeomorphic to two disconnected circles.   

The $K3_R$ can be viewed, therefore, as one or two circles
fibered over the equator given by real $z$.  Note however, that
the number of circles can change.  This happens if as we change $z$ the
discriminant $\Delta =0$.  Let us 
assume that $\Delta$ has a single zero at $z=z_0$. 
This implies that $\Delta$ should change sign and therefore the real fiber
over the $z$ in a neighborhood of $z=z_0$ will have to 
interpolate between a fiber with two components to a fiber with only one or
vice versa depending on the way $\Delta$ changes. It is clear that 
$f(z_0) < 0$  and $g(z_0) \neq 0$. This is true because if $g(z_0)$ and
$f(z_0)$ were zero then $\Delta$ would have at least a zero of order two.
This makes clear that there are two possibilities characterized by the sign
of $g$ at $z=z_0$.

For $g$ positive the transition takes place when the two circles approach
each other, join at one point and then open up to give a single circle. For
$g$ negative one of the circles shrinks to a point and then disappear.  
All this is shown in Figure 7.

\bigskip
\centerline{\epsfxsize=0.75\hsize\epsfbox{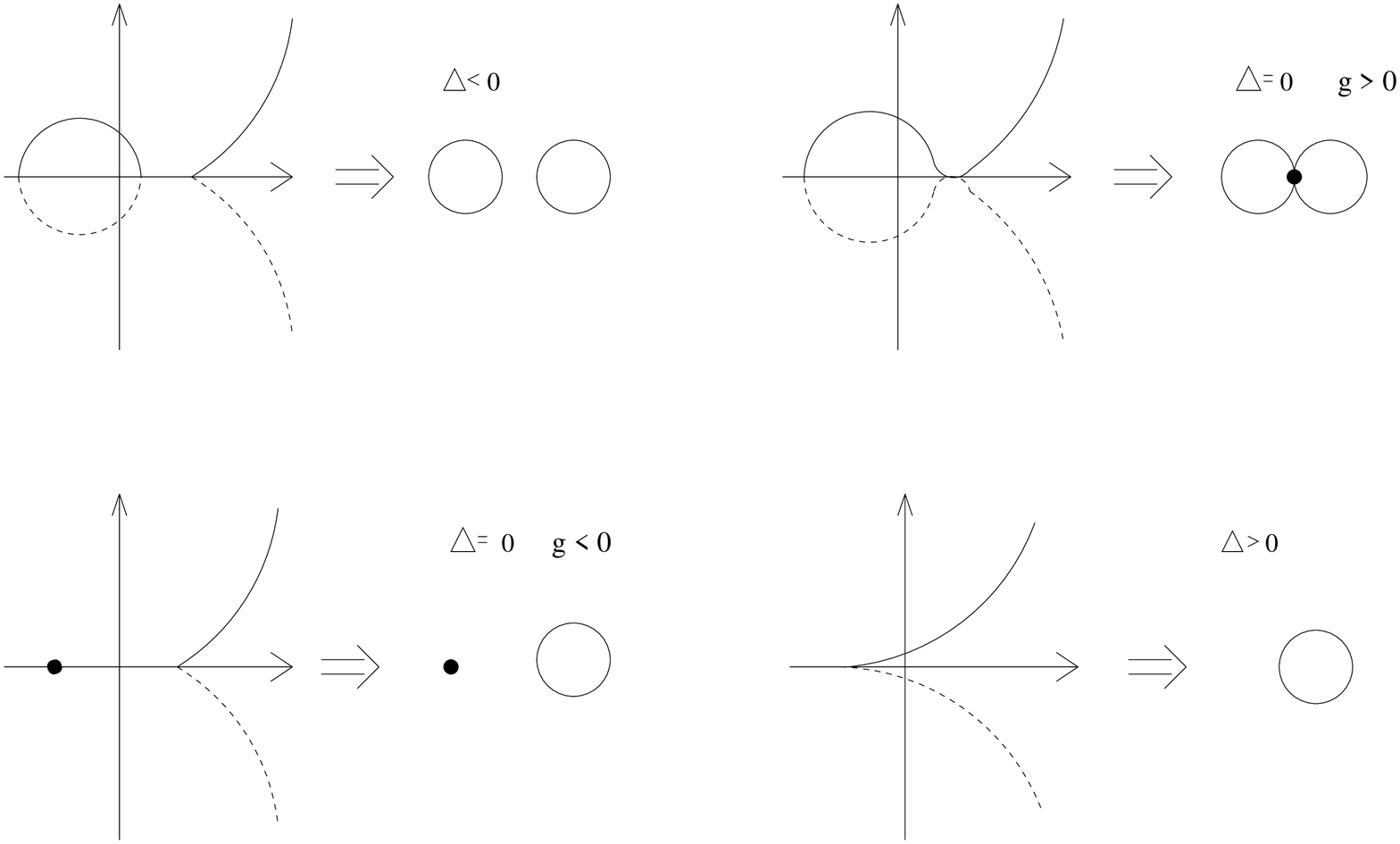}}
\noindent{\ninepoint\sl \baselineskip=8pt {\bf Figure 7}: {\sl
Classification of the possible Real fibers according to the sign of $\Delta$
for $\Delta\neq 0$ and to the sign of $g$ for $\Delta =0$ . }}
\bigskip

It is important to notice that the circle containing the point $\infty$ in
the $(x,y)$ plane is always present for any $z$. If we vary $z$ over ${\bf
R}$ and add $z=\infty$ we will necessarily get at least one Riemann surface
of genus $\ge$ 1.  

Now we are ready to describe the possible transitions involving two
consecutive single zeros of $\Delta$. There are only four possibilities if
we start and end with single components, i.e., with $\Delta >0$.
Three of them are shown in Figure 8$A$, 8$B$ and 8$C$. The fourth is just a
reflection of Figure 8$A$. In Figure 8$D$ we have shown a transition where
we start and end with $\Delta <0$  and the points with $\Delta = 0$ have
$g<0$ and $g>0$ respectively from left to right. In this case it is easy to
show that $f(z)$ must have two real zeros (denoted by white dots) between
the two branes (denoted by black dots). 

It is clear now that transitions of the $B$ type will
increase the genus of our ``basic'' Riemann surface and transitions of the
$C$ type will leave invariant the topology of the basic Riemann surface but 
will add
an extra component that can only be a sphere!

Finally, the transitions of the $A$ and $D$ type do not change the topology 
of the
Real $K3$ but will play an important role later on.

\bigskip
\centerline{\epsfxsize=0.75\hsize\epsfbox{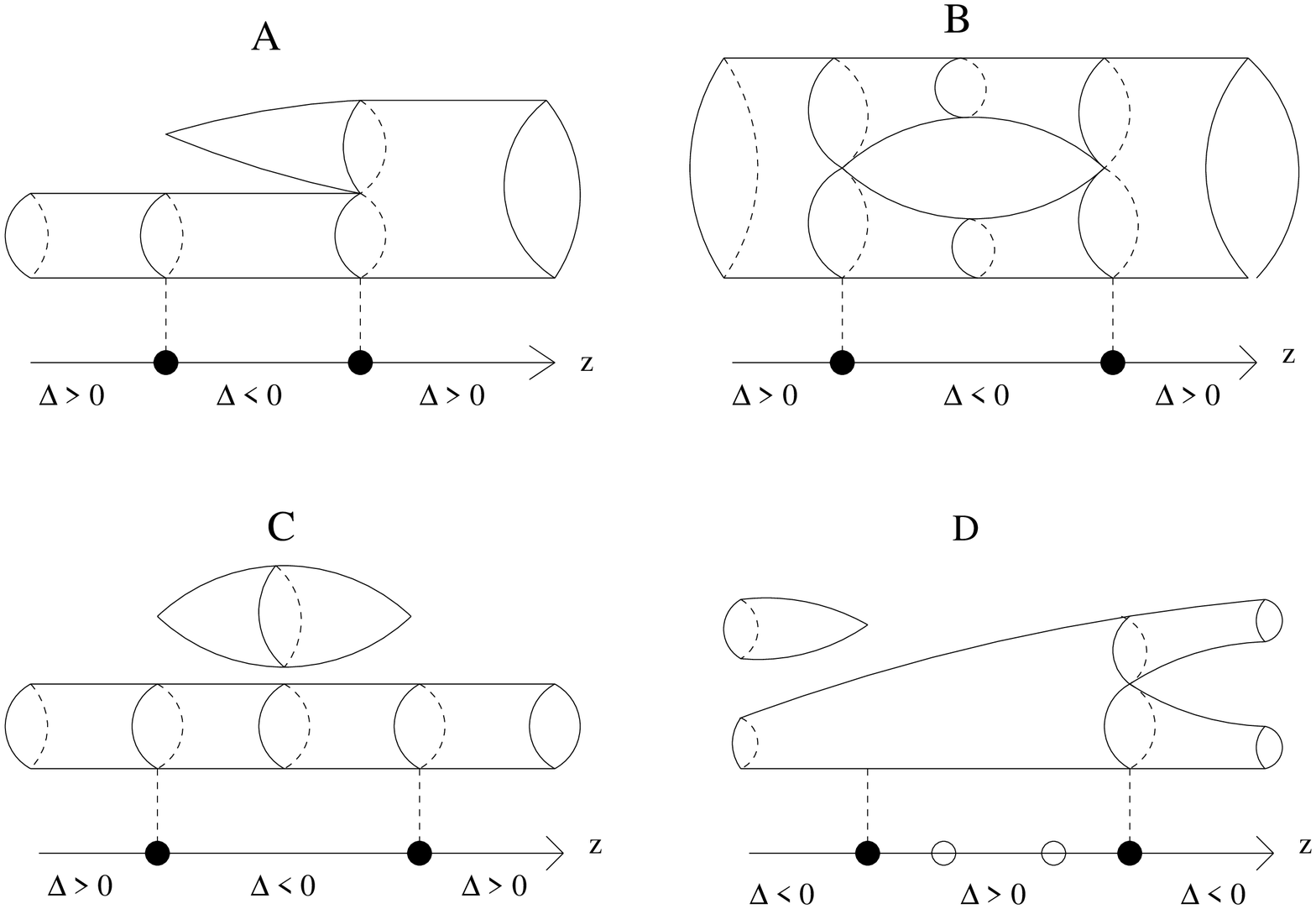}}
\noindent{\ninepoint\sl \baselineskip=8pt {\bf Figure 8}: {\sl Some possible
sections of a real $K3$ in an interval containing two single singular fibers.}}
\bigskip

\newsec{General Aspects of Real $K3$ and the moduli space in 9 dimensions}
In this section we discuss some general aspects of real $K3$ and
how it relates to the moduli space in 9 dimensions.  We will see
explicitly how the type I' branes (and their generalization)
have a natural interpretation in this language.   We will divide
our discussion into two parts: qualitative and quantitative.

\subsec{Qualitative analysis}

We have argued in the previous section that the real $K3$
should have two components, one with genus 10 and another with
genus 0.  In other words
as we go along the real $z$ direction, we get the splitting
and joining described in previous section, which makes up
a sphere and a genus 10 Riemann surface.  The simplest possibility
is that shown in figure 9.  Note that the Riemann surface with
genus 10 has nine holes, plus one hole going around the z-equator.
Also note that in the region in the z-axis where the real sphere
arises, between $X1$ and $X2$ the genus 10 Riemann surface cannot have a hole.  
This is because over each point $z$ we can have at most two circles of
real $K3$.  Out of the 24 points on the $z$ sphere with vanishing
discriminant $\Delta$ in the case shown here we have accounted for 20
of them:  $X1,X2$ where the sphere is formed and  from the points
$1,...,18$ where the 9 holes of the genus 10 Riemann surface
are carved out.  This leaves us with 4 extra branes which must
be in the bulk.  Since $f$ and $g$ are real polynomials in $z$,
so is $\Delta$, which implies that all the roots come
in complex conjugate pairs.  Thus the 4 roots which are not
real come in two complex conjugate pairs.

\bigskip
\centerline{\epsfxsize=0.5\hsize\epsfbox{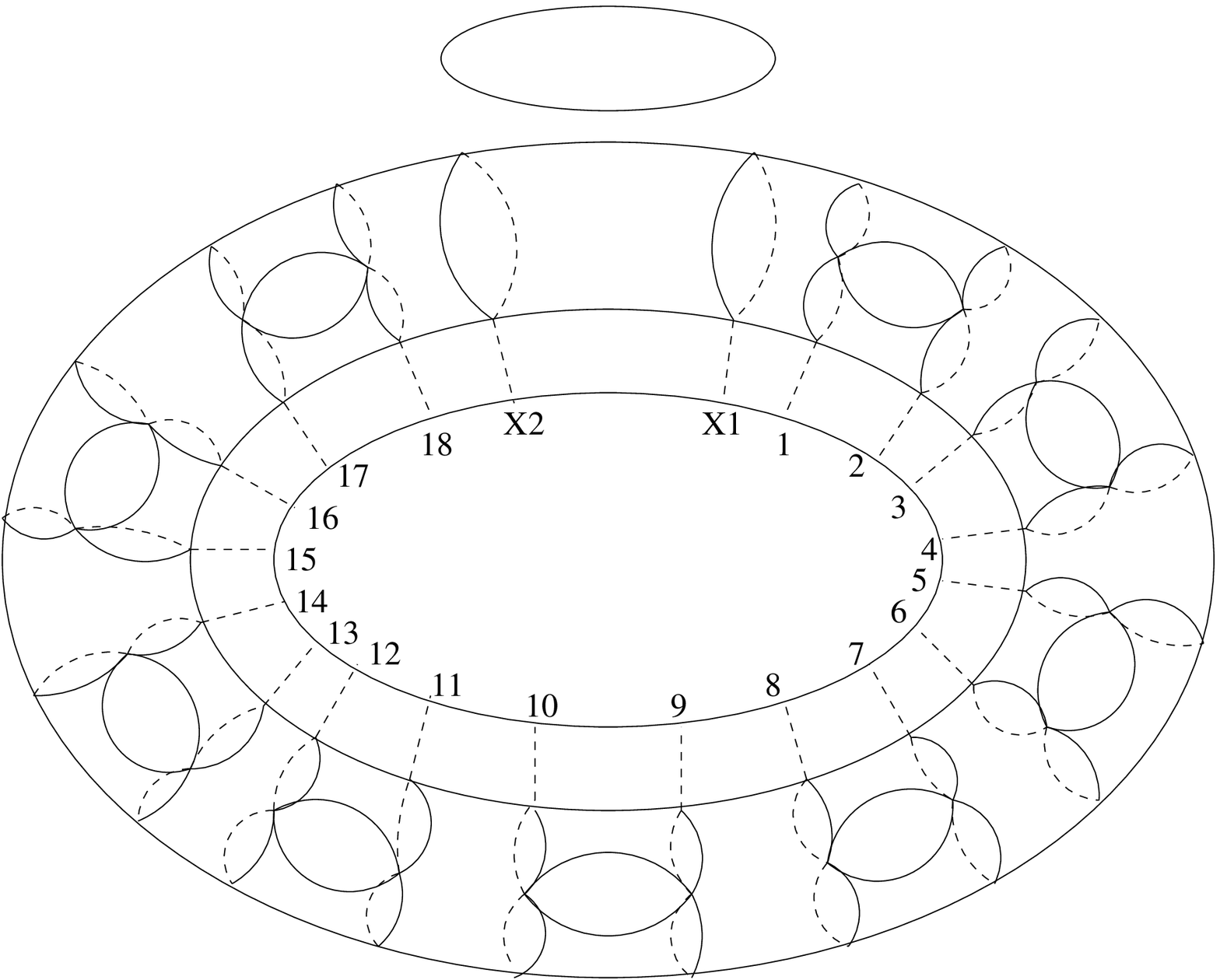}}
\noindent{\ninepoint\sl \baselineskip=8pt {\bf Figure 9}: {\sl Real $K3$
corresponding to the case with 20 real roots.}}
\bigskip

It is natural to ask if the figure 9 is the most general
possibility compatible with having a genus 10 Riemann surface
and a sphere.  In fact it is not.  Even though one may think
that bringing the branes from the bulk down to the real axis
changes the topology, it is possible to preserve
the condition that the genus is 10.  In fact we can have
configurations of two branes as shown in figure 8$A$ which
does not change the topology.  We will argue later that
if two pairs come down to zero it can intersect the real $z$
axis only
between the points 2 and 3 or between the points 16 and 17.  
So altogether we have 4 possibilities: All 4 extra branes off the
real axis, one or the other pair on the real $z$ axis
(Figure 10), and both pairs
on the real $z$ axis (Figure 11).  This matches very nicely the 
4 possibilities
predicted from the analysis of Wilson lines for $SO(32)$:  As
we discussed before the first brane being positive or negative
relative to the orientifold position are distinct possibilities.  The same
being true for the last brane.  Thus the four possibilities
we are encountering from the real $K3$ geometry matches beautifully
this aspect of the choices of inequivalent $SO(32)$ Wilson loop
values.

\bigskip
\centerline{\epsfxsize=0.5\hsize\epsfbox{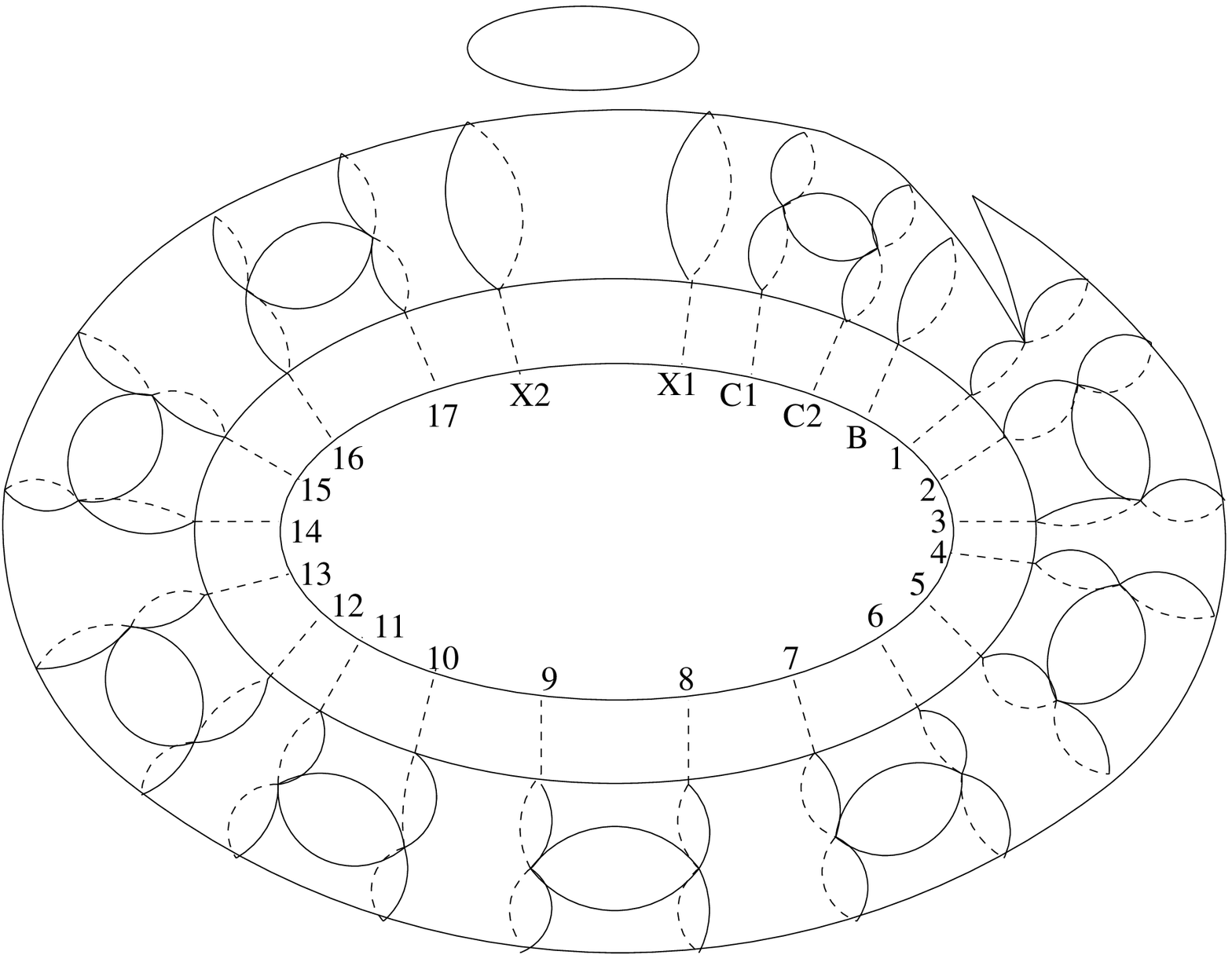}}
\noindent{\ninepoint\sl \baselineskip=8pt {\bf Figure 10}: {\sl Real $K3$
corresponding to the case with 22 real roots.}}
\bigskip

It is helpful to give a description of the basis of two cycles
of $K3$ projected onto the real $K3$, in the basis used
to describe the moduli space of type I' and the modular
group for the $\Gamma^{17,1}$ lattice.   This is most conveniently
done for the case where all four branes are on the real axis. In this
case the 2-cycles of $K3$, which project onto cycles (circles) 
on the real $K3$ are shown in figure 11.  In this form
their intersection is identical with the intersection expected for the
$\Gamma^{17,1}$ lattice (this is also the fastest way
of understanding why the extra branes from the bulk should
land at those positions on the real axis, though we give
other arguments at the beginning of the quantitative section based on the
${\hat E}_9 \times {\hat E}_9$ configuration studied in \wolfe\ and its
descendant). The two sphere in the real
$K3$ is to be identified
with the class in $\Gamma^{1,1}$ with self-intersection $-2$
(and is the direct analog of the base in the context of F-theory
dual of heterotic string).

Let us now argue why the extra branes can intersect
the real z-axis only as described in Fig. 10 and Fig. 11.
Let us suppose that in Figure 10, the two branes denoted by $B1$ and $1$
were located at any other position. Let us start by locating them between
$7$ and $8$. In this case we would have two sets of relative local fibers,
one with 11 fibers and the other with 8 fibers, if we bring them together we
would have an $A_{10} \oplus A_7$ singularity that clearly can not be
embedded in $\Gamma^{17,1}$. In the same way it is possible to show that at
any location we would get singularities that can not be realized if we want
to preserve the decomposition $\Gamma^{18,2}=\Gamma^{17,1}\oplus\Gamma^{1,1}$. 

\bigskip
\centerline{\epsfxsize=0.5\hsize\epsfbox{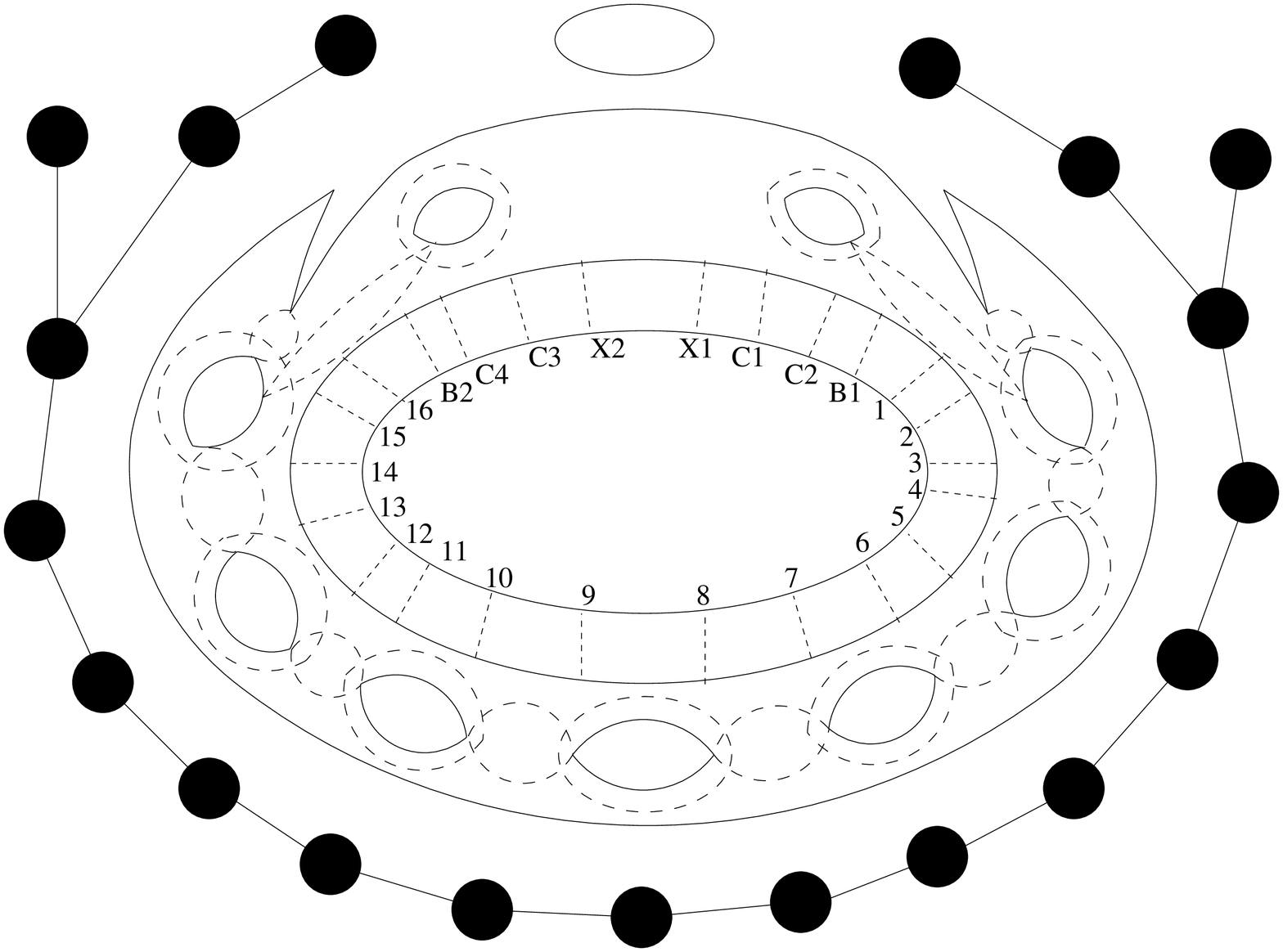}}
\noindent{\ninepoint\sl \baselineskip=8pt {\bf Figure 11}: {\sl Real $K3$
corresponding to the case with 24 real roots. The circles correspond to the
projection of the 2-cycles of $K3$ on the Real $K3$ giving the intersection
structure of $\Gamma^{17,1}$}}
\bigskip

{\bf Counting Parameters}

We have been considering real $K3$ in the form
$$y^2=x^3+f(z)x +g(z)$$
where $f$ and $g$ are real polynomials in $z$ of degree
$8$ and $12$ respectively. Thus to specify real $K3$ we have
$9+13$ parameters going into definition of $f$ and $g$
minus $3$ for $SL(2,R)$ symmetry which is the symmetry
preserving the real structure acting on $z$, and an overall
rescaling of the equation (the rescaling of $x$ and $y$
are frozen because we have chosen the coefficients of $y^2$ and
$x^3$ to be one in the above equation). This gives us a total
of $18$ real parameters. This is exactly the right number expected
based on the fact that we need to describe 16 Wilson lines and
the radii of two circles, the eighth and the ninth circle, as measured
say from the heterotic string side.  
This is of course consistent, as we discussed before with the
splitting of the lattice
$$\Gamma^{18,2}\rightarrow \Gamma^{17,1}+\Gamma^{1,1}$$
where 17 parameters (16 Wilson lines and the ninth radius)
go into defining the moduli of $\Gamma^{17,1}$ and the
eighth radius defines the moduli of $\Gamma^{1,1}$.  In fact
in principle we can read off the exact point we are
on the moduli, by simply measuring the volume (using the
metric on $K3$) of the corresponding 2-cycles which correspond
to elements in $\Gamma^{17,1}$ and those of $\Gamma^{1,1}$
(which gives the $P_R$ components of the corresponding
elements, from which we can reconstruct the Lorentzian
rotations).
However, here we wish to develop some intuition in particular
for the limit corresponding to going to 9 dimensions.

In going to 9 dimensions, we need to decompactify
the eighth circle, which means taking the corresponding
heterotic radius to infinity.  As shown in equation \limfor\
this means that the size of two sphere should be much
bigger than the size of the elliptic class dual to it.  
The basic intuition we have, is in the case of F-theory,
to which our situation is equivalent up to a change
in complex structure.  In that case the limit one takes
is the elliptic fiber going to zero size.  Moreover,
in that limit the metric on $K3$ becomes independent of the
position on the elliptic fiber (i.e. has an approximate $U(1)\times
U(1)$ symmetry.  Similarly here, we should first identify
the analog of the elliptic fiber
and then take the limit in moduli where it goes to zero size.
 More precisely
 we require that  the integral 
$$\int_E dz{dx \over y} \rightarrow 0$$
We now need to get a better understanding of the elliptic class
$E$ dual to our sphere.  It should intersect it at a point.  More
precisely, for every point on the real two sphere, there should exist
exactly one $E$ class, with a canonical BPS cycle chosen by minimizing
the volume.  

\bigskip
\centerline{\epsfxsize=0.5\hsize\epsfbox{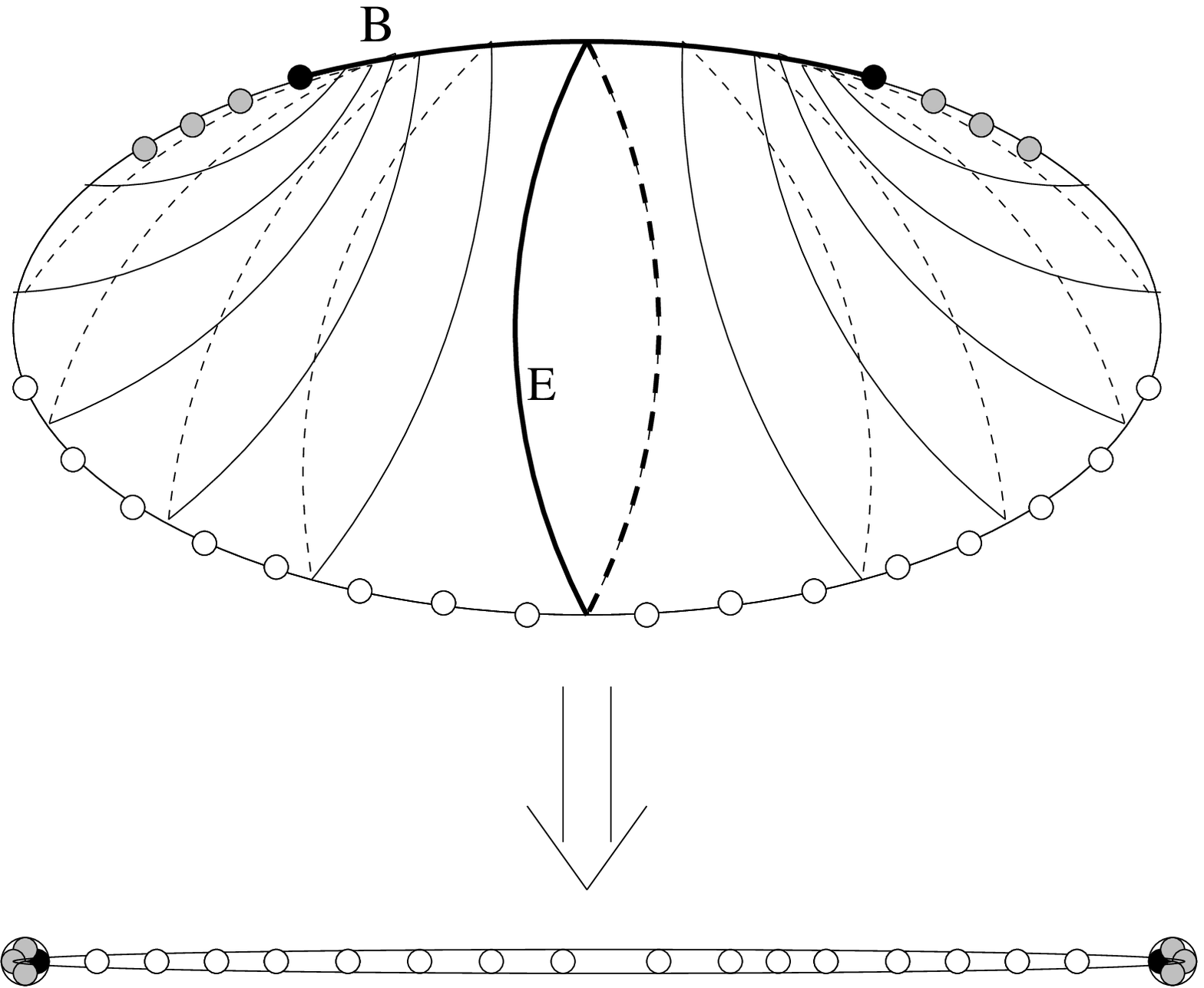}}
\noindent{\ninepoint\sl \baselineskip=8pt {\bf Figure 12}: {\sl New elliptic 
fibration. The circles depicted here are the projection
of the new $E$ class representative on the $z$-sphere that shrink to
zero. $B$ is the projection of the new $P^1$ base on the old $P^1$. White
dots are relative local branes, black and grey dots are non local to the
white branes.}}
\bigskip

The real two sphere is a circle fibered over an interval
on the real $z$-line where at the two ends it shrinks.  This
circle can be identified with one of the two circles in the
original elliptic fibration of F-theory.  
The elliptic class $E$ dual to our real sphere, should have
a cycle intersecting the cycle of the real
sphere at one point, and corresponding to a cycle
on the $z$-sphere, crossing the interval at one point transversally
as shown for example in figure 12, where the image of $E$ on the
sphere is drawn as circles.  The case depicted
in figure 12 is when all the 24 branes are on the real axis.
Note that we can also view the elliptic class as a choice
of a circle in the $z$-plane which intersects
the interval $B$ transversally, and for which there is an invariant monodromy
direction, which intersects the cycle corresponding
to real sphere on the original elliptic fiber of F-theory  at one point. 
As we change
the intersection point of $E$ with the interval $B$ the 
image of $E$ crosses some of the branes on the real $z$-axis.
We wish to argue that precisely those that it crosses are branes of the
same $(p,q)$ type, as that of the cycle in the elliptic fiber
which represents
the cycle of  $E$.  This follows from the fact that as we cross
the brane, represented by a class $\beta$, then the class $\alpha$
of $E$ changes by
$$\alpha \rightarrow \alpha -(\alpha \cdot \beta ) \beta$$
where the dot product is in the 1-cycles of the elliptic fiber.
Since the infinitesimal change of the cycle should not affect the $E$ cycle
globally (in particular
we could choose the same cycle
near $B$), this implies that $\alpha \cdot \beta =0$.  In particular
all the cycles that $E$ crosses are all of the same type!
This is beginning to sound like type I' as in type I'
only the branes of the same type are allowed.  However
this also implies that the E cycle cannot cross all the branes,
because they are not all local relative to one another.  Indeed
what we will find is that precisely for the case we have depicted
in figure 12 exactly 16 of them are of the same type and
the $E$ can pass through them, and the last and first branes
on either side (depicted by gray dots in the figure) will be
the boundary of where the $E$ cycle reaches.  These
limiting cases would correspond to when the $E$ cycle crosses the interval
$B$ at one of its boundary points.

Now the limit of going to $9$ dimensions is clear:  We simply
have to take the brane configurations corresponding to having zero
size for {\it all} the cycles represented by $E$ on the $z$-plane
to be of zero size.  This in particular means, in the case
where all the 24 branes are on the real axis, that the first
four and the last four approach each other (this must also
necessarily shrink all the other cycles as they are all represented
by the same integral).  Note that now we are also left with
an effectively one dimensional object, with 16 branes on it,
the boundaries of which are identified with the first and last
branes (depicted by the gray dots). See Figure 12. Note that similar
limits were taken in \aspinwall\ but in order to get a 10 dimensional
picture of the heterotic strings.

{}From the viewpoint of counting
parameters,
in this limit we see that we have
naively 20 real parameters left:  16 branes in the middle, 2 positions
of where each group of 4 branes has collapsed and 2 relative
$SL(2,R)$ invariants from cross rations of each of the two groups.
The overall $SL(2,R)$ gets rid of 3 of them and we are left with
17 parameters which we can take to be 16 positions of the branes,
and one parameter controlling the type I' coupling at 
one of the orientifolds given by
one of the cross ratios (the other one being fixed in terms of the
rest). 

Here we have mainly concentrated on the case where all the 24 branes
were on the real line, but we know that 2 or 4 of them could be
off of it.  This should correspond to bringing one brane
to the orientifold and crossing it. This means taking
one of the white nodes in Fig. 12  and bringing it close
to the last curve of the $E$ foliation.  As that meets
the first gray node, they can pair up to go to the complex
plane. This corresponds to the fold disappearing off the genus
10 surface. In that case the second gray brane becomes the visible
brane, i.e. becomes `white' , as the last foliation of $E$ gets
pushed further back (see Fig.13). 

\bigskip
\centerline{\epsfxsize=0.5\hsize\epsfbox{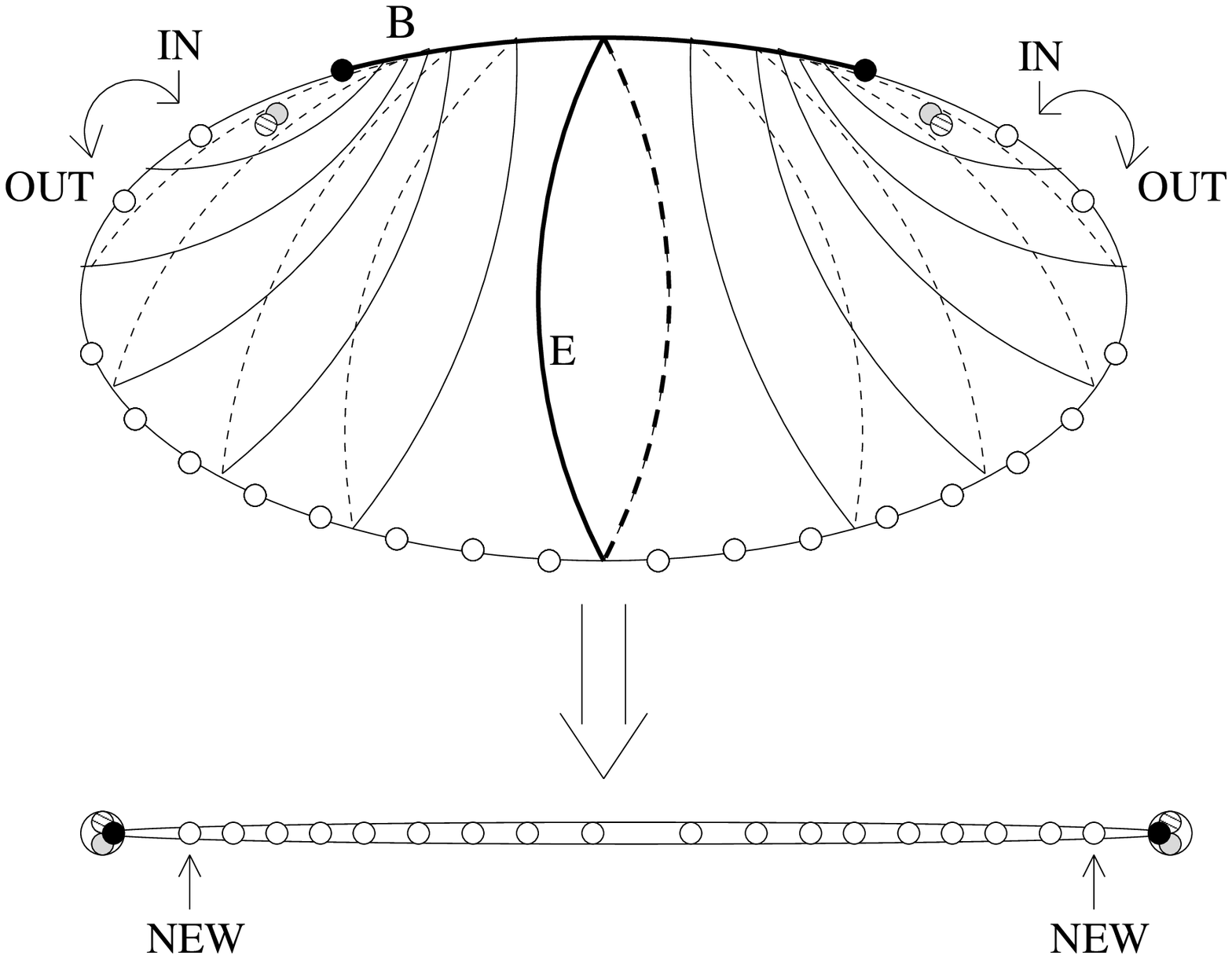}}
\noindent{\ninepoint\sl \baselineskip=8pt {\bf Figure 13}: {\sl New elliptic 
fibration. The circles are the $E$ class representative that shrink to
zero. $B$ is the projection of the new $P^1$ base on the old $P^1$. White
dots are relative local branes, black and grey are non local to the white
branes. There are two branes that can be IN or OUT the minimal $E$ cycles
 giving
NEW white branes when they are OUT.}}
\bigskip

 This we will demonstrate
using monodromy arguments below.  So far, even though two
branes have gone to the complex plane, we are still in the
regime of perturbative type I', though
in the regime where $-\theta_2<\theta_1<0$.  Again the last curve $E$, which
shrinks to zero size in going to 9 dimensions, has four branes in it,
and thus one real parameter describes their relative moduli, which
is to be identified with the coupling at the orientifold.  Now as we bring
the last gray brane towards the physical brane beyond some point
the curve $E$ will not include it anymore, which corresponds
to pulling one extra brane out (see Fig.13). This is exactly
what we were anticipating from the analysis we made of Type I'
perturbation breakdown.  Moreover, note that now inside the
last curve of $E$ we have 3 branes, and thus no relative moduli.
This means that there is no degree of freedom associated with the
orientifold, and in particular the coupling must be frozen there.
We will demonstrate below, that the value is given by $1/g =0$,
completing what we expected from all the regimes of moduli
of heterotic strings.

\subsec{Quantitative analysis}

In this part we will start by giving the monodromy analysis of the
configuration of branes that we studied in the qualitative analysis. 
This will show that the statements
we made about the monodromy along the cycle denoted by $E$ are
correct. Then we will go to the quantitative description of the 9
dimensional limit and its connection to type I' like descriptions.

\bigskip 

{\bf Monodromy Analysis}

Luckily the monodromy of various singularities relevant
for us have been extensively studied in
\wolfe , whose results we will borrow
(for detailed conventions we refer the reader to their paper).

In F-theory the singular fibers are associated with 7-branes of 
the $(p,q)$ type (once we choose a convention of branch cuts
on the ${\bf P}^1$) \foot{$p$ and $q$ are coprimes and a brane $(p,q)$ is
identified with a brane $(-p,-q)$}. 
Let us denote the brane of $(p,q)$ type by $X_{[p,q]}$.
The monodromy around any $X_{[p,q]}$ fiber is given by 
\eqn\mono{K_{[p,q]}=  \left( \matrix{\hfill 1+pq \hfill  &&& 
\hfill -p^2 \hfill \linesp 
                                          \hfill q^2 \hfill &&& 
                                          \hfill 1-pq \hfill } \right) }
 
Note that when we get $n$  branes
of the same type 
together we get an $A_{n-1}$ singularity.  Now let us 
consider the case where we have all the branes on the real
axis. We identify this with the configuration obtained in \wolfe\ by
starting with ${\hat E}_8,~ {\hat E}_8,~A,~A$ (where ${\hat E}_8$ is the
affine version of $E_8$) and then going 
down to ${\hat E}_2,~{\hat E}_2,~A^{16}$. 
Where $A=X_{[1,0]}$ and ${\hat E}_2$ will be given below.  

The brane assignments in this geometry have been given. If we
set the 16 relative local branes to be the $A$ branes, then the
non-trivial structure for the branes denoted by $B1~C2~C1~X1$ (and similarly
for $B2~C4~C3~X2$) are given by 
$$ {\hat E}_2 = BCCX_{[3,1]}$$
where $B=X_{[1,-1]}$ and $C=X_{[1,1]}$. This makes clear the notation chosen to denote the branes in Figure
11. In Appendix B we give a review of
the properties of the ${\hat E}_2$ configuration that lead to the
identification made.

This assignment of brane type can be replaced by any of its equivalent
configurations\foot{By equivalent we mean that they 
are related by moving branch cuts. In general,
global $SL(2,Z)$ conjugations are also allowed but in our case we restrict
the possibilities only to those of the $T$ type since we do not want to
change the type of the 16 $A$ branes. See \wolfe\ for the general case
and more details.},
\eqn\bcbc{BCCX_{[3,1]} \sim BCBC} 
\eqn\bccx{BCCX_{[3,1]} \cong X_{[-1,-1]}X_{[3,1]}X_{[3,1]}X_{[5,1]} 
\sim CX_{[3,1]}X_{[3,1]}X_{[5,1]}}
where in the first case the branch cut of the $X_{[3,1]}$ brane was moved
through the $C$ brane next to it so that it becomes a $B$ on the other side
of the $C$ brane. In the second case $\cong$ means a global conjugation by
$T^2$ and after that we used that $X_{[-1,-1]} \sim X_{[1,1]}=C$. 

If we compute the monodromy \foot{Following the conventions of \wolfe\
, for a configuration of branes $X_{[p_1,q_1]}\ldots X_{[p_n,q_n]}$ the
total monodromy is given by $K_{[p_n,q_n]}\ldots K_{[p_1,q_1]}$} 
around any ordered collection of the branes one
finds that it does not have an invariant direction, except when we take the
monodromy by the whole group, in which case the monodromy is given by
$T^8$ where $T$ and $S$ are the generators of $SL(2,Z)$. Each of the other 16
branes being of the $A$ type has a $T^{-1}$ monodromy. Therefore, each time
the $E$ cycle passes through one of them, the monodromy matrix shifts its
power by $-1$, until it reaches $T^{-8}$ around the last cycle on the other
side.  Note the power of monodromy (taking into account orientations)
is correlated with the $D8$ brane charge we wish to assign to the orientifold
for type I'.
  
We also discussed the possibility of moving one $A$ brane towards one of
the group of 4 branes contained in the last $E$ cycles. Once we do that there
is one more possible equivalent configuration given by,
$$ ABCCX_{[3,1]} \sim AABX_{[2,1]}X_{[5,1]}$$
It is important to follow the chain of equivalence relations that led to the
above result,
\eqn\cxxx{ABCCX_{[3,1]} \cong ACX_{[3,1]}X_{[3,1]}X_{[5,1]} \sim
CX_{[2,1]}X_{[3,1]}X_{[3,1]}X_{[5,1]}}
\eqn\axxxx{\sim CAX_{[2,1]}X_{[3,1]}X_{[5,1]}\sim
AX_{[0,1]}X_{[2,1]}X_{[3,1]}X_{[5,1]}}
\eqn\aabxx{\sim AX_{[0,1]}AX_{[2,1]}X_{[5,1]}\sim AABX_{[2,1]}X_{[5,1]}}
where in \cxxx\  $\cong$ is that of \bccx\ and $\sim$ is moving
$A$ through the $C$ brane becoming an $X_{[2,1]}$ brane. In \axxxx\ the
first $\sim$ is  moving the left $X_{[3,1]}$ through $X_{[2,1]}$ becoming an $A$
brane and the second is moving the $C$ through the $A$ becoming a
$X_{[0,1]}$ brane. Finally, in \aabxx\ the first $\sim$ is moving the remaining $X_{[3,1]}$ through
$X_{[2,1]}$ becoming an $A$ brane and the last $\sim$ is moving the
$X_{[0,1]}$ through the $A$ brane becoming a $B$ brane.

This makes clear that the two $A$ branes in the final result are the same
as the two $C$ branes of the initial configuration and that the initial
$A$ brane is now a $X_{[2,1]}$ brane. 

Let us use this chain to explain the transition between the different Real
$K3$'s discussed earlier in this section. Consider in Figure 11 the first $A$ brane
approaching $B1$ that according to the result of \bccx\ can be thought of as
a $C$ brane.  These two form an $H_0 = AC$ configuration that has no gauge
symmetry associated to it and then take off the real line. This is
illustrated in \cxxx . Now the $C2$ brane -- given by $X_{[3,1]}$ in \cxxx\
-- becomes an $A$ brane in \axxxx\ and it is free to move in the
interval. The remaining four branes still have a total $T^8$ monodromy. We
simply have exchanged one $A$ brane by another $A$ brane. This is the same
as exchanging the last brane with its mirror brane in the Type I'
language. Now however, we can also push the $C1$ brane pass the $X_{[2,1]}$
monodromy and as before, it converts the $X_{[3,1]}$ brane again to an $A$
brane. This is given in \aabxx . Now if we consider the foliation of the
$E$ curve to pass just through the $B~X_{[2,1]}~X_{[5,1]}$ branes, then we
obtain the monodromy $T^9$. This is again correlated with the charge at the
orientifold expected when we pull an extra brane off.

\bigskip
\bigskip

{\bf Type I' like descriptions}

In the F-theory configuration studied by Sen \sen\ , where the gauge group
was $SO(8)^4$, it was possible to show that the Type IIB description was
nothing but the orientifold  $T^2/(-1)^{F_L}\cdot \Omega \cdot {\bf
Z_2}$. It was also shown that upon
T-duality on each of the circles of the $T^2$ this orientifold was
equivalent to Type I on $T^2$. Suppose now that one of the circles of the
Type I theory is very small, then we have to go to the T-dual description that is
Type I'.  Therefore it is natural to expect that the F-theory description
in some regimes can be connected to Type I' upon a single T-duality. Of
course, this is only possible if on the F-theory side we are working with a
configuration that admits a natural $S^1$ action and in the qualitative
analysis we found that this is the case.

In order to study the 9 dimensional limit and get explicit results to match
with the Type I' description, we will consider explicit expressions for
the metric in the limit corresponding to
large eighth radius, and try to get the relevant physical quantities in
9 dimensions.

The crucial point that we will try to argue now is that in the 9
dimensional limit, the generic fiber of the original F-theory
compactification is degenerating and moreover the complex structure
will grow like $\tr \sim R_8$.

In F-theory, as it was pointed out in \mov , the middle monomial of $f(z)$
and $g(z)$ will control the complex and k\"{a}hler structure of the 
dual heterotic
$T^2$. In the limit to 9 dimensions both moduli blow up and therefore we
expect that the corresponding monomials will dominate the rest at generic
points on the $z$ sphere.  (See section 8.1 for more details)

This means that in order to get information from the general behavior 
of the periods of $B$
and $E$ we can consider the elliptic equation to be of the form $y^2 = x^3
+ \alpha z^4 x + \beta z^6$. The corresponding periods are obtained, as it
was explained in the qualitative part, by integrating the holomorphic 2-form
over the corresponding cycle\foot{Here $E$ and $B$ denote the new elliptic
fiber and the new base respectively, but also they are used to denote the projection of
the corresponding 2-cycles on the F-theory base.}. 
$$ \Gamma_E = \int_E { dx \over y }dz \;\;\;\; {\rm and} \;\;\;\; \Gamma_B =
\int_B {dx \over y}dz $$
rescaling $x\rightarrow z^2 x$ we get the following result,
$$\Gamma_E = \int_a {dx \over \sqrt{x^3 + \alpha x + \beta }}\int_E
{dz\over z} = 2\pi i \left( {w_1 \over 2}\right) $$
where the first integral is over the $a$ cycle of the generic fiber of
F-theory. This integral is nothing but the first half period ${w_1 \over
2}$ and the integral over $z$ is independent of the 1-cycle $E$ and gives
only $2\pi i$ since the $E$ cycle winds around the origin once.

The period over $B$ is given by 
$$\Gamma_B = \int_b {dx \over \sqrt{x^3 + \alpha x + \beta }}\int_B
{dz\over z} = {w_2 \over 2} \ln ({z_{X2} \over z_{X1}})$$
where the first integral is over the $b$ cycle and gives us the second 
half period.

But we know that $\Gamma_{E} \sim {1\over R_8}$ and $\Gamma_{B}\sim
R_8$. This immediately implies that $w_1 \sim {1\over R_8}$ and $w_2 \ln ({z_{X2}
\over z_{X1}})\sim R_8$. This tells us that 
$$ \tau = {w_2 \over w_1} \sim R_8$$
at least. But it can not have higher powers of $R_8$ since $\tau$ has to
reduce to the heterotic complex structure in the 10 dimensional limit. This
implies that $w_2 \sim 1$ and therefore 
\eqn\save{\ln \left( {z_{X2} \over z_{X1}} \right) \sim R_8}
Now we are ready to continue with the analysis of the limit to 9 dimensions.

The metric for the sphere in the type IIB compactification equivalent to
 F-theory on $K3$ was given as part of the 10 dimensional metric in
\metric\ .

Certainly this metric is very hard to handle under general considerations
but here we will make use of some approximations that will become exact in the
strict limit to 9-dimensions.

Using that ${\rm Im}(\tau )\gg 1$ it is possible to write,
\eqn\approx{\eta (\tau )= q^{1/24} \;\;\;\; {\rm and} \;\;\;\;
 q^{-1} = j(\tau ) = {f^3 \over
\Delta }\;\;\; \Rightarrow \;\;\; {\eta^2 \over \Delta^{1/12}} =
 {1\over f^{1/4}} }
 
Therefore the metric reduces to,
$$ds^2 = k {\rm Im}(\tau ) \left| {dz \over f^{1/4}(z)} \right|^2 $$  
This computation only makes sense in the regions where the only monodromies
that $\tau$ can find are of the form $T^n$ for some integer $n$. But we saw
that this is the case in the regions that we expect to see in the 9
dimensional limit.

Moreover, now we can make use of our knowledge about the zeros of
$f(z)$. From Figure 8$D$ we can conclude that in the case where the two
branes are real $f(z)$ should have four real zeros inside the minimal cycle
of the $E$ class. If the two branes are complex, then we will see in the
Appendix B that $f(z)$ still have four zeros (only two of them are real)
inside the minimal $E$ cycle even in the case where the latter contains only
three branes.    
Therefore we will have control over $f(z)$ in the 9
dimensional limit when the two minimal $E$ cycles will effectively shrink
to a point.
The final point we need to work out is the behavior of ${\rm
Im}(\tau )$.

A very convenient choice of coordinates over the $P^1$ is one that will
parametrize in a natural way the circle action -- that become an actual
symmetry in the limit -- induced by moving along the $E$ class
representatives and the position along the $B$ cycle that is the
projection of the new $P^1$ on the F-theory sphere. See Figure 12 and 13.

Let us start by locating the two groups of four branes one near $z=0$ and
the other near $z=\infty$. This means that,
\eqn\cut{|z_i| >{\rm Max}(|X1|,|C1|,|C2|,|B1|) \;\;\;\;\; |z_i| < {\rm Min}(|X2|,|C3|,|C4|,|B2|)} 
where we have denoted by $z_i$ with $i=1,\ldots ,16$ the position of the 16
$A$ branes. The branes are ordered by $|z_i| \le |z_j|$ if $i < j $.   

Let $|X1| > {\rm Max}(|C1|,|C2|,|B1|)$  and  $|X2| < {\rm Min}(|C3|,|C4|,|B2|)$,
therefore we have, $|X2| \ge |z_{16}| \ge \ldots \ge |z_1| \ge |X1|$.

The polynomial $f(z)$ can be written as follows,
$$f(z) = C_2 \prod^4_{i=1}(z-{\hat z}_i)(z-{\hat {\tilde z}}_i)$$  
where $|\hat{z}_i| > |X2|$ and $|{\hat{\tilde z}}_i| < |X1|$ for all $i=1,\ldots 4$. 
If we concentrate in the region $|X1|<|z|<|X2|$ that will be the physical
region after the limit, then $f(z)$ can be written approximately  as,
$$f(z) = C_2 \left( \prod^4_{i=1}{\hat z}_i\right) z^4$$
This implies that the metric is roughly $ds^2 \sim \tr \left| {dz \over z}\right|^2$, this
motivates a conformal transformation from the sphere $z$ to a cylinder by,
$$z = e^w$$
In this new coordinates we can write the discriminant as follows,
$$\Delta = C_1 e^{12(w+w_{cm})}\prod^{24}_{i=1}\sinh ({1\over 2}(w-w_i))$$
where $w_{cm}={1\over 24}\sum^{24}_{i=1}w_i$ and $w_i$ are the position of
the 24 branes in the new coordinates.

{}From \save\ we see that the natural coordinates to introduce in order to
parametrize the region between $X1$ and $X2$ are $w = R_8 (y_1 + i y_2)$
where $y_1,y_2 \in {\bf R}$.

Now, in the case where $R_8 \gg 1$ we have that,
$$ \sinh \left( \half (w-w_i)\right) \rightarrow {\rm (Phase)}e^{R_8 |y_1-y_1^i|/2}$$
Therefore, the discriminant can be written as,
$$\Delta = C_1 e^{R_8(12(y + y_{cm}) + {1\over
2}(y_{x2}+y_{C3}+y_{C4}+y_{B2})-{1\over 2}(y_{x1}+y_{C1}+y_{C2}+y_{B1})+{1\over 2}\sum^{16}_{i=1}|y-y_i|)}$$
In this expression we have replaced $y_1 ={\rm Re}(y)$ by $y$ itself in order to
save some notation. The $C_1$ constant contains all the phase factors
that were produced in addition to a normalization constant that will be
fixed later.

In this coordinates $f(z)$ also takes a special form,
$$f(z)=C_2 e^{R_8 (4 y + \sum^4_{i=1}\tilde{y}_i)}$$
Now we are ready to compute $\tr$. This can be done by using \tauf .
$$j(\tau )=1728 {4 f^3 \over \Delta }= ({\rm Phase})e^{R_8 ( C_3 + 3
\sum^4_{i=1}\tilde{y}_i -(y_{x2}+y_{C3}+y_{C4}+y_{B2}) -{1\over
2}\sum^{16}_{i=1}y_i -{1\over 2}\sum^{16}_{i=1}|y - y_i| )}$$
where $C_3$ is defined to contain all the constants that might arise from
$f(z)$ and $\Delta$. And the factor in front of the exponential contains all
the phases from the different terms. This phase factor will not affect
$\tr$ and reveals the independence of $\tr$ from the radial direction of
the cylinder. 

We will use that $j(\tau ) \sim q^{-1}$.  This approximation will become exact
in the limit $R_8 \rightarrow \infty$ and it is valid in all the regions we are considering, even close to
the end points . Now we have to distinguish between the three regimes
that were analyzed in previous discussions.

{\bf Case I}: All 24 branes are real.

Using the above approximation we get that,
$$Im(\tau ) ={1\over 2\pi}R_8(C_3+3
\sum^4_{i=1}\tilde{y}_i -(y_{x2}+y_{C3}+y_{C4}+y_{B2}) -
 {1\over 2}\sum^{16}_{i=1}y_i - {1\over 2}\sum^{16}_{i=1}|y - y_i|)$$ 
Now let us use that in the limit, $\tilde{y}_i\sim y_{C3}\sim y_{C4} \sim
y_{B2} \sim y_{X2}$ in order to replace all of them by $y_{X2}$, therefore
we get that,
$$Im(\tau ) ={1\over 2\pi}R_8\left( C_3+8\left[ (y_{X2} - y_{cm}) - {1\over
16}\sum^{16}_{i=1}|y - y_i|\right] \right)$$
by the symmetry of the problem, we can always choose $y_{cm} ={1\over
16}\sum^{16}_{i=1}y_i$ to be smaller or equal to zero without lost of
generality. (This is analog to the freedom of 
reflecting the interval coordinates in
Type I' in order to put the branes as close as possible to the origin). 

Now we see that $\tr \left|_{(y=y_{X2})}=R_8 C_3\right.$ 
and $\tr\left|_{(y=-y_{X2})}=R_8(C_3-16y_{cm})\right.$
where $C_3$ has to be fixed by an actual computation of the average of
$\tr $ around the minimal $E$ cycle containing the four branes. This value
is controlled by the cross ratio of the four branes positions that is an
$SL(2,{\bf R})$ invariant and it is the only modulus that survives the limit.  
This computation tells us that $\tr$ is positive in the physical region,
signalling that our computation is valid.

It is important to notice that the appearance of a piece wise linear
function is due to the fact that the logarithmic behavior of the 2
dimensional ``electric'' potential of the $A$ branes -- that are point like
charges in the sphere -- is smoothed out to a linear function that
changes slope at the position of the charge when the limit is taken and the
sphere reduces to a 1-dimensional object.    

{\bf Case II}: A pair of branes are complex and 22 branes are real.
If we are in the regime of parameters where there are still
4 branes enclosed by the last $E$ curve (on the side where
the pair of branes have become complex), then the story is as
in case I.  However, if the last $E$ curve encloses only
three branes, the story changes.
Following the same analysis as in case I we get, 
$$Im(\tau ) ={1\over 2\pi}R_8(C_3+3
\sum^4_{i=1}\tilde{y}_i -(y_{x2}+y_{C3}+y_{C4}) - 
{1\over 2}\sum^{16}_{i=0}y_i - {1\over 2}\sum^{16}_{i=0}|y - y_i|)$$ 
Notice that now the sums run from $0$ to $16$ since a {\bf new} $A$ brane
is in the bulk and its position is denoted by $y_0$.  
The answer in this case is 
$$Im(\tau ) ={1\over 2\pi }R_8\left( C_3+9 y_{X2} - {17 \over 2}y_{cm}  - {1\over
2}\sum^{16}_{i=1}|y - y_i|\right)$$
Now we see that 
$$\tr\left|_{(y=y_{X2})}=R_8 (C_3+{1\over 2}y_{X2})\right.~~{\rm and}~~ \tr\left|_{(y=-y_{X2})}= R_8(C_3 + {1\over
2}y_{X2} - {17\over 2}y_{cm})\right.$$

In this case the value of $C_3$ can be fixed completely because we only have
three branes inside the minimal cycle and therefore no parameter left after
the limit.

{\bf Case III}: Two pairs of branes are complex and 20 branes are real.
Again if the last $E$ curves on either side enclose 4 branes, or
one contains 4 and the other 3, we are reduced to cases I and II above.
However if the last $E$ curve encloses only 3 branes on each side
the story changes.   We now have
two more `visible' branes in the bulk (whose positions are denoted by
$y_0$ and $y_{17}$) and 
the same analysis as above results in
$$Im(\tau ) ={1\over 2\pi}R_8(C_3+3
\sum^4_{i=1}\tilde{y}_i -(y_{x2}+y_{C3}+y_{C4}) -
 {1\over 2}\sum^{17}_{i=0}y_i - {1\over 2}\sum^{17}_{i=0}|y - y_i|)$$ 
and using the approximation $\tilde{y}_i \sim y_{C3} \sim y_{C4} \sim y_{X2}$
we get,
$$Im(\tau ) ={1\over 2\pi}R_8\left( C_3+9\left[ (y_{X2} - y_{cm}) - {1\over
18}\sum^{17}_{i=0}|y - y_i|\right] \right)$$

Now we have that $\tr \left|_{(y=y_{X2})}=R_8 C_3 \right.$ and $\tr
\left|_{(y=-y_{X2})}=R_8(C_3 -18 y_{cm}) \right.$. In this case the two values
can be computed. Using the counting of parameters, we can set only 17
positions of the branes to be independent and therefore $y_{cm}$ is frozen
to zero. Moreover, as discussed in Case II, the value of $\tr$ at the ends
is also frozen since we only have three branes and no cross ratio can be
constructed, hence, no modulus survives the limit.

The computation of the constant $C_3$ can be done explicitly in the Cases
II and III, since all we have to do is to choose a configuration where the
three branes are isolated and compute $\tr$ as they collapse to a point. 
The configuration we are talking about is called $\hat{\tilde E}_0$
\bartetsen\ \wolfe\
and it is known to be out of the reach of Kodaira singularities at finite
distance in moduli space in 8 dimensions. However, the limit to 9
dimensions is at infinite distance and the collapse to 
a point becomes meaningful.
In the appendix A it is shown that following the analysis of \bartetsen\ the
$\hat{\tilde E}_0$ configuration can be properly isolated and the value we
are looking for is given by $\tr = \half$. 

In Case II, this implies that $R_8 (C_3 + {1\over 2}y_{X2}) =\half$ and in
Case III we get that $R_8 (C_3)=\half $.
 
Therefore, in the limit to 9 dimensions we get in the first case that $C_3
+{1\over 2}y_{X2}\rightarrow 0$ and in the second $C_3 \rightarrow 0$ since
they go as ${1\over R_8}$.

Summarizing the results of this part, we have found that 
the metric in each case
is given by
$$ ds^2 = k Im(\tau )\left| {dz \over z}\right|^2 +
\eta_{\mu\nu}dx^{\mu}dx^{\nu} = kR_8^2 Im(\tau )\left( dy_1^2 + dy_2^2\right) +
\eta_{\mu\nu}dx^{\mu}dx^{\nu} $$ 
where $k$ is an overall constant that is the usual F-theory - heterotic
duality contains information about the 8-dimensional heterotic coupling
since the latter is related to the overall volume of the $P^1$ base.  
And $\tr$ is given by,
$$Im(\tau ) = \left\{ \matrix{\hfill {1\over 2\pi}R_8\left( C_3+8\left[ (y_{X2} - y_{cm}) - {1\over
16}\sum^{16}_{i=1}|y - y_i|\right] \right)  \hfill
&&& \hfill  {\rm Case}\;{\rm I}   \hfill
\linesp \hfill {1\over 2\pi }R_8\left( {17\over 2}\left[ y_{X2} - y_{cm}  - {1\over
17}\sum^{16}_{i=1}|y - y_i|\right]\right)  \hfill &&& \hfill  {\rm
Case}\;{\rm  II}  \hfill \linesp
\hfill  {1\over 2\pi}R_8\left( 9\left[ y_{X2} - {1\over
18}\sum^{17}_{i=0}|y - y_i|\right] \right)  \hfill &&& \hfill  {\rm
Case}\;{\rm  III}   \hfill }  \right. $$

Up to terms of order $0$ in $R_8$ that will become irrelevant in the limit
$R_8 \rightarrow \infty$

\bigskip

{\bf Connection to Type I' and Type I' like descriptions}

Now we are ready to find the connection of our descriptions to Type I' in
the case where we expect Type I' to be valid using the observation made
from Sen's analysis.

In the coordinates for the cylinder we find that the 10 dimensional metric is given by
$$ds^2 = kR_8^2 \tr ( dy^2_1 + dy^2_2 ) + \eta_{\mu\nu}dx^{\mu}dx^{\nu}$$
where $\mu ,\nu = 0, \ldots , 7$. (Here we are back to the notation $y=y_1+iy_2$)

This metric is in the Einstein frame. This is the frame where the
$SL(2,{\bf Z})$ U-duality group of IIB theory is manifest. 

The idea is to perform a T-duality along the circles that 
the limit to 9-dimensions has
created for us.   
Therefore, we first have to go to the string frame by
using that $G^{(E)} = e^{-{\phi_{{\rm IIB}}\over 2}}G^{(S)}$.
 By definition $Im(\tau )=
e^{-\phi_{{\rm IIB}}}$ and therefore we get,
\eqn\finite{ds^2_{(s)} = kR_8^2 (\tr )^{1/2}(dy^2_1 + dy^2_2) + (Im(\tau
))^{-1/2}(\eta_{\mu\nu}dx^{\mu}dx^{\nu})}

In the decompactification limit, the $y_2$ dependence of $Im(\tau )$ drops
out as it was shown before. This makes manifest that the circle action
predicted in the qualitative part has become exact.

Before performing the T-duality we should give make $k$ a bit more
explicit. $k$ can be computed -- as we will do in the
$E_8\times E_8$ example -- by the fact that the volume of the $P^1$ of
F-Theory should be equal to the heterotic coupling in 8 dimensions.

The volume is easy to compute since the metric is just a cylinder with
varying radius in $y_1$. Therefore, using the metric we have,
$${\rm Vol}(P^1) = 2\pi k R_8 \int^{y_{x2}}_{-y_{x2}}F(y)dy$$
where we have defined $F(y) = {2\pi \tr \over R_8}$. Therefore the integral
is finite in the limit $R_8\rightarrow \infty$.

Using now that ${\rm Vol}(P^1)=e^{\phi_{h8}}$ and that
$e^{\phi_{h8}}={e^{\phi_{h(10)}}\over (R_9 R_8)^{1/2}}$ we get that the
singular behavior of $k$ is $k\sim R_8^{-5/2}$. This result tells us that
the metric \finite\ behaves as follows, 
$$ds^2 \sim F(y_1)^{1/2}(dy^2_1+dy_2^2)+R_8^{-1/2}F(y_1)^{-1/2}(\eta_{\mu\nu}dx^{\mu}dx^{\nu})$$
Let us now perform the T-duality in the 8-th direction. This will only
change the radius of the 8-th circle to its inverse. Remember that $y_2\sim
y_2 + {2\pi \over R_8}$. Introducing a new variable $x^8$ and rescaling the
other $x^\mu$ the metric can be written schematically as follows,
$$ds^2 = F(y_1)^{1/2}dy_1^2 + F(y_1)^{-1/2}(\eta_{\mu\nu}dx^{\mu}dx^{\nu})$$
where now $\mu ,\nu = 0, \ldots , 8$.

Finally, we would like to bring the metric to a conformally flat
metric. This can be done by a simple change of coordinates since the metric only depends on $y_1$ and it is
already in the desired form in 9 of the dimensions. 
The change of coordinates will have to be such that,
\eqn\flatcoor{F(y_1)^{1/2} dy^2_1 = F(y_1)^{-1/2} dx^2_9} 
Once this is done the metric looks like $ds^2 =F(y_1)^{-1/2}
(\eta_{MN} dx^M dx^N)$ where $(M,N = 0,\ldots , 9)$.

The final step is to express $F(y)$ in terms of $x_9$. The result of doing
this is simply that, up to a constant\foot{In a more detailed computation,
as will be done for some examples in section 8.1, all the constants that we
have not written can be computed and reproduces precisely the results
expected from Type I'.}, 
$$ F(y) = (z(x_9))^{2/3} $$ 
where $z(x_9)$ are the functions in \zeto\ for the case with 24 real branes
and a generalization like \zeta\ for the case with 20 real branes.

For the example of $8.2$ we will be able to compute all the relevant
quantities of Type I' from this description, i.e., $B$, $C$ and the
position of the branes. 

The other computation that can be done explicitly is the appearance of the dilaton
of Type I' as the T-dual of the IIB dilaton.

Here we only need to use that the T-dual IIB dilaton ${\tilde\phi}$ is given by,  
\eqn\cross{ e^{{\tilde\phi}} = {\cal R}e^{\phi_{IIB}}}
where the dual radius ${\cal R}$ is given by $F(y)^{-1/4}R_8$. But
$e^{\phi_{IIB}} = (\tr )^{-1} = R_8^{-1}F(y)^{-1}$. Therefore we get,
$$ e^{{\tilde\phi}} = F(y)^{-5/4} = (z(x_9))^{-5/6} $$
that is precisely the behavior expected in Type I' as was shown in
in \abc .

\newsec{EXAMPLES}
In this section we first review the 9 dimensional limit
of $E_8\times E_8$ theory by taking the appropriate
limit of F-theory.  We then show how the real resolution
of $E_8 \times E_8$ gives rise to the real $K3$ structure
anticipated on general grounds.

\subsec{9 dimensional limit of $E_8 \times E_8$}

In the quantitative analysis of section 7 we described how the connection
between Real $K3$ configurations and Type I' like descriptions can
be achieved. It is the aim of this part of the section to apply all the
result we got to a simple case where we expect the usual Type I' description and to show
how non trivial quantities match with full precision. The point we are
going to study is the $E_8 \times E_8$. The relevant Type I'
description was given in section 2.1 with the important results at the end
of the section. 

We will start by considering the 8 dimensional description of the $\E$
theory with zero Wilson lines in terms of F-theory.

The moduli space of the heterotic compactification is given in terms of two
complex parameters, $U$ denoting the complex structure and
$T$ denoting the complexified K{\"a}hler structure of the $T^2$,
 and one real parameter given by the 8 dimensional coupling constant. 

The F-theory model in this case is given in terms of the following elliptic 
equation,
\eqn\ell{y^2=x^3+\a z^4 x + z^5 (1+\b z + z^2 )}

The defining polynomials are $f(z)=\a z^4$ and $g(z)= z^5 (1+ \b z +
z^2)$. $\a $ and $\b $ are complex parameters that should be mapped to the
$U$ and $T$ and the volume of the $P^1$ base will be related to the
heterotic coupling constant.

The explicit map was found in \lust\ and it is given by 
\eqn\map{j(iU)j(iT)=-1728^2 \left( {\a \over 3} \right)^3} 
\eqn\mep{(j(iU)-1728)(j(iT)-1728)=1728^2 \left( {\b \over 2} \right)^2}

In \map\ we can see that the $({\rm PSL}(2,Z) \times {\rm PSL}(2,Z))/Z_2$
 symmetry of
the heterotic compactification is explicit in the map.

Now let us take the decompactification limit on the heterotic side, this
means that we are taking a rectangular ($\t = 0$) torus with zero B-field
($B_{12}=0$) and $R^h_8>>R^h_9>R^h_{9c}$. With this choice we are breaking
the full  $({\rm PSl}(2,Z) \times {\rm PSL}(2,Z))/Z_2$ symmetry but at the 
end we will recover
the $Z_2$ symmetry of the $S^1$ heterotic compactification to 9 dimensions. 

In this limit, we can approximate $j(iU)=e^{2 \pi U}$ and $j(iT)=e^{2 \pi
T}$.

The discriminant of \ell\ is given by 
\eqn\dis{\Delta = 4 f^3+ 27 g^2 = z^{10}\left( 4 \a^3 z^2 + (1+ \b z +
z^2)^2\right) }
The roots of \dis\ can easily be found by noting that \dis\ can be factorized
in two quadratic pieces,
\eqn\fac{\Delta=27z^{10}\left( z^2 +\left( \b + 2\left( -{\a \over
3}\right)^{3/2} \right) z+1
\right) \left( z^2 + \left( \b
- 2\left( -{\a \over 3}\right)^{3/2}\right) z+1 \right) }
Using the map \map\ we find that in the limit the roots behave as follows,
\eqn\roots{z_1=e^{-\pi (T-U)} \;\;\;\; z_2=e^{\pi (T-U)} \;\;\;\; 
z_{X1} = -e^{-\pi
(T+U)} \;\;\;\; z_{X2} = -e^{\pi (T+U)}}
It is important to notice that the four roots came out real, two on $R^+$
and two on $R^-$ while the $E_8$ singularities are located at $z=0$ and at
$z=\infty$.

We can also obtain these results directly from our discussion
of the relation between the periods of the base and elliptic
fiber and the radius of the eighth dimension derived in \limfor .
The periods we are interested in can be written in the following form,
\eqn\periodone{\Gamma = \int_{\gamma } |p - \tau q| \left| {\eta^2 (\tau) \over
\Delta^{1/12}}\right| |dz| }
where $\gamma $ is the relevant 1-cycle on the F-theory $P^1$
base shown in Figure 12. This formula for the periods is the same 
that would give the mass of
a $(p,q)$ BPS open string stretched along $\gamma $. 
It is possible to choose basis in which $X1$ and $X2$ are $(3,1)$ branes and
$z_1$ and $z_2$ are $(1,0)$.

Here we will only assume that $Im(\tau )\gg 1$. This implies that the
period formula reduces to,
\eqn\periodtwo{\Gamma = \int_{\gamma } |p - \tau q||\alpha|^{-1/4} \left| {dz \over
z}\right| }
It is possible to remove the $|\alpha|^{1/4}$ factor since it is an overall
normalization of the periods and we will be computing only ratios.
Let us compute first the period corresponding to $\gamma = E$ and
the cycle $(1,0)$,
\eqn\periodthree{{\cal E}_8 =\int_{|z|=1}\left|{dz \over z}\right| = 2\pi }

Next, let us compute the period corresponding to $\gamma =B$ and
the cycle $(1,0)$, 
\eqn\periodfour{{\cal B}_9 =\int_{z_1}^{z_2}\left|{dz \over z}\right| = 
2 ln(z_2)}
where use have been done of the fact that $z_1 = 1 / z_2$. 
{}From the heterotic string we learn that the ratio of these to periods
should be equal to $T-U$.

This immediately tells us that,
\eqn\result{{{\cal B}_9 \over {\cal E}_8 }= T-U = { \ln (z_2) \over \pi}
\;\;\;\; \Rightarrow \;\;\;\; z_2 = e^{\pi (T-U)} }

In order to compute $z_{X1}$ we need the value of $\tau$ and following a
procedure similar to that of the previous section we get,
\eqn\tauper{\tau = {i \over 2\pi }\left( 3 \ln (\alpha ) - \ln (z_{X2}) -
\ln (z_2) \right)   }

{}From the elliptic equation we have that we can choose $z_2 \in {\bf R}^+$
and $z_{X2} \in {\bf R}^-$ and $\alpha \in {\bf R}^-$. 

This implies that $|\tau |= {1\over 2\pi} \ln \left( {\alpha^3
 \over z_{X2} z_2 } \right)$. 
Now we can compute the two other periods corresponding to $\gamma =
 E$ and the cycle $(3,1)$ and to $\gamma = B$ and the cycle $(3,1)$,
\eqn\periodfive{{\cal E}_9 =\int_{|z|=1}|\tau | \left|{dz \over z}\right| =
\ln \left( {\alpha^3 \over z_{X2} z_2 } \right) }

Finally, after a some computations,
\eqn\periodsix{{\cal B}_8 = \int_{z_{X1}}^{z_{X2}}|\tau | \left|{dz \over
z}\right| = {1\over 2\pi}\left( 2\ln (-z_{X2})\ln (-\alpha^3)- 3\ln^2
(-z_{X1}) - \ln^2 (z_2)   \right)  }

Using the ratio ${\cal E}_9/{\cal E}_8 = U$ it follows from
\periodthree\ , \periodfive\ and \result\ that, 
\eqn\resulttwo{{\alpha^3 \over z_{X2}} = e^{\pi (T+U)}}

The last independent ratio we have is,
\eqn\resultthree{{{\cal B}_8 \over {\cal E}_8} = TU  }

Using \result\ , \resulttwo\ and \resultthree\ we get our final result,
\eqn\finalr{z_{X2} = - e^{\pi (T+U)} \;\;\;\;\;\;\;\;  z_1 = e^{\pi (T-U)} }
It is important to remark that the periods for 
this example can also be computed
explicitly without the use of the 9-dimensional limit in terms of
hypergeometric functions \lerche\ , but in this example we were
 interested in showing
how the approximation works and become exact in the limit in a simple case. 
We have thus found perfect agreement with the results of \lust .

The metric on the $P^1$ is given by \metric . Let us restrict our 
attention to the region of the $P^1$ given by
$\Gamma = \{ z \in C \;\; / \;\; e^{-\pi (T+U)}<|z|<e^{\pi (T+U)} \} $. It
is not hard to check that $|j(\tau )|>>1 \;\;\; \forall \;\;\; z\in
\Gamma$. This justifies the use of the approximations of all the previous
sections and we write the metric as,
\eqn\newm{ds^2 = k (Im(\tau ))|\a |^{-1/2} \left| {dz \over z} \right|^2 }  
If $\tau$ were constant, this metric would describe a cylinder, so it is
natural to change coordinates given by the conformal map from the complex
plane to a cylinder of radius $2\pi$, namely, $z=e^w$ as it was suggested
in the previous section. Now we introduce $w=R_8 (y_1 + iy_2)$. In the
notation of the previous section we have,
$$ y_{X2} = \pi (R_9 + {2\over R_9}) ~~~~~  y_{9} = \pi (R_9 - {2\over
R_9}) ~~~~ y_{8} = - y_{9}  ~~~~ y_{X1} = - y_{X2} $$
Now we can compute $\tr $ with the following result,
$$ \tr = {R_8 \over 2\pi }\left[ y_{X2} - \half |y+y_9| -\half |y-y_9|
\right] $$
This is already looking like the structure in \fred\ .

Next we have to compute the volume of the F-theory sphere in this limit.
This is easily done using the metric at hand and integrating only on the
region between $-y_{X2}$ and $y_{X2}$. The result is given by,
$$ {\rm Vol}(P^1) = 8\pi^2 k R_8^2 |\alpha |^{-1/2} $$
using that ${\rm Vol}(P^1) = e^{\phi_{h8}} = {e^{\phi_{h10}}\over (R_8
R_9)^{1/2}}$ we can compute $k$ explicitly with the following result,
$$k = {1\over 8\pi^2} e^{\phi_{h10}}|\alpha |^{1/2}R_8^{-5/2}R_9^{-1/2}$$
The metric \newm\ that is in the Einstein frame is finally given by,
$$ ds^2 =  {1\over 8\pi^2} e^{\phi_{h10}}R_8^{-1/2}R_9^{-1/2}\tr (dy^2_1+
dy^2_2) $$
The next step is to go to the string frame, but for this we have to
consider the full 10 dimensional metric that is given in \metric .

We want to make explicit all the $R_8$ dependence, therefore let us define
$F(y) = {2\pi \over R_8} \tr $ and $y_2 = R_8 \theta $ so that $\theta \sim
\theta + 2\pi$. And the result is given by,
$$ds^2 ={ e^{\phi_h} \over R_9^{1/2}} F(y)^{1/2}(dy^2_1 + R_8^{-2}d\theta^2
)+ R_8^{-1/2}F(y)^{-1/2}(\eta_{\mu\nu}dx^{\mu }dx^{\nu })$$

At this point we should expect to be able to get some of the results of
section 2.1, in particular, we can compute the proper distances from
$y_9$ to $y_{X2}$ and from $-y_9$ to $y_9$.

Let us start with $y_9$ to $y_{X2}$,
$$\Phi_1 = \int^{y_{X2}}_{y_9}\sqrt{G_{99}(y)}dy = {\lambda_h^{1/2}  \over
R^{3/2} } $$

The second proper distance is given by,
$$\Phi_2 = \int^{y_9}_{-y_9}\sqrt{G_{99}(y)}dy = \lambda_h^{1/2}{(R^2-2)\over R^{3/2}}$$   
Remembering that in the map \map\ and \mep\ we are working with $E_8\times
E_8$ variables we find that we have been able to reproduce \properone\ and
\propertwo\ solely from F-theory.

Now we are ready to perform a T-duality in the $\theta$ direction and now
it is clear that this direction is going to be non compact in the limit
$R_8 \rightarrow \infty$, therefore we introduce $x^8$ and by some finite rescalings that do not
depend on $y_1$ we can write the metric as follows,
\eqn\metricfinal{ds^2 ={ e^{\phi_h} \over R_9^{1/2}}\left[ F(y_1)^{1/2}dy^2_1 + \gamma^2
F(y_1)^{-1/2}(\eta_{\mu\nu}dx^{\mu }dx^{\nu }) \right] }
where $\gamma$ is a generic constant that can be used to fix the range of
the only compact coordinate left. Note that this does not affect the proper
distances we computed earlier.

Finally we would like to bring the metric to the conformal gauge by
defining $F(y_1)^{1/2}dy_1 =\gamma dx_9$. Imposing that $y_1 = -y_{X2}$
implies $x = 0$  and that $y_1 = y_{X2}$ corresponds to $x = 2\pi$ we
can fix the integration constant and $\gamma$.

Now, if we define $x=x_1$ to correspond to $y = -y_9$ we get that,
$$x_1 = 2\pi {4 \over  (3 R_9^2 + 2)} $$
and this is exactly the expression found in \radii .

The metric now can be written as,
$$ ds^2 = {\lambda_h \over R^{1/2}}\gamma^2
F(y(x_9))^{-1/2}(\eta_{MN}dx^Mdx^N)$$
but it turns out that $F(y(x)) = \gamma^{2/3} z(x)^{2/3}$ and $\gamma =
R^{-3/2}(3R^2 + 2)$, therefore we get for the metric the following,
$$ds^2 = \lambda_h {(3R^2 + 2)^{5/3} \over
R^3}z(x)^{-1/3}(\eta_{MN}dx^Mdx^N)$$
this is again in agreement with \metricee .

Finally we want to compute the dual of the IIB coupling. According to
\cross , we only need to compute the dual radius, i.e., the radius of the
8-th direction after the T-duality. This radius can be read off from \metricfinal\
and it is given by ${\cal R} = {R^{1/4}\over \lambda^{1/2}}F(y)^{-1/4}R_8$.

Using \cross\ and that $e^{\phi_{\rm II}} = (\tr )^{-1} =
R_8^{-1}F(y)^{-1}$ and the expression of $F(y)$ in terms of $x$  we get the
following for the dual coupling,
$$ e^{\tilde \phi }= \lambda^{-1/2}R^{3/2}(3R^2 + 2)^{-5/6}z(x)^{-5/6}$$
that reproduces \edil . 

Therefore we see that we have been able to recover all aspects of the Type I'
description in 9 dimensions by taking the decompactification
limit of F-theory
from 8 dimensions.


\subsec{``Real Resolution of $E_8 \times E_8$''}

Up to now we have shown that the real $K3$ encoding the 9 dimensional
heterotic behavior consists of a Riemann surface of genus 10 and a
sphere. This was done by using tools of real algebraic geometry. It is the
aim of this part to show how the Real $K3$ mentioned before arises directly
from a full resolution of the $E_8\times E_8$ point by turning on 
``real'' Wilson
lines on the heterotic side. This means that we are turning on only the
Wilson lines in the 9-th direction.

In order to go ahead in spite of not having the explicit map between the
Wilson lines and the coefficients of the polynomials of the elliptic
equation we will take the following strategy. We will start with 
the point in the heterotic
$E_8\times E_8$ where all Wilson lines are zero. The map in this case is
known explicitly. The next step is a resolution from $E_8 \rightarrow E_7$
on both singularities, 
in this case the number of parameters is
still small and the deformations are uniquely determined. Beyond this
point the global analysis of the $K3$ is not possible at least with the
techniques we have at hand. Therefore we perform a local analysis of each
of the $E_7$ singularities. The small resolutions are achieved via the
invariant theory for Weyl groups as was done in \morrkatz . 

{\bf Global analysis }

Let us start with the $E_8 \times E_8$ point on the F-theory side. Using
$SL(2,C)$ we can set one $E_8$ singularity at $z=0$. This means using table
2 that,
\eqn\fyg{f(z)=a_4 z^4 + \cdots + a_8 z^8 \;\;\;\;\;\;\;\;\;\; g(z)=b_5 z^5 +
 \cdots + b_{12}z^{12} }  

It is also possible to set the other $E_8$ at $z=\infty$ and therefore we
should impose,
\eqn\fygs{f(z)=a_4 z^4  \;\;\;\;\;\;\;\;\;\; g(z)=b_5 z^5 + b_6 z^6 + b_{7}z^{7} }  
This is obtained by changing variables $z=1/w$ and using the global
rescaling invariance of the elliptic equation.   

Now we can use our last $SL(2,C)$ degree of freedom by a rescaling of $z$
in order to make $b_5 = b_7 = b$, and finally we use the global rescaling of
the elliptic equation \foot{The rescaling is given by $y\rightarrow \lambda^3
y$, $x\rightarrow \lambda^2 x$, $f\rightarrow \lambda^4 f$, $g\rightarrow
\lambda^6 g$ } to set $b = 1$. 

This gives us the $E_8\times E_8$ polynomials to be,
\eqn\fygt{f(z)=\alpha z^4  \;\;\;\;\;\;\;\;\;\; g(z)=z^5 (1 + \beta z + z^2)}

The resolution to $E_7 \times E_7$ can only affect the $f$ polynomial since
the condition on the order of $g$ is the same for $E_7$ as that for $E_8$. We
have to lower the order by one unit of the zeros of $f$ at the location of
the $E$ singularities. This gives us,
\eqn\fyg{f(z)=z^3 (a_1 + \alpha z + a_2 z^2) \;\;\;\;\;\;\;\;\;\; g(z)=z^5 (1 + \beta z + z^2) }

{\bf Local analysis } 

Any $ADE$ singularity on an elliptic $K3$ can be modeled locally by means
of an $ALE$ space. The $ALE$ spaces can be thought of as algebraic
varieties in $C^3$ given by the following equations,

{\centerline{
\vbox{\offinterlineskip
\hrule
\halign{&\vrule#&
\strut\quad\hfil#\quad\cr
height2pt&\omit&&\omit&\cr
& \hfill   Type   \hfill &&  \hfill  $ALE$ singularity  \hfill & \cr 
\noalign{\hrule}
height2pt&\omit&&\omit&\cr
& \hfill   $A_n$   \hfill &&  \hfill  $ xy+ z^{n+1} = 0 $  \hfill & \cr 
\noalign{\hrule}
height2pt&\omit&&\omit&\cr
& \hfill   $D_n$   \hfill &&  \hfill  $ x^2 + yz^2-z^{n+1} = 0 $  \hfill &  \cr 
\noalign{\hrule}
height2pt&\omit&&\omit&\cr
& \hfill   $E_6$   \hfill &&  \hfill  $ -x^2 - x y^2 +y^3 = 0 $  \hfill &  \cr 
\noalign{\hrule}
height2pt&\omit&&\omit&\cr
& \hfill   $E_7$   \hfill &&  \hfill  $ -x^2 - y^3 +16 y z^3 = 0 $  \hfill &  \cr 
\noalign{\hrule}
height2pt&\omit&&\omit&\cr
& \hfill   $E_8$   \hfill &&  \hfill  $ -x^2 + y^3 - z^5  = 0 $  \hfill & \cr 
height2pt&\omit&&\omit&\cr}
\hrule}
}}
\noindent{\ninepoint\sl \baselineskip=8pt {\bf Table 3}: {\sl Equations in
$C^3$ defining $ALE$ spaces with the singularity located at the origin.}}
\bigskip

The key idea in the resolution using the Cartan
generators is to realize that the coefficients in the deforming polynomials
should be generators of the algebra of polynomials invariant under the
Weyl group of the corresponding root system. 

As an example we will consider the $A_n$ case. Here we will follow \morrkatz\
closely and the details of the other cases can be found there.

Let $e_i \;\; i=1, \ldots , n+1$ be orthonormal basis of $R^{n+1}$, then
the root system of $A_n$ is generated by the following set of simple roots
$v_i = e_i - e_{i+1}$. We define a set of distinguished functionals $t_1,
\ldots , t_{n+1}$ given by,
\eqn\tis{t_i = - v^*_{i-1}+v^*_i  \;\;\;\;\;\;\;\; (1\le i \le n+1)}
where $v^*_i$ are the dual of the root basis.

In this case we have to consider the Weyl group of $A_n$ that is nothing
but the symmetric or permutation group $S_{n+1}$. Therefore the basis of
invariant polynomials of $t_i$ can be obtained by considering the
coefficients of the following auxiliary polynomial,
\eqn\aux{P(U) = \prod^{n+1}_{i=1}(U+t_i) = U^{n+1} + \sum^{n+1}_{i=1}s_i U^{n+1-i}} 

This is the way to define the i-th symmetric functions $s_i = s_i (t_1
\ldots t_{n+1})$ that are in this case the basis we were looking for.

Now the resolution of the $A_{n}$ singularity can be achieved by the
following deformations of the expression in Table 3:
\eqn\defor{ xy + P(z) = 0 } 

Going back to our case, we are interested in the resolutions of $E_7$. It
turns out that the deformation takes the form \morrkatz\ ,
\eqn\resolve{-y^2-x^3+16xz^3+\ep_2 x^2z+\ep_6 x^2 + \ep_8 xz+ \ep_{10}z^2 + \ep_{12}x+\ep_{14}z+\ep_{18}=0}
where $\ep_i$ are the basis of the algebra of polynomials invariant under the
Weyl group action of $E_7$ and are given as functions of $s_1 , \ldots ,
s_7$. For the explicit form of the $\ep_i$ see Appendix 2 of \morrkatz .

The $E_7$ $ALE$ singularity has the same structure as an elliptic equation,
therefore we can easily  read off the $f(z)$ and $g(z)$ after the appropriate
changes of variables. The answer is given by,
\eqn\fff{f(z)=-16 z^3-{1\over 2}\ep^2_2z^2-({2\over 3}\ep_2\ep_6+\ep_8)z
-{1\over 3}\ep_6^2-\ep_{12}}
\eqn\ggg{ \matrix{\hfill g(z)= {16\over 3}\ep_2z^4+({2\over 27}\ep^3_2+ 
{16\over 3}
\ep_6)z^3 + (\ep_{10}+{2\over 9}\ep_6\ep_2^2+{1\over 3}\ep_8\ep_2)z^2 + 
 \hfill  \linesp 
                      ({2\over 9}\ep^2_6 \ep_2+\ep_{14}+
                      {1\over 3}\ep_{12}\ep_2+{1\over
3}\ep_8\ep_6)z+({2\over 27}\ep^3_6+{1\over 3}\ep_{12}\ep_6+\ep_{18})                   
                                  \hfill } }

Our next step is clearly to perform a full resolution of the $E_7$
singularity. We will see with an example that choosing a generic ``real''
Wilson line, i.e., real $t_i$'s, we get that the surface has
three holes and no spheres. Together with the global analysis this means
that the total smooth space consists of a Riemann surface of genus 10 and
a sphere.

Let us try now to consider examples where we can illustrate all the
transitions between the different regimes discussed in section 7 and
the different embeddings of some groups in the real $K3$. 
Let us consider the following family of deformations $(t_1,\ldots
t_7) = (-1,1,\ldots ,t)$. This correspond to the following elliptic
equation,
\eqn\family{f(z)=-16 z^3 -{1\over 3}(784+592t+136t^2)z^2-24(27+t)(t+2)^2 z-
27(t+2)^4}
\eqn\gfamily{ \matrix{\hfill g(z)=({448\over 3}+{128\over 3}t + 16t^2)z^4+
{1 \over
9}({49088\over 3} + 18304t+7184 t^2 + {2752 \over 3}t^3)z^3  \hfill  \linesp 
 + 8(25t^2 +
130t+196)(t+2)^2z^2+72(2t+7)(t+2)^4z +54 (t+2)^6 \hfill } }
The discriminant is given by,
\eqn\discr{ \matrix{\hfill  \Delta = z^5(-16384z^4+a(t)z^3+b(t)z^2+2304
(35t^4+182t^3+119t^2-372t-36)(t+2)^2z  \hfill  \linesp               
+ 41472(t+3)(t-1)t(t+2)^4)\hfill} } 

It is easy to see that there is always an $A_4$ singularity at $z=0$ but it
becomes $A_5$ at $t=0$ and $t=1$. Less evident is that at $t=-1$ we get
instead of $A_5$ an extra $A_1$ singularity. Finally we see that at $t=2$ we get a $D_5$ singularity. 

The transitions are shown in Figures 14 and 15 with the corresponding Dynkin
Diagram of $E_7$ showing the unbroken piece for each value of $t$. 

\bigskip
\centerline{\epsfxsize=0.7\hsize\epsfbox{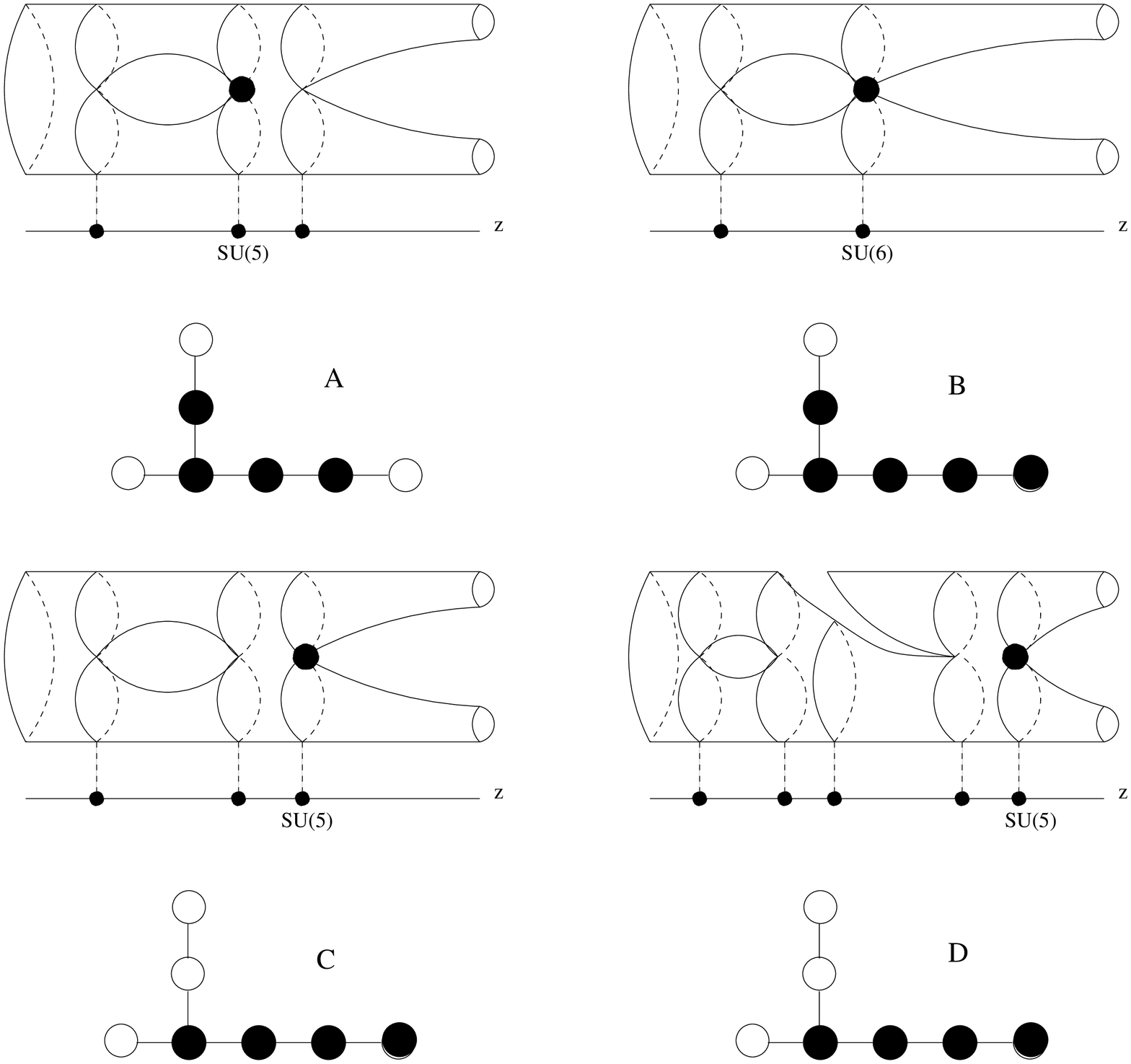}}
\noindent{\ninepoint\sl \baselineskip=8pt {\bf Figure 14}: {\sl One
parameter family of Real resolutions of $E_7$ showing the transition when two branes become real
(C-D) and embeddings of $SU(5)$ and $SU(6)$ in $E_7$}}
\bigskip

In Figure 14A we see that the $SU(5)$ is embedded in the way corresponding
to the non-trivial value for the $Z_2$ Wilson line. In Figure 14B the brane
at the right joined the $SU(5)$ to give an $SU(6)$ gauge group. In Figure
14C one of the original branes in the $SU(5)$ separates from the $SU(6)$
giving again $SU(5)$ but with a different embedding. In Figure 14D we see
the result after the two branes that were complex land on the real axis
without modifying the gauge group. In Figure 15A one of the new branes
joins the $SU(5)$ group to give an $SU(6)$ with the embedding corresponding
to the trivial choice of $Z_2$ Wilson line. In Figure 15B one of the branes
separates to the left leaving an $SU(5)$ again with the trivial $Z_2$
Wilson line. In Figure 15C the two branes to the left of the ``Hook'' join
to give a extra $SU(2)$. Finally we see that when the $SU(5)$ branes, the
``Hook'' and one of the two branes of the left side join we get an $SO(10)$
gauge group. 

Even though this family cannot show it, if we start from Figure 14B it is
possible to get the final brane on the left to join the group in order to
give an $SU(7)$. This $SU(7)$ is one of the groups that requires one
``extra'' brane from the view point of Type I' since the Type I'
description of $E_7$ involves only 6 D8-branes at the orientifold with
infinite coupling.
  
\bigskip
\centerline{\epsfxsize=0.7\hsize\epsfbox{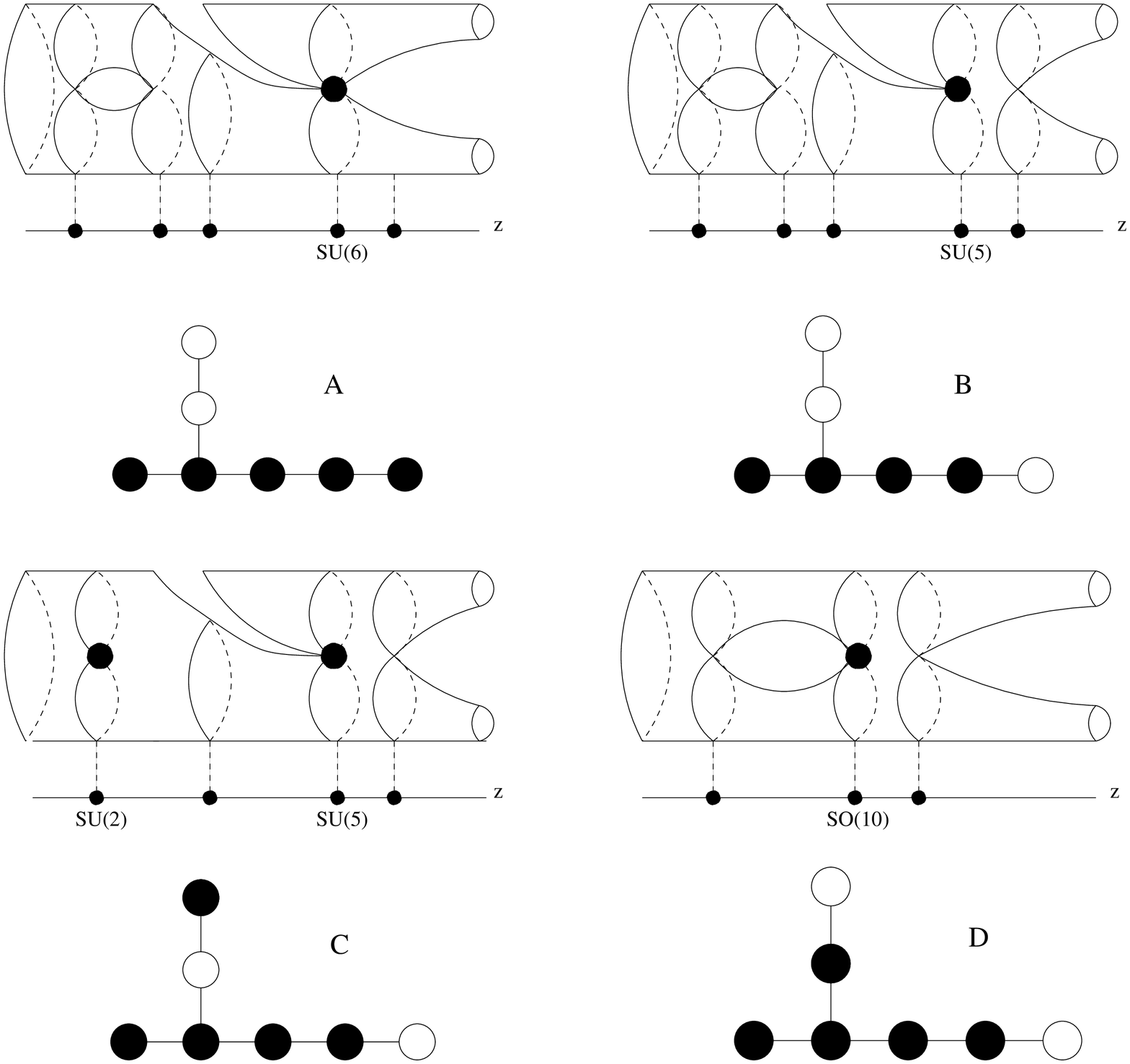}}
\noindent{\ninepoint\sl \baselineskip=8pt {\bf Figure 15}: {\sl Embeddings
of $SU(6)$, $SU(5)$, $SU(5)\times SU(2)$ and $SO(10)$ in $E_7$.}}
\bigskip

{\bf Acknowledgements:}

We would like to thank D. Allcock, O. Bergman,
A. Chari, K. Hori, J. Maldacena, D. Morrison,
T. Pantev, S. Sinha,
A. Strominger and B. Zwiebach for valuable discussions.

The research of F.C. was supported by a fellowship from CONICIT and 
Universidad Sim\'{o}n Bol\'{\i}var. The research of C.V. is partially supported by 
NSF grant PHY-98-02709.

\appendix{A}{Map between the heterotic theories at the $E_8\times E_8$ point}

In section 2.1 we described the $E_8\times E_8$ point of the heterotic
$SO(32)$ as an enhancement of the $SO(14)\times SO(14)$ point. This
description led to a Type I' scenario with infinite coupling at both
orientifolds. However, as mentioned in section 2.1 this point is most
naturally described in term of heterotic $E_8\times E_8$ variables and the
map between them is the aim of this appendix.

The moduli in the $E_8\times E_8$ theory consists of the 9 dimensional
radius $R_{E8}$ and the 10 dimensional coupling $\lambda_{E8}$. The Wilson
lines are all zero. On the other hand, the moduli in the $SO(32)$ theory
consists of the radius $R_{SO}$, the coupling $\lambda_{SO}$ and the Wilson
lines are $\theta_I = (0^7, \half -\lambda , \lambda , \half^7)$ where
$\lambda$ and $R_{SO}$ are related by 
\eqn\radiwil{ R^2_{SO} = 2 \lambda (\half -\lambda)} 
The way to map the two theories at more complicated points than the one we
are discussing is to find the $SO(17,1)$ rotation that connects
them. For an example interpolating between the two theories see \ginsparg .

Here however, due to the fact that $R_{SO}$ should be given only in terms
of $R_{E8}$, a single relation will be enough. In particular, we can
compare the masses of the BPS states responsible for the extra $SU(2)$
enhancement of symmetry in both theories and then get the map.

On the $E_8\times E_8$ theory, the $SU(2)$ is achieved at the critical
radius and therefore the new states should be the usual states of winding
and momentum number $\pm 1$ and neutral with respect to the $E_8\times
E_8$.

The mass is given by \foot{Setting $\alpha'=2$}   
$$ M^2_{BPS} = P^2_R = \left( {1\over R_{E8}} - {R_{E8}\over 2} \right)^2 $$

As we expect, the mass is zero at the critical radius for zero Wilson lines
$R^2_{E8} = 2$

On the $SO(32)$ theory, the states becoming massless at the $SU(2)$ point
are just off diagonal vector bosons of the original $SO(32)$ group with
charges (or root vectors) $P = \pm (e_8 - e_9)$ where $e_i$ are orthonormal
vectors in $R^{16}$ where the root system lives. This states have no
winding or momentum. Using the mass formula,
$$ M^2_{BPS} = R_R^2 = \left( {\theta_I P_I  \over R_{SO} } \right)^2 =
{4\over R^2_{SO}}\left( \lambda - {1\over 4}\right)^2 $$
Here we are summing over $I$. We see that for $\lambda = {1\over 4}$ these
states are massless as we expect.

Finally, using \radiwil\ in order to express everything in terms of
$R_{SO}$ and then equating the two masses we get,
$$ R^2_{SO} = {R^2_{E8} \over 4(1+R^2_{E8})^2}  $$

Now it is possible to find the relation between the two coupling constants
in 10 dimensions. This is achieved by equating the couplings in 9 dimensions
and using that in $S^1$ compactifications $\lambda^2_{9} = {\lambda^2_{10}\over
R}$.

The result given in terms only of $R_{E8}$ is,
$$ \lambda_{E8} = (R_{E8}^2 + 2)^{1/2}\lambda_{SO}$$
%

\appendix{B}{Description of ${\hat E}_2$}

The aim of this appendix is to show the computation that was used in 
section 7 for the
value of $\tr$ at the location of the three branes left inside the minimal
$E$ cycle.

It was shown in \bartetsen\ \yamada\ that the configurations of branes
given by $\hat{E}_n$ $(1 \le n \le 8)$, $\hat{\tilde{E}}_0,
\hat{\tilde{E}}_1$ can be properly isolated. This means that there exists
polynomials\foot{Here we will use the definition of \bartetsen\ for $f(z)$
and $g(z)$ where the
discriminant is $\Delta = - f^3 + g^2$ and not $\Delta = 4 f^3 + 27 g^2$ as
we were using in the rest of this work.} $f(z)$ and $g(z)$ such that the
relevant  branes are around $z=0$ and the others are at $z=\infty$.

Let us consider the two parameter family of polynomials giving the properly 
isolated
$\hat{E}_2$, $\hat{E}_1$, $\hat{\tilde{E}}_1$ and $\hat{\tilde{E}}_0$.

The family is given by \bartetsen\ \yamada ,
\eqn\etwof{f(z) = z^4 + z^3 + {1\over 4 }s z^2 + {1\over 16}t z}
\eqn\etwog{g(z) = z^6 + {3\over 2}z^5 + {3\over 8}(1+s)z^4 + {1\over 32}(3t
+ 6 s -1)z^3 + {3\over 128}(1-2s  + s^2 + 2 t)z^2}
\eqn\etwogtw{ -{3\over
256}(s-1)(s-t-1)z +{(14+18s^2-2s^3+12t+3t^2-6s(5+2t))\over 2048}}

For generic $(t,s)$ we have an $\hat{E}_2$ configuration.    
The discriminant given by $\Delta = - f^3 + g^2$, is a polynomial of degree
five and the coefficient of $z^5$ term is given by,
\eqn\coeff{(-2+2s-t)(3-4s+s^2+t)}
revealing the existence of two branches. It is also important to mention
that the discriminant has an overall factor of $(s-1)$ signaling that $s=1$
is at infinite distance away in the moduli space. In figure 16A we show
$K3_R$ around the five branes of $\hat{E}_2$.

\bigskip
\centerline{\epsfxsize=0.90\hsize\epsfbox{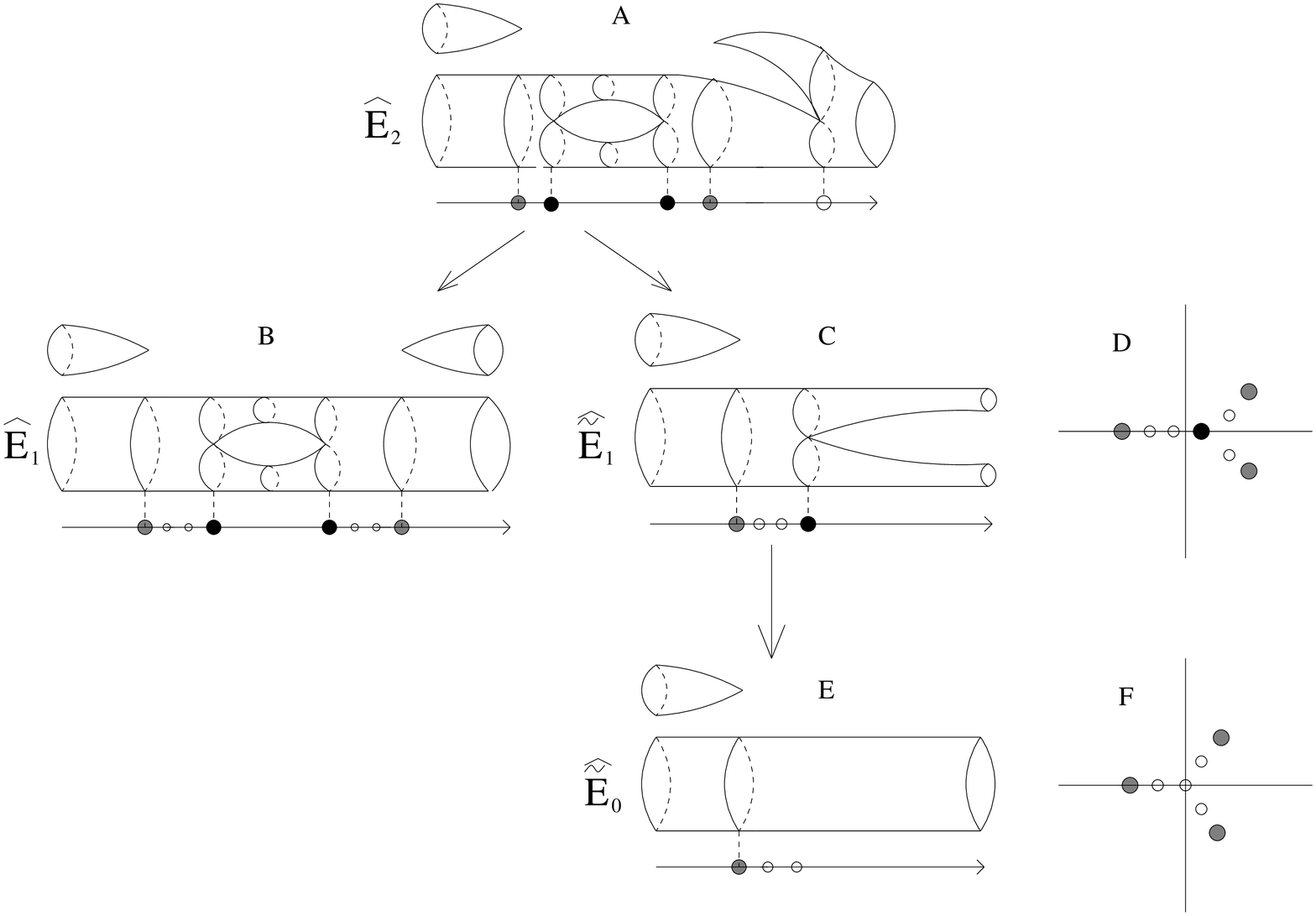}}
\noindent{\ninepoint\sl \baselineskip=8pt {\bf Figure 16}: {\sl $K3_R$
in the vicinity of brane configurations $\hat{E}_2$, 
$\hat{E}_1$, $\hat{\tilde{E}}_1$
and $\hat{\tilde{E}}_0$. Besides the $\hat{\tilde{E}}_1$
and $\hat{\tilde{E}}_0$ also the locations of the branes and
zeros of $f(z)$ (small white dots) are shown in the complex $z$-plane.}}
\bigskip

The first branch $t=2(s-1)$ gives the $\hat{E}_1$ configuration. The region
in the $s$ line we are interested in is given by $1 < s < -{1\over
2}+\sqrt{3}$. The lower bound is the one discussed before, and the upper
bound is an $SU(2)$ wall. In figure 16B we show how $K3_R$ looks like
around the four real branes of $\hat{E}_1$ after the $A$ brane of
$\hat{E}_2$ has escaped to infinity. We see that the two branes that form the
$SU(2)$ are $C1$ and $C2$ of section 7. It is also possible to check that
$f(z)$ has four real zeros indicated by small white dots in figure 16B.   

The second branch $t= -3+4s - s^2 = (s-1)(3-s)$ gives $\hat{\tilde{E}}_1$.
In this case, the brane $A$ and $B$ come off the real line and the branes
$C2$ escapes to infinity. The resulting $K3_R$ around the remaining two
branes on the real axis is shown in figure 16C. 
The region of $s$ is given by $1 < s < {3\over 2}$. The upper bound
comes from the $z^4$ coefficient of the discriminant that vanishes at
$s={3\over 2}$ signaling that the $C1$ brane moves all the way to infinity
leaving us with a $\hat{\tilde{E}}_0$ configuration.

Finally the $\hat{\tilde{E}}_0$ configuration has no parameters left since
we have set $s={3\over 2}$. It is possible to see that upon a shift in $z$
and a rescaling, the discriminant (a cubic)
for this configuration is given by $z^3 - 1$ and $f(z) = z(z^3 - {8\over
9})$ (See figure 16D). This shows the statement made in section 7 about 
the zeros of $f(z)$ being enclosed by the last $E$ cycle. 

The fact that the $\hat{\tilde{E}}_0$ has a $Z_3$ symmetry tells us that in
the limit to 9 dimensions in which the $E$ cycle shrinks to a point this
configuration will collapse to $z=0$. The
effect of the limit and of the remaining branes can only affect this
configuration in a global $z$ rescaling or a rescaling of $f(z)$ and
$g(z)$ given before. But non of them affects the value of $j(\tau )$ at
$z=0$ that always vanishes since $f(z=0)=0$ and $\Delta (z=0)\neq 0$. This
implies that 
$$\tau |_{(z=0)} = {\sqrt{3}\over 2} + i {1\over 2} $$    

{}From this we can get the result we were looking for the computation of
the boundary values of $\tr$ in the Case II and Case III of the
quantitative part of section 7, namely, $\tr = {1\over 2}$.

\listrefs
\end